\documentclass[%
reprint,
preprintnumbers,
amsmath,amssymb,longbibliography,
aps
prd
]{revtex4-1}
\pdfoutput=1
\usepackage{graphicx}
\usepackage[utf8]{inputenc}
\usepackage{flushend}
\usepackage{dcolumn}
\usepackage{bm}
\usepackage{soul}
\usepackage{balance}
\usepackage[normalem]{ulem}
\usepackage[colorlinks = true,
            linkcolor = blue,
            urlcolor  = blue,
            citecolor = blue,
            anchorcolor = blue]{hyperref}
\usepackage{verbatim,subfigure}
\usepackage{color,ulem}
\usepackage[english]{babel}
\usepackage{MnSymbol,wasysym}
\usepackage[utf8]{inputenc}

\input Starburst.fd
\newcommand*\initfamily{\usefont{U}{Starburst}{xl}{n}}\initfamily 

\newcommand{\beq}{\begin{eqnarray}}
\newcommand{\eeq}{\end{eqnarray}}
\usepackage{amsmath}
\usepackage{tikz}
\usetikzlibrary{decorations.pathmorphing}
\usetikzlibrary{shapes.misc}
\tikzset{cross/.style={cross out, draw=black, minimum size=8*(#1-\pgflinewidth), inner sep=0pt, outer sep=0pt},
cross/.default={1pt}}
\usetikzlibrary{patterns,math}
\definecolor{applegreen}{rgb}{0.55, 0.71, 0.0}
\usepackage{contour}
\usepackage{xparse}

\definecolor{darkviolet}{rgb}{0.58, 0.0, 0.83}
\definecolor{mygreen}{rgb}{0.0, 0.5, 0.0}

\begin{document}
\preprint{\texttt{IFT-UAM/CSIC-26-118, APCTP Pre2026 - 012}}

%
\title{Unfolded Krylov complexity: universal chaotic dynamics without false positives}

\author{Johanna Erdmenger,$^{1}$ Kyoung-Bum Huh,$^{2}$ Hyun-Sik Jeong$^{3,4}$ and Juan F. Pedraza$^{5}$}
\affiliation{\vspace{4pt} $^{1}$Institute for Theoretical Physics and Astrophysics and W\"urzburg-Dresden
Cluster of Excellence ctd.qmat, Julius-Maximilians-Universit\"at W\"urzburg,
Am Hubland, DE-97074 W\"urzburg, Germany}
\affiliation{$^{2}$Wilczek Quantum Center, School of Physics and Astronomy, Shanghai Jiao Tong University, Shanghai 200240, China}
\affiliation{$^{3}$Asia Pacific Center for Theoretical Physics, Pohang 37673, Korea}
\affiliation{$^{4}$Department of Physics, Pohang University of Science and Technology, Pohang 37673, Korea}
\affiliation{$^{5}$Instituto de F\'isica Te\'orica UAM/CSIC, Calle Nicol\'as Cabrera 13-15, 28049 Madrid, Spain\vspace{-1.6mm}}

\begin{abstract}
A central challenge in diagnosing quantum chaos is to distinguish genuine many-body scrambling from kinematic effects of the spectrum. Krylov state complexity, or spread complexity, has emerged as a powerful diagnostic, with its characteristic growth, peak, and relaxation often taken as signatures of chaos. However, previous work has shown that this criterion can give false positives: saddle-dominated integrable systems may display prominent peaks even without random-matrix level correlations. We argue, based on complementary numerical and analytical evidence, that this ambiguity can be resolved by unfolding the spectrum prior to constructing the ensuing Krylov dynamics. By removing the non-universal smooth density of states while retaining microscopic spectral correlations, unfolding suppresses spurious peaks in integrable systems while preserving the universal spectral signatures of chaotic systems. Analytically, the formulation of the Lanczos iteration in terms of orthogonal polynomials yields an exact complexity kernel with a robust near-diagonal structure whose fine-grained features reflect the underlying spectral correlations. Moreover, for the logarithmic model, unfolding can be performed exactly, mapping the spectrum to a uniform lattice and yielding an analytic spread complexity that removes the false-positive peak. These findings establish unfolded Krylov complexity as a more reliable probe of genuine many-body scrambling.
\end{abstract}

\maketitle
\section{Introduction}

Recent developments at the interface of quantum information, many-body physics, and holography have sharpened the search for reliable diagnostics of quantum chaos~\cite{Shenker:2013pqa,Shenker:2014cwa,PhysRevD.94.126010,Cotler:2016fpe,deBoer:2017xdk,Stanford:2019vob,Altland:2026tog}. A central challenge in this effort is to distinguish genuine many-body scrambling from dynamics that merely mimic chaos. Motivated by the proposal that black holes are the fastest scramblers in nature~\cite{Sekino:2008he}, out-of-time-order correlators (OTOCs)~\cite{larkin1969quasiclassical,berman1978condition} have become standard probes of operator growth and information scrambling. In chaotic systems, OTOCs display early-time exponential growth governed by a quantum Lyapunov exponent, $\lambda_L$, which under general assumptions obeys the Maldacena-Shenker-Stanford (MSS) bound, $\lambda_L\leq 2\pi k_B T/\hbar$~\cite{Maldacena:2015waa}. This regime has consequently become a widely used signature of quantum chaos.

However, exponential growth alone does not imply genuine many-body scrambling. Unstable saddles in semiclassical phase space can generate local exponential sensitivity even in integrable systems. This occurs, for example, in the Lipkin-Meshkov-Glick (LMG) model~\cite{Lipkin:1964yk, Xu:2019lhc} and the inverted harmonic oscillator (IHO)~\cite{Hashimoto:2020xfr}, despite the absence of universal spectral correlations associated with random matrix theory (RMT)~\cite{mehta2004random}. Such \emph{saddle-dominated scrambling} can therefore produce false po\-si\-tives. The distinction from genuine many-body chaos is particularly sharp at the spectral level: the Bohigas-Giannoni-Schmit (BGS) conjecture~\cite{Bohigas:1983er,Bohigas:1984aa,Guhr:1997ve,Bohigas} associates quantum chaos with Wigner-Dyson correlations, while the Berry-Tabor conjecture~\cite{Berry1977Level} predicts Poisson statistics for integrable systems. A local phase-space instabi\-lity can thus mimic early-time growth without the level repulsion and spectral rigidity characteristic of RMT.

Krylov complexity provides a complementary diagnostic of quantum dynamics~\cite{Parker:2018yvk,Balasubramanian:2022tpr,Caputa:2024vrn}. It has recently found extensive applications in many-body quantum dynamics and holography; for comprehensive reviews, see \cite{Nandy:2024evd,Rabinovici:2025otw,Jeong:2026gdc}. Rather than probing correlation functions, it characterizes how an operator or state spreads through the Krylov basis generated by repeated action of the Liouvillian or Hamiltonian under the Lanczos algorithm~\cite{Lanczos:1950zz}. Krylov operator complexity was introduced as a probe of operator growth~\cite{Parker:2018yvk}, while its state counterpart, spread complexity, tracks the spreading of states~\cite{Balasubramanian:2022tpr}. However, in saddle-dominated systems, both notions can produce false positives, exhibiting chaos-like behavior despite integrability: operator Krylov complexity may display the characteristic exponential growth~\cite{Bhattacharjee:2022vlt}, while spread complexity can develop chaos-like peaks before reaching saturation~\cite{Huh:2023jxt}. Thus, neither feature alone constitutes a reliable diagnostic of RMT-type spectral chaos.

For operator Krylov complexity, Ref.~\cite{PhysRevLett.134.050402} addresses this issue through a multiseed construction, whose late-time plateau distinguishes chaotic from integrable Hamiltonians. We instead focus on spread complexity, whose dependence on the Hamiltonian spectrum and the spectral weights of the initial state makes it particularly sensitive to spectral correlations. In chaotic systems, its characteristic growth-peak-saturation structure mirrors the dip-ramp-plateau structure of the spectral form factor (SFF)~\cite{Balasubramanian:2022tpr,Erdmenger:2023wjg}.
Specifically, analyzing transition probabilities into Krylov states reveals that the spread complexity peak in chaotic systems is associated with fine-grained spectral correlations~\cite{Erdmenger:2023wjg}. These probabilities inherit the SFF ramp structure, driving a peak before saturation. This persistent ramp serves as a key signature of spectral rigidity, a hallmark of quantum chaos. In contrast, uncorrelated spectra lack a spectral ramp, approaching saturation monotonically. 

Yet, saddle-dominated integrable systems challenge this classification by displaying a similar peak-relaxation structure, even though local phase-space instabilities near an unstable saddle do not generate the RMT-type level repulsion predicted by the BGS conjecture. This tension raises a natural question: which spectral features are actually responsible for such peak-relaxation dynamics in spread complexity? The Lanczos coefficients are determined by moments of the spectral measure and are therefore sensitive both to microscopic level correlations and to the smooth, non-universal density of states. Strong variations in the latter, for instance near spectral edges or saddle-induced level accumulation, can modify these moments and, in turn, generate nontrivial Krylov dynamics without RMT correlations. False positives may therefore originate from the smooth spectral density rather than its fluctuating component.

This observation points naturally to spectral unfolding, which separates the smooth background density of states from its microscopic fluctuations. Given energy levels $E_i$, the unfolded levels are defined as $\xi_i=\bar{\eta}(E_i)$, where $\bar{\eta}(E)$ is the smooth cumulative density, representing the average number of levels below energy $E$. The resulting spectrum has unit mean level spacing, with macroscopic variations in the mean density of states $\bar{\rho}(E)$ removed. In spectral studies of quantum chaos, unfolding is standard for quantities such as level spacing distributions and spectral form factors~\cite{Bohigas1989Random,Guhr1999Random,Gomez2002Spectral,Morales2011Spectral}. Here, we apply the same principle to spread complexity by constructing the Krylov dynamics from the unfolded levels $\{\xi_i\}$. We refer to the resulting quantity as \emph{unfolded Krylov complexity}. Importantly, unfolding removes the smooth one-point density while retaining the local fluctuations underlying RMT universality. Unfolded Krylov complexity therefore isolates the influence of microscopic spectral correlations from that of macroscopic spectral structure. We can then ask whether the characteristic signatures of chaos persist once the non-universal smooth density is removed.

We address this question by first considering benchmark systems with well-understood spectral statistics, including RMT ensembles, quantum billiards, the Heisenberg XXZ chain, and the Sachdev-Ye-Kitaev (SYK) model. In each case, unfolding preserves the distinction between `generic' RMT-universal and integrable behavior, while retaining the characteristic peak-relaxation profile and, where applicable, the Dyson-class hierarchy. It also aligns late-time saturation with the Heisenberg scale of the unfolded spectral form factor, $t\simeq 2\pi$. We then turn to non-generic systems in which conventional spread complexity yields false positives. As a clean toy example, we first consider the logarithmic spectrum motivated in Ref.~\cite{Das:2023yfj}, which isolates the effect of a non-uniform density of states, before examining the physical saddle-dominated IHO and LMG models. In all three cases, unfolding suppresses the spurious peak and replaces it with oscillatory behavior, supporting the conclusion that the false signal originates from the smooth spectral density rather than from microscopic level correlations.

To clarify analytically how spectral information enters Krylov dynamics, we exploit the equivalence, via Favard's theorem, between the Lanczos recursion for a finite discrete spectral measure and the three-term recurrence of orthogonal polynomials. This allows us to express the Krylov amplitudes as polynomial wavefunctions and the complexity in terms of an exact two-energy kernel. Using the Christoffel-Darboux structure, we obtain an analytic expression valid for arbitrary finite discrete spectra. The resulting kernel exhibits a robust banded structure around the principal diagonal, while its finer off-diagonal features reflect the underlying spectral correlations. This framework makes explicit how spectral statistics are imprinted on spread complexity and sets the stage for an analytic treatment of particular systems.

The logarithmic model provides a transparent realization of this mechanism. Its exponentially growing smooth density of states produces both a spurious ramp in the SFF and a pronounced peak in spread complexity, despite the absence of RMT correlations. Here, unfolding can be performed exactly, mapping the spectrum onto a uniform lattice. The corresponding orthogonal polynomials are discrete Chebyshev polynomials, allowing the unfolded Lanczos coefficients and spread complexity to be obtained analytically. The resulting expression agrees with the numerics and contains no false-positive peak, demonstrating that removing the smooth density of states eliminates the apparent chaotic signature.

Taken together, our numerical and analytical results support spectral unfolding as a way to sharpen spread complexity as a diagnostic of quantum chaos. By removing the smooth, non-universal density of states, unfolding suppresses false-positive peaks while preserving signatures associated with RMT universality. Unfolded Krylov complexity thus provides a cleaner distinction between microscopic spectral correlations characteristic of chaos and saddle-driven dynamics that can mimic them.

The remainder of this paper is organized as follows. In Sec.~\ref{sec2}, we review spectral unfolding and spread complexity. In Sec.~\ref{sec3}, we study unfolded Krylov complexity in RMT ensembles and representative generic systems, while Sec.~\ref{sec4} focuses on non-generic systems ex\-hi\-biting false-positive signatures of chaos. In Sec.~\ref{sec5}, we develop the orthogonal-polynomial framework and derive the complexity kernel for general discrete spectra. In Sec.~\ref{sec6}, we specialize this framework to unfolded spectra, derive the associated discrete Chebyshev structure, and apply it to the logarithmic model. We conclude in Sec.~\ref{sec7} with a short discussion and outlook. Some technical details are relegated to the appendices.

\vspace{2mm}
\paragraph*{Note added.}
While this manuscript was being finalized, the independent preprint~\cite{Basu:2026gvl} appeared, deve\-loping a closely related analytical approach to spread com\-plexity based on orthogonal polynomials and unfolded spectral statistics. The two analyses are consistent where they overlap, while our work focuses on using spectral unfolding to distinguish genuine many-body quantum chaos from saddle-dominated integrable systems.

\section{Preliminaries}\label{sec2}
\setcounter{paragraph}{0}

We begin by reviewing spectral unfolding and spread complexity, emphasizing how the latter can be constructed directly from the unfolded spectrum.

\subsection{Unfolding procedure}

In spectral studies of quantum chaos, random matrix theory provides a framework for characterizing universal spectral correlations~\cite{Bohigas:1983er,Berry1985-mx,Muller:2004nb}. A standard spectral diagnostic is the nearest-neighbor level-spacing distribution $P(s)$, which distinguishes generic integrable systems obeying Poisson statistics~\cite{Berry1977Level} from chaotic systems exhibiting Wigner-Dyson statistics~\cite{Bohigas:1983er,Bohigas:1984aa,Guhr:1997ve,Bohigas}.

A raw energy spectrum generally contains both a smooth, non-universal density of states and microscopic level fluctuations. Since RMT universality concerns the latter, these contributions must be cleanly separated before comparing spectral statistics. Otherwise, variations in the local mean spacing can obscure level repulsion and distort the inferred statistics. Explicitly, the density of states can be decomposed as $\rho(E)=\bar{\rho}(E)+\delta\rho(E)$, where the smooth component $\bar{\rho}(E)$ captures macroscopic, system-specific features, such as the leading Weyl terms associated with area, perimeter, and curvature in billiards, or the extensive thermodynamic density of states in many-body systems. The fluctuating component $\delta\rho(E)$, in contrast, encodes the microscopic spectral correlations underlying RMT universality~\cite{Gutzwiller1990,PhysRevC.3.117,PhysRevE.63.026204,PhysRevE.84.016203}.

Spectral unfolding removes the smooth spectral variation by mapping the original eigenvalues onto a scale with unit local mean spacing~\cite{Bohigas1989Random}. Denoting the smooth density of states by $\bar{\rho}(E)$, the corresponding cumulative density is
\begin{align}
    \bar{\eta}(E_i)
    =
    \int_{-\infty}^{E_i}
    \bar{\rho}(E)\,\mathrm{d}E\,,
\end{align}
and the unfolded eigenvalues are defined by
\begin{align}
    \xi_i=\bar{\eta}(E_i)\,.
\end{align}
The resulting spectrum has unit mean spacing, so that the normalized nearest-neighbor spacings are simply $s_i=\xi_{i+1}-\xi_i$. In practice, the smooth density $\bar{\rho}(E)$, or equivalently $\bar{\eta}(E)$, must generally be estimated numerically, for instance through a polynomial fit \footnote{The extracted fluctuations can depend on the smoothing prescription, including the polynomial degree, motivating alternative unfolding procedures~\cite{Guhr1999Semiclassical,Gomez2002Misleading,Morales2011Improved,ABULMAGD2014185,10.1119/1.2198883}.}.

\subsection{Spread complexity from unfolded spectra}

We now review spread complexity in a form that makes its dependence on the spectrum explicit. Starting from a normalized initial state $|\psi(0)\rangle\equiv|K_0\rangle$, the Lanczos algorithm~\cite{Lanczos:1950zz} generates an orthonormal Krylov basis $\{|K_n\rangle\}$ and coefficients $\{a_n,b_n\}$ satisfying~\cite{Lanczos:1950zz,RecursionBook}
\begin{align}
H|K_n\rangle
=
a_n|K_n\rangle
+
b_{n+1}|K_{n+1}\rangle
+
b_n|K_{n-1}\rangle\,.
\end{align}
Since the Lanczos construction depends only on the spectral measure associated with the Hamiltonian $H$ and the initial state, it is convenient to work directly in the energy eigenbasis, where
\begin{align}
D_E=\operatorname{diag}(E_1,E_2,\ldots,E_{N_{\rm max}})\,.
\end{align}
The same Lanczos recursion then takes the form
\begin{align}\label{D_reculsion}
D_E|K_n\rangle
=
a_n|K_n\rangle
+
b_{n+1}|K_{n+1}\rangle
+
b_n|K_{n-1}\rangle\,,
\end{align}
which is also advantageous numerically~\cite{parlett1998symmetric,Hashimoto:2023swv}.

This representation makes the unfolded construction immediate: we replace the raw eigenvalues $E_i$ by the unfolded levels $\xi_i$ and define
\begin{align}
D_\xi
=
\operatorname{diag}(\xi_1,\xi_2,\ldots,\xi_{N_{\rm max}})\,.
\end{align}
Applying the Lanczos algorithm to $D_\xi$ defines the unfolded Krylov dynamics considered throughout this work. Expanding the time-evolved state as
\begin{align}
|\psi(t)\rangle=\sum_n\psi_n(t)|K_n\rangle\,,
\end{align}
the Krylov amplitudes obey
\begin{align}
i\,\partial_t\psi_n(t)
=
a_n\psi_n(t)
+
b_{n+1}\psi_{n+1}(t)
+
b_n\psi_{n-1}(t)\,,
\end{align}
with $\psi_n(0)=\delta_{n0}$. The spread complexity is then
\begin{align}
C(t)=\sum_n n\,|\psi_n(t)|^2\,.
\end{align}

Following~\cite{Balasubramanian:2022tpr}, we take the thermofield double (TFD) state as the initial state, which is particularly convenient for our purposes. For the sequence ${\xi_i}$, it is given by
\begin{align}
|\psi(0)\rangle
=
\frac{1}{\sqrt{Z(\beta)}}
\sum_i
e^{-\beta\xi_i/2}
|i\rangle_L\otimes|i\rangle_R\,,
\end{align}
where $Z(\beta)=\sum_i e^{-\beta\xi_i}$. We focus on the infinite-temperature limit, $\beta=0$, for which the TFD state is maximally entangled and all spectral weights are uniform. This choice is especially useful in this context because it isolates the role of the spectrum itself, without introducing additional energy dependence through the initial-state weights~\cite{Balasubramanian:2022tpr,Huh:2023jxt,Baggioli:2024wbz}.

\section{Unfolded Krylov complexity in generic systems}\label{sec3}
\setcounter{paragraph}{0}

With the above definitions in place, we now test whether unfolding preserves the characteristic spread complexity signatures of systems with well-understood spectral statistics. We begin with RMT ensembles, which provide controlled benchmarks, and then turn to quantum billiards, the Heisenberg XXZ chain, and the SYK model. In each case, we compare the raw and unfolded dynamics, with integrable spectra as a reference. This allows us to assess whether the characteristic RMT-universal features survive the unfolding procedure.

\subsection{Random matrix ensembles}

Before considering specific Hamiltonians, we establish a benchmark using canonical random matrix ensembles. Following previous studies of spread complexity in RMT~\cite{Caputa:2022yju,Erdmenger:2023wjg}, we consider the Gaussian Orthogonal Ensemble (GOE), Gaussian Unitary Ensemble (GUE), and Gaussian Symplectic Ensemble (GSE), together with Poisson spectra as an integrable baseline.

The joint probability density of the $N$ eigenvalues of the Gaussian ensembles is~\cite{mehta2004random}
\begin{align}
\rho_\mathcal{D}(E_1,\ldots,E_N)
=
\mathcal{C}
\prod_{1\leq i<j\leq N}
|E_i-E_j|^\mathcal{D}
\exp\left(-A\sum_{i=1}^N E_i^2\right),
\end{align}
where $\mathcal{D}=1,2,$ and $4$ is the Dyson index for the GOE, GUE, and GSE, respectively. Here, $\mathcal C$ is a normalization constant, while $A$ sets the overall energy scale.

A defining spectral signature of these ensembles is level repulsion. For the unfolded levels, with spacings $s_i=\xi_{i+1}-\xi_i$, the nearest-neighbor spacing distribution is well approximated by the Wigner surmise~\cite{Wigner_1951},
\begin{align}
P_\mathcal{D}(s)=C_\mathcal{D},s^\mathcal{D} e^{-A_\mathcal{D} s^2},
\end{align}
where $C_\mathcal{D}$ and $A_\mathcal{D}$ are fixed by normalization and the unit mean-spacing condition $\langle s\rangle=1$. By contrast, Poisson spectra obey $P_{\rm P}(s)=e^{-s}$ and exhibit no level repulsion.

\begin{figure}[t!]
 \centering
     \vspace{1.2mm}
     {\includegraphics[width=4.1cm]{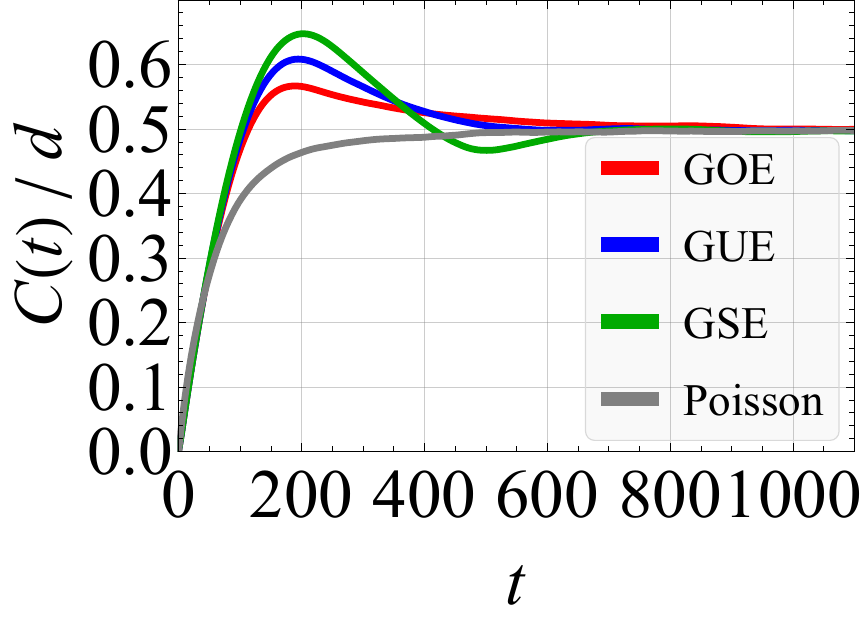}}
     \vspace{1.2mm}
     {\includegraphics[width=4.1cm]{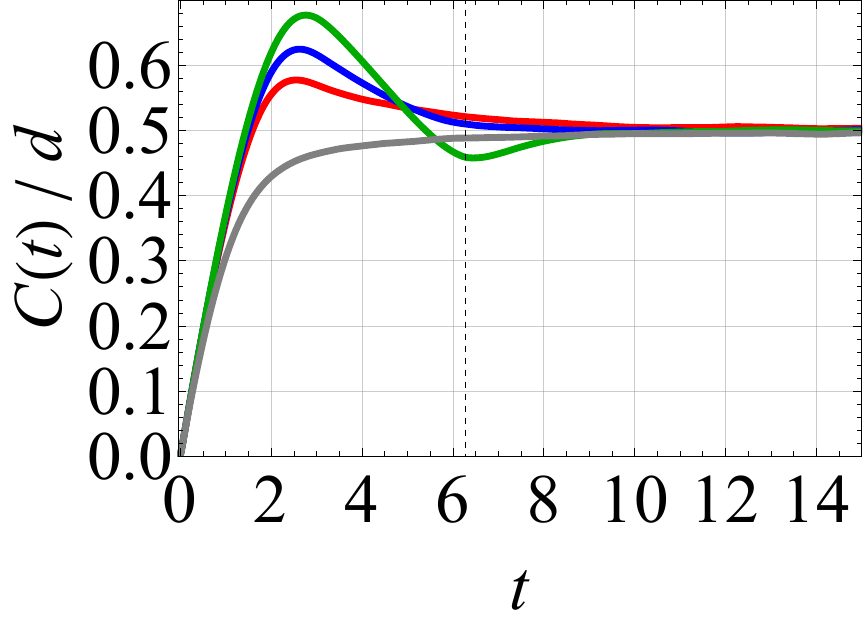}}
     \vspace{1.2mm}
     {\includegraphics[width=4.2cm]{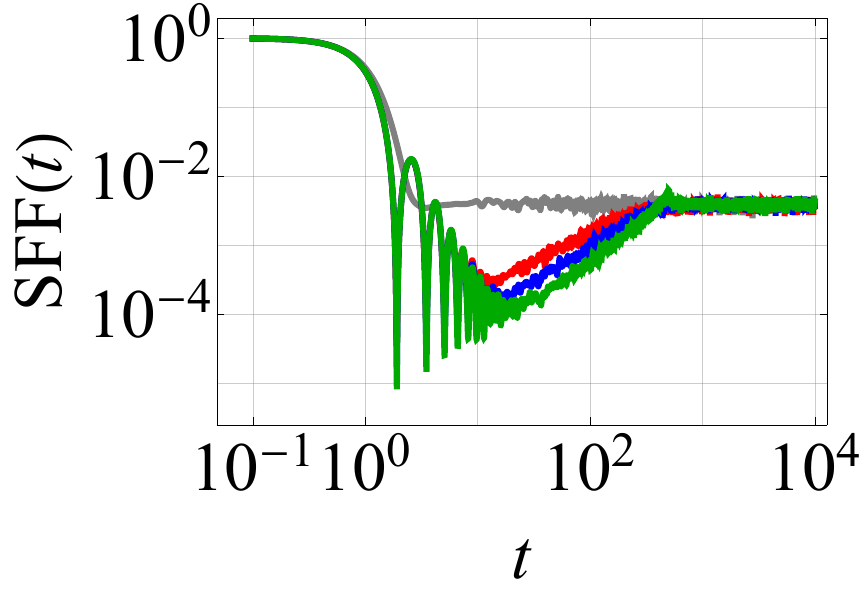}}
     \vspace{1.2mm}
     {\includegraphics[width=4.2cm]{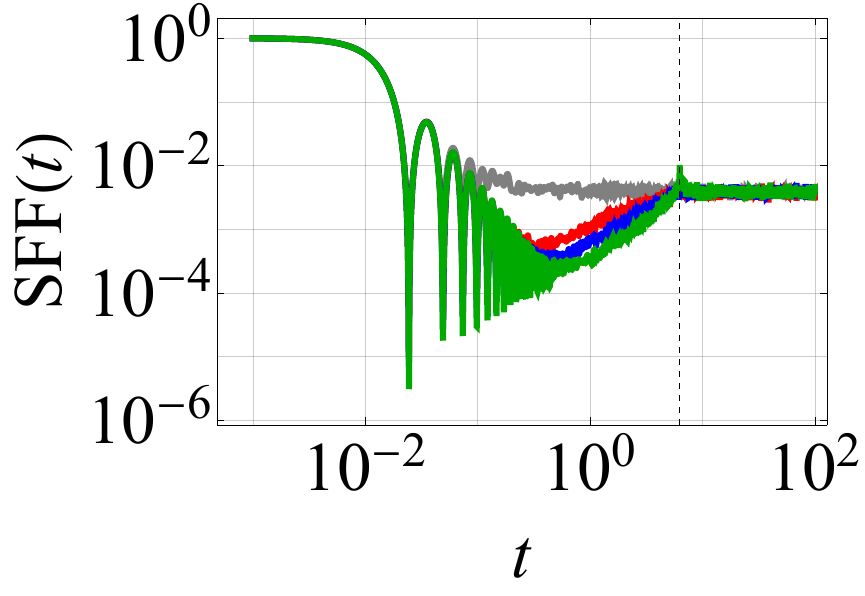}}
\vspace{-5mm}
\caption{Spread complexity and SFF for RMT spectra. The top and bottom panels show the spread complexity and SFF, respectively, with the left and right columns corresponding to the raw and unfolded spectra. The curves represent the GOE (red), GUE (blue), GSE (green), and Poisson (gray) ensembles. The vertical dashed lines indicate the Heisenberg time $t_{\rm H}=2\pi$. Unfolding preserves the Dyson-class hierarchy and the distinction between RMT and Poisson statistics while placing their late-time dynamics on a common scale.}
\label{Fig.RMT}
\end{figure}

As shown in the top panels of Fig.~\ref{Fig.RMT}, unfolding leaves the characteristic RMT structure of spread complexity qualitatively intact. In particular, the GOE, GUE, and GSE retain their peak-relaxation profiles and their characteristic hierarchy of peak heights, while the Poisson spectrum remains clearly distinct. Thus, removing the smooth density of states does not erase the microscopic correlations responsible for the RMT behavior.

Unfolding also fixes the mean spectral scale and therefore provides a common temporal normalization. To make this connection explicit, we compare with the spectral form factor (SFF) of the unfolded spectrum,
\begin{align}\label{eq:SFF}
\mathrm{SFF}(t)
=
\frac{|Z(\beta+it)|^2}{|Z(\beta)|^2}
=
\frac{1}{Z(\beta)^2}
\sum_{m,n}
e^{-\beta(\xi_m+\xi_n)}
e^{i(\xi_m-\xi_n)t}\,.
\end{align}
For unfolded RMT spectra, the ramp-to-plateau transition occurs exactly at the Heisenberg time $t_{\rm H}=2\pi$~\cite{PhysRevB.111.165108}. Fig.~\ref{Fig.RMT} shows that the onset of late-time saturation of the unfolded spread complexity occurs on the same scale.

This correspondence is also reflected in their long-time limits. For the maximally entangled TFD state at $\beta=0$, the long-time averages satisfy~\cite{Cotler:2016fpe,Rabinovici:2020ryf,Rabinovici:2022beu,Erdmenger:2023wjg}
\begin{align}
\lim_{T\rightarrow\infty}
\frac{1}{T}\int_0^T \mathrm{SFF}(t)\,dt
=
\frac{1}{1+2C(t\rightarrow\infty)}\,,
\end{align}
where $C(t\rightarrow\infty)=(d-1)/2$ and $d$ denotes the dimension of the Hilbert-space. Unfolding therefore removes model-dependent spectral scales while preserving the universal correlations governing the late-time dynamics.

\subsection{Quantum billiards}

\subsubsection{Stadium billiards}
We next test whether the RMT benchmark extends to concrete Hamiltonian systems. As a first example, we consider prototypical quantum billiards~\cite{Sinai1970,Bunimovich1974,Bunimovich1979,BunSinChe91,PhysRevA.17.773}, contrasting the integrable circular billiard with the chaotic Bunimovich stadium. In both cases, the quantum dynamics follows from a two-dimensional Schr\"odinger equation with a hard-wall potential,
\begin{align}
    \hat{H} &= \hat{p}_x^2 + \hat{p}_y^2 + V(\hat{x},\hat{y})\,,\\
    V(x,y) &=
    \begin{cases}
    0\,, & (x,y)\in\mathcal A\,,\\
    \infty\,, & \text{otherwise}\,,
    \end{cases}
\end{align}
where $\mathcal A$ denotes the billiard interior. The geometry of the billiard consists of two semicircles of radius $R$ joined by straight segments of length $2a$, giving a total area $\pi R^2+4aR$, which we fix to unity. The ratio $a/R$, in turn, controls the classical dynamics: $a/R=0$ gives the integrable circular billiard, while $a/R=1$ corresponds to the chaotic Bunimovich stadium.

\begin{figure}[t!]
 \centering
     \vspace{1.2mm}
     {\includegraphics[width=4.1cm]{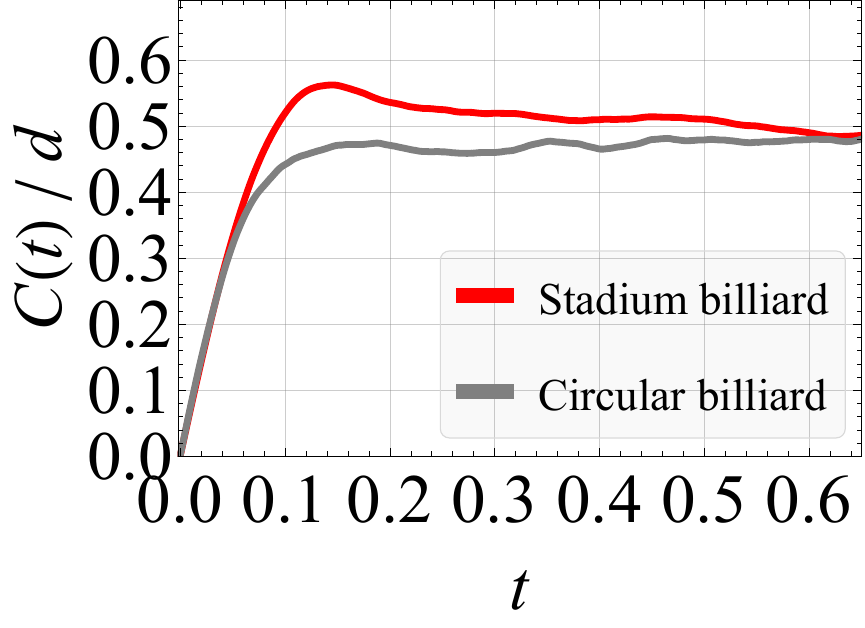}}
     \vspace{1.2mm}
     {\includegraphics[width=4.1cm]{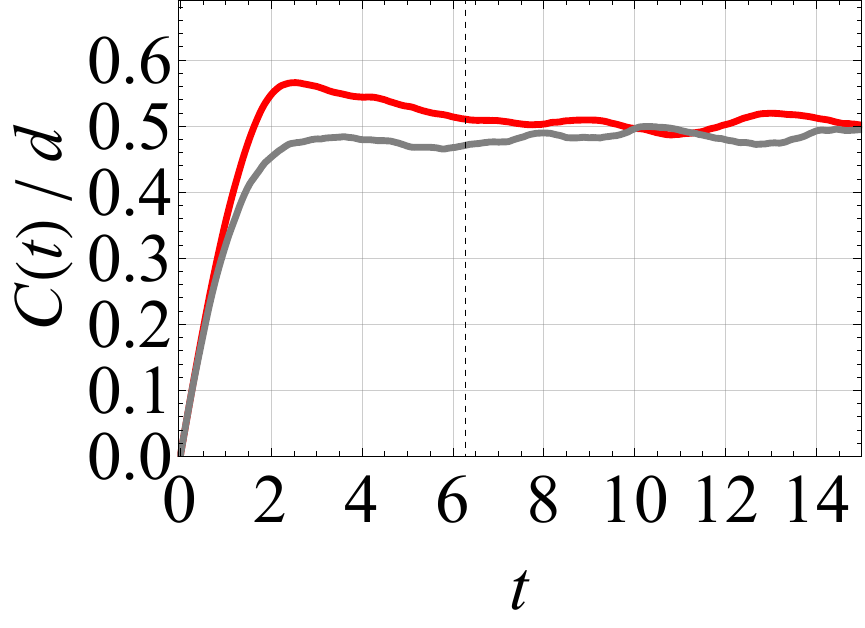}}
     \vspace{1.2mm}
     {\includegraphics[width=4.1cm]{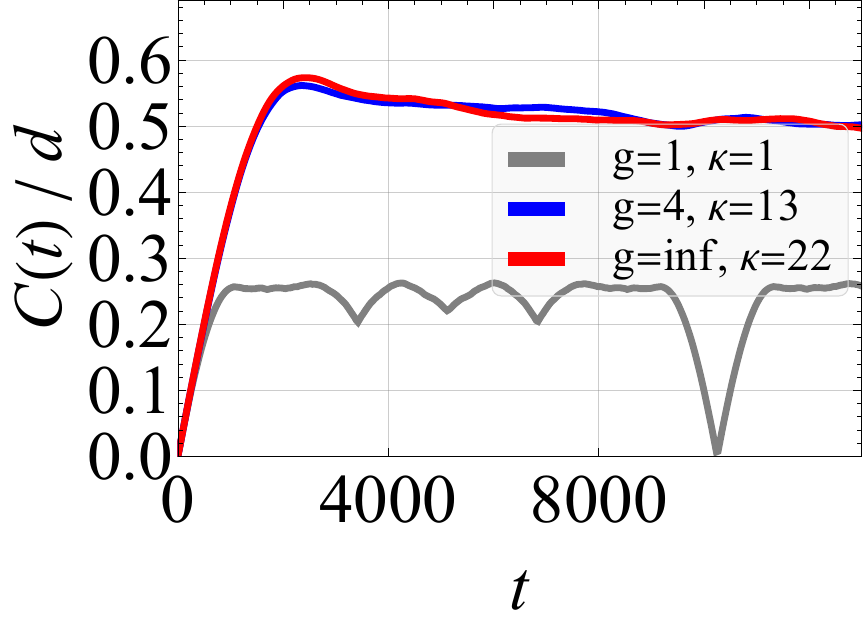}}
     \vspace{1.2mm}
     {\includegraphics[width=4.1cm]{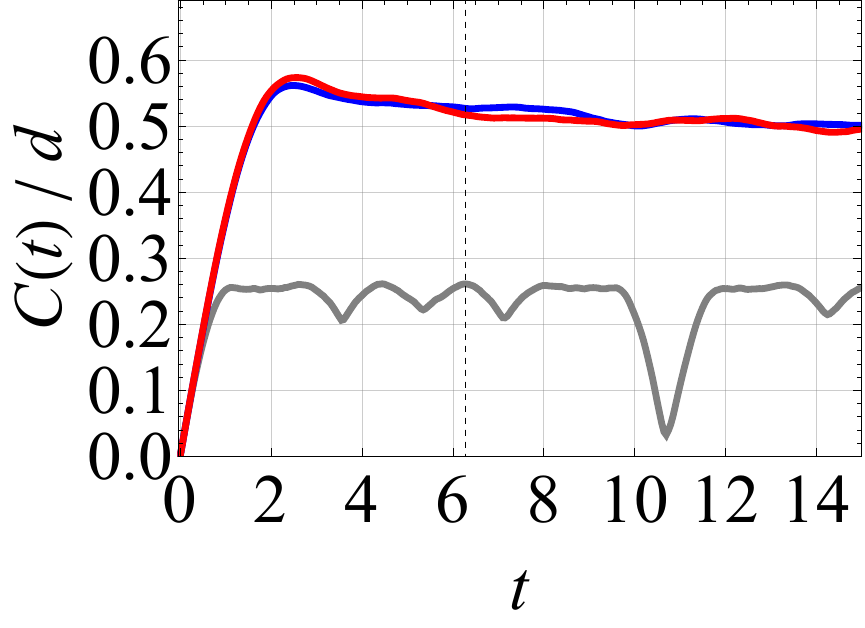}}
\vspace{-5mm}
\caption{Spread complexity in representative quantum billiards. The left and right columns show the raw and unfolded dynamics, respectively, for the stadium and circular billiards (top) and triangular billiards (bottom). In each case, unfolding preserves the distinction between chaotic (or pseudo-integrable) and integrable behavior while placing the saturation timescale on a common spectral scale.}
\label{Fig.biliards.KC}
\end{figure}
Following Refs.~\cite{Hashimoto:2023swv,Camargo:2023eev}, we retain the lowest $N_{\rm trunc}=100$ energy levels in our numerics. As shown in the top row of Fig.~\ref{Fig.biliards.KC}, unfolding preserves the qualitative distinction between the two regimes. The chaotic stadium retains its characteristic peak-relaxation profile, while the integrable billiard remains clearly distinct. Moreover, the normalized chaotic complexity $C(t)/d$ approaches the expected late-time value $1/2$~\cite{Caputa:2022yju,Erdmenger:2023wjg}. This provides initial evidence that unfolding removes the non-universal spectral scale without distorting chaotic dynamics.

\subsubsection{Triangular billiards}

The stadium--circle comparison above provides a clean benchmark between chaotic and integrable dynamics. To test whether unfolding remains effective in the absence of classical exponential instability, we next consider two-dimensional triangular quantum billiards~\cite{PhysRevLett.92.244102,Balasubramanian:2024ghv}. Unlike the stadium billiard, where nearby trajectories separate exponentially, triangular billiards have vanishing classical Lyapunov exponents, while their quantum spectral statistics can nevertheless range from integrable to RMT-like behavior depending on the internal angles.

For rational triangles, the angles can be written as $\{\alpha,\beta,\gamma\}=\{\frac{p_1}{q_1},\frac{p_2}{q_2},\frac{p_3}{q_3}\}\pi$, with $p_i,q_i\in\mathbb{Z}^+$ coprime integers. This rationality condition ensures that successive reflections generate a compact translation surface, whose topology is characterized by its genus~\cite{Zemlyakov:1975},
\begin{align}
g = 1 + \frac{\mathcal{N}}{2}\sum_{i=1}^{3} \frac{p_i - 1}{q_i}\,,
\label{eq:genus}
\end{align}
where $\mathcal{N}$ is the least common multiple of ${q_i}$ and gives the number of triangular copies required to close the translation surface. Triangles with $g=1$ are classically integrable, while those with finite $g>1$ are pseudo-integrable: they retain vanishing Lyapunov exponents but their dynamics takes place on higher-genus translation surfaces rather than invariant tori. If one or more angles are irrational, the translation surface has infinite genus, $g=\infty$, and the dynamics is non-integrable despite the absence of exponential trajectory separation~\cite{Balasubramanian:2024ghv}. Triangular billiards thus provide a setting to test whether unfolded spread complexity probes quantum spectral correlations independently of classical exponential instabilities.

\begin{table}[t!]
\centering
\vspace{1.5mm}
\renewcommand{\arraystretch}{1.55}
\begin{tabular}{c @{\hspace{2em}} ccc}
\hline\hline
$\kappa$ & Angles $\{\alpha, \beta, \gamma\}$ & Genus $g$ & Regime \\
\hline
1  & $\{\frac{1}{3}, \frac{1}{3}, \frac{1}{3}\}\pi$ & 1 & Integrable (I) \\
4  & $\{\frac{2}{5}, \frac{2}{5}, \frac{1}{5}\}\pi$ & 2 & Pseudo-integrable (PI) \\
13 & $\{\frac{1}{4}, \frac{1}{6}, \frac{7}{12}\}\pi$ & 4 & Pseudo-integrable (PI) \\
22 & $\{\frac{3-\sqrt{5}}{5}, \frac{3}{5}, \frac{-1+\sqrt{5}}{5}\}\pi$ & $\infty$ & Non-integrable (NI) \\
\hline\hline
\end{tabular}
\caption{Representative triangular billiards categorized by the index $\kappa$, internal angles $\{\alpha, \beta, \gamma\}$, genus $g$ of the invariant translation surface, and dynamical regime~\cite{Balasubramanian:2024ghv}.}
\label{tab:triangles}

\end{table}

To compare these regimes, we adopt the classification of Ref.~\cite{Balasubramanian:2024ghv} and consider the representative geometries listed in Table~\ref{tab:triangles}. We normalize each billiard to unit area and retain the lowest $N_{\rm trunc}=1000$ energy levels. The bottom row of Fig.~\ref{Fig.biliards.KC} shows the corresponding spread complexity for triangles labeled by $\kappa=1$, $13$, and $22$, spanning the integrable, pseudo-integrable, and non-integrable regimes. As in the stadium--circle comparison, unfolding preserves the qualitative distinction between them. The integrable billiard ($\kappa=1$) retains its suppressed complexity growth and pronounced oscillations, whereas the pseudo-integrable ($\kappa=13$) and non-integrable ($\kappa=22$) triangles preserve the characteristic ramp-peak-slope-plateau profile. Thus, even in the absence of classical exponential instability, unfolding places the dynamics on a common spectral scale without erasing the distinctions between integrability, pseudo-integrability, and quantum-chaotic behavior.

\begin{figure}[t!]
\centering
\vspace{1.2mm}
{\includegraphics[width=4.1cm]{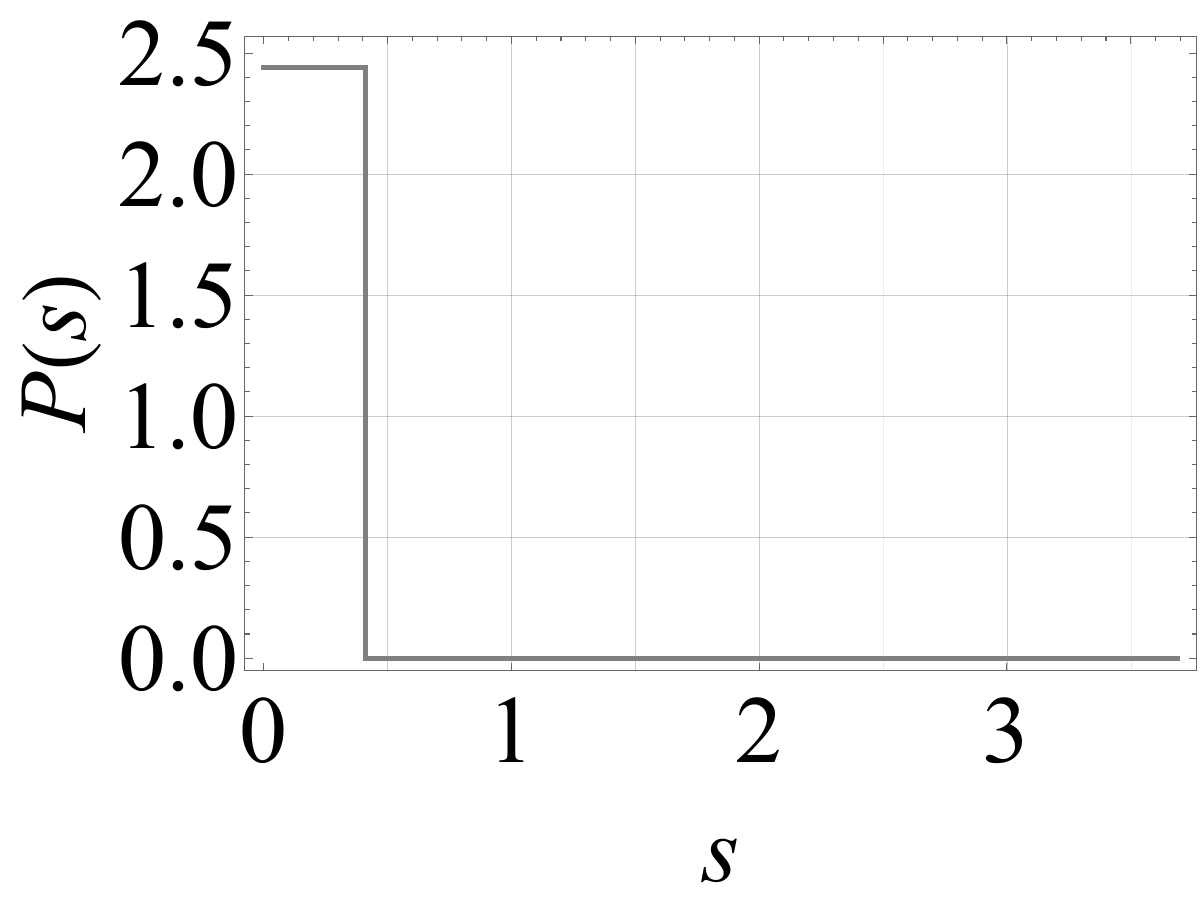}}
\vspace{1.2mm}
{\includegraphics[width=4.1cm]{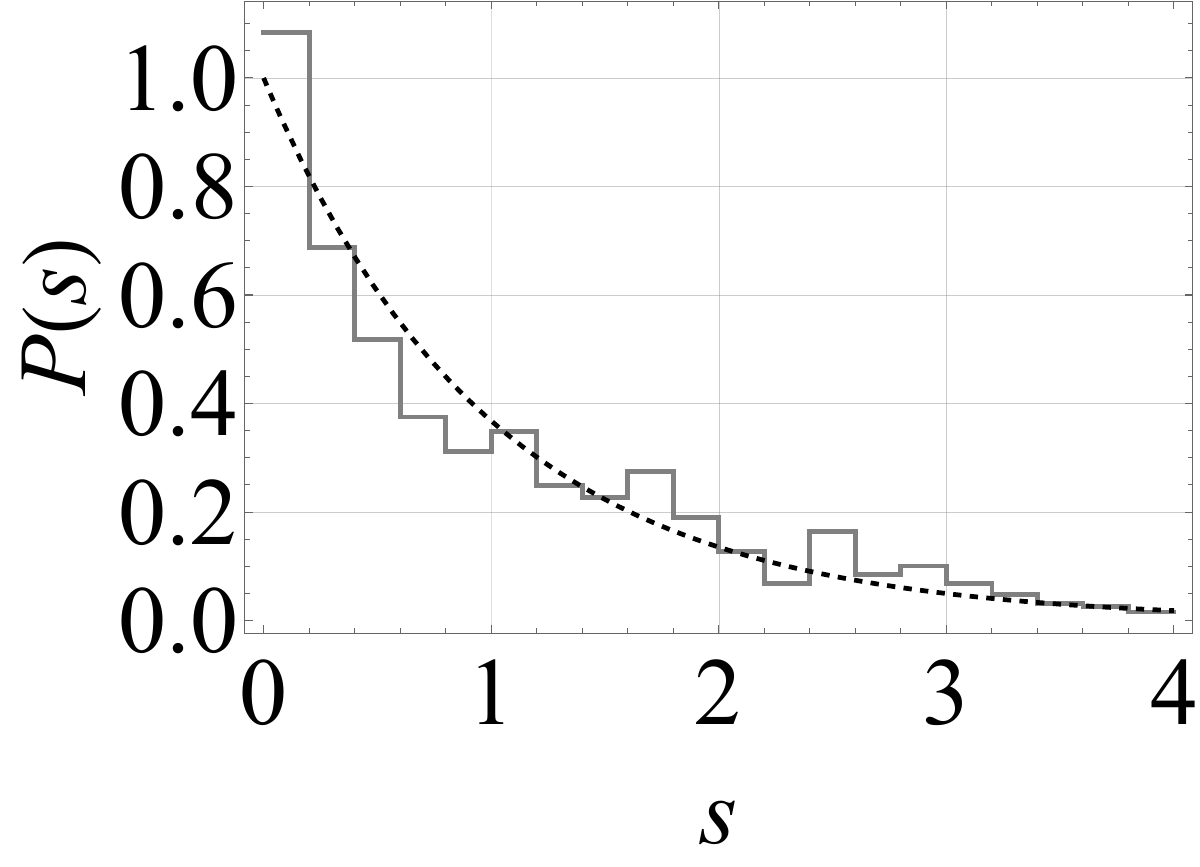}}
\vspace{1.2mm}
{\includegraphics[width=4.1cm]{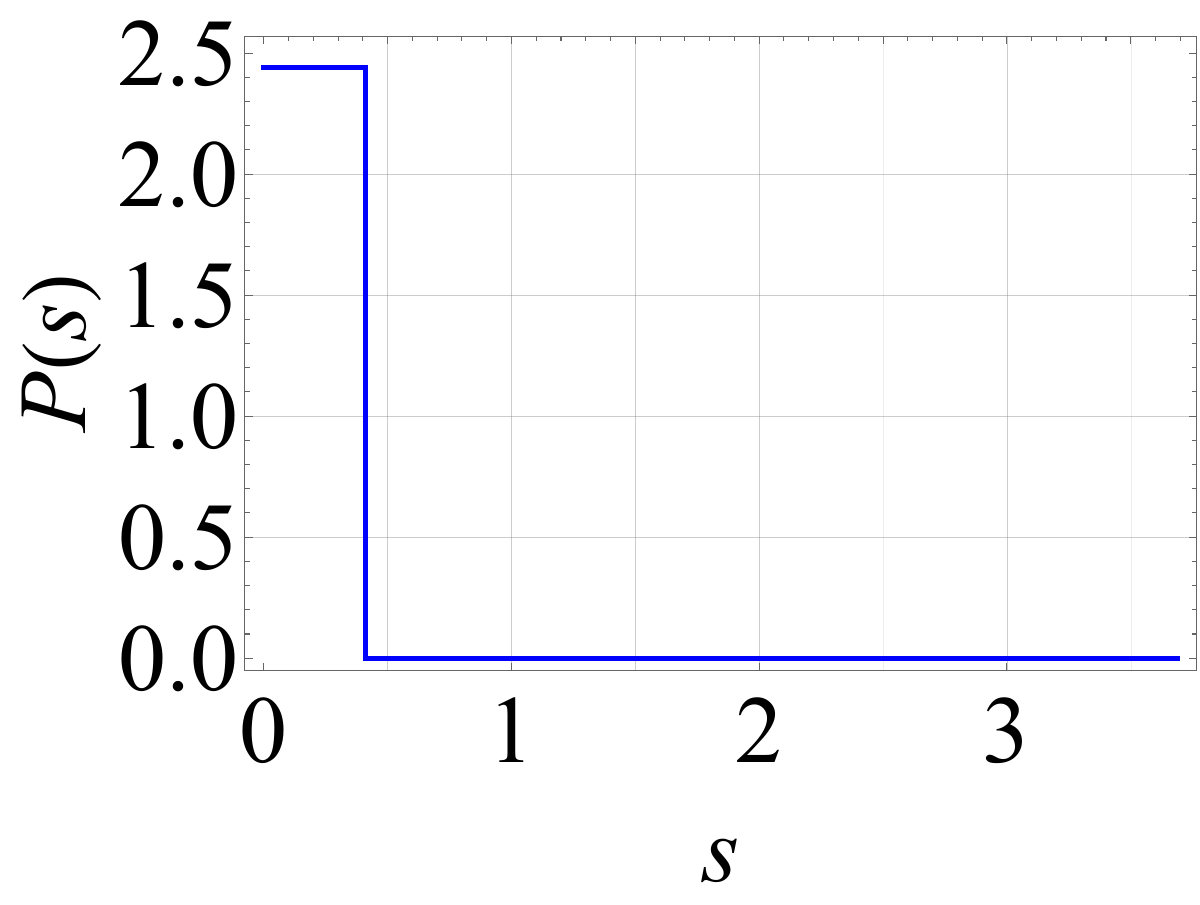}}
\vspace{1.2mm}
{\includegraphics[width=4.1cm]{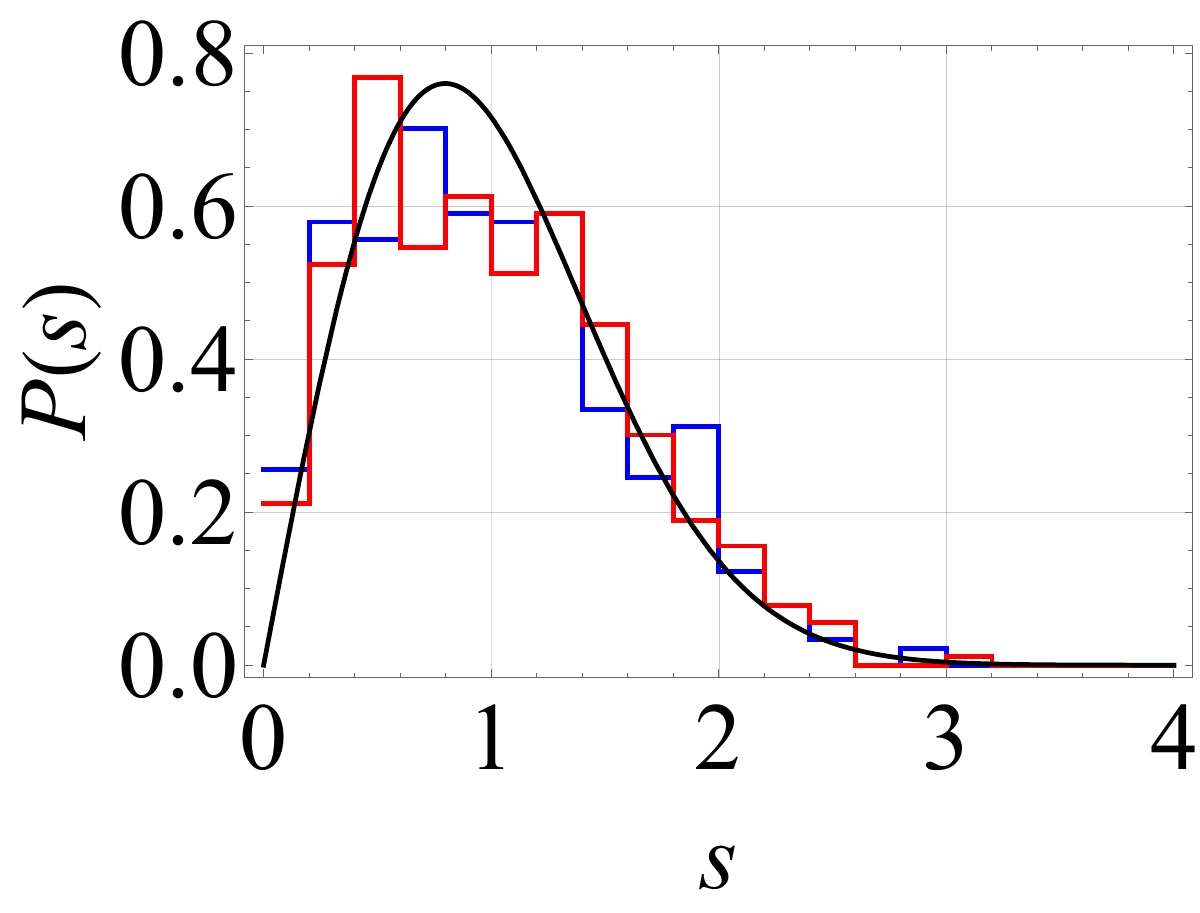}}
\vspace{1.2mm}
{\includegraphics[width=4.1cm]{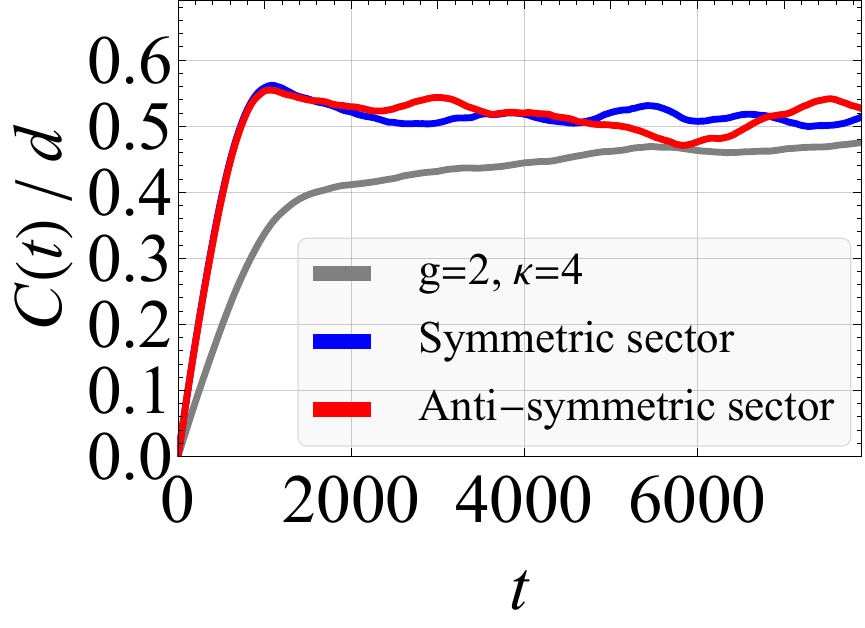}}
\vspace{1.2mm}
{\includegraphics[width=4.1cm]{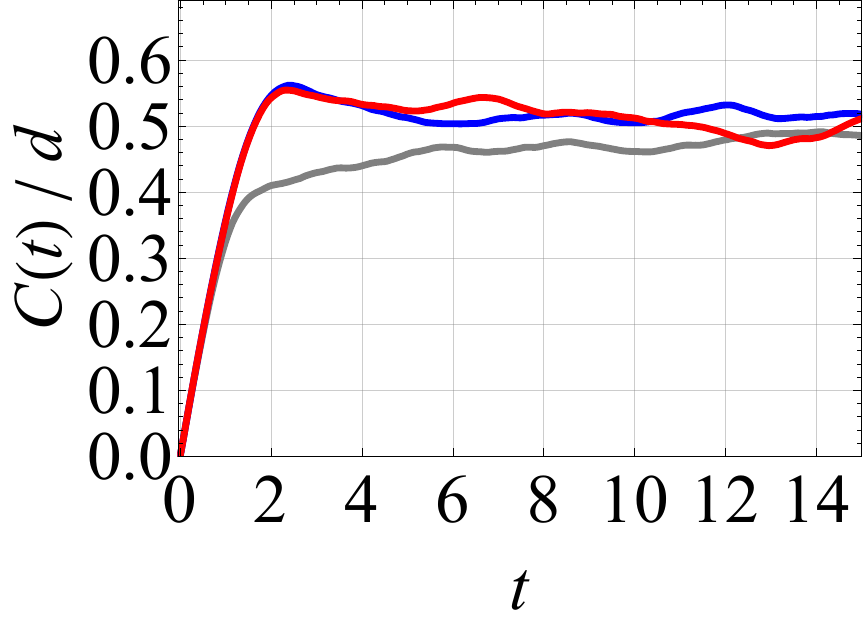}}
\vspace{-5mm}
\caption{Level-spacing statistics and spread complexity for the $\kappa=4$ isosceles triangular billiard in Table~\ref{tab:triangles}. The left and right columns show the raw and unfolded data, respectively, for the unresolved level-spacing distribution (top), the symmetry-resolved level-spacing distributions (middle), and spread complexity (bottom). Blue and red curves represent the symmetric and antisymmetric sectors, respectively, while gray curves correspond to the unresolved spectrum. In the level-spacing panels, solid and dashed black curves denote the GOE and Poisson distributions, respectively.}
\label{Fig.sector.KC}
\end{figure}

A further subtlety arises with spatial symmetries. To illustrate this, we consider the reflection-symmetric isosceles triangle with $\kappa=4$ ($g=2$), listed in Table~\ref{tab:triangles}. Reflection symmetry decomposes the Hilbert space into symmetric and antisymmetric parity sectors. As shown in the top row of Fig.~\ref{Fig.sector.KC}, the full, unresolved spectrum exhibits apparent Poisson-like level statistics because it combines statistically independent levels from the two sectors. Once the symmetry is resolved, however, each sector displays GOE-like level repulsion, as seen in the middle row. Correspondingly, the spread complexity within each sector develops the characteristic ramp-peak-slope-plateau profile shown in the bottom row. This illustrates the importance of symmetry resolution in Krylov dynamics~\cite{Caputa:2025ozd,Das:2026gko}: mixing independent sectors can obscure the spectral correlations and complexity signatures associated with quantum ergodicity. Comparing the raw and unfolded results further shows that unfolding fixes the mean spectral scale while preserving both the symmetry-resolved level correlations and their imprint on the Krylov dynamics.
In essence, the absence of a peak for the unresolved spectrum (gray curve of Fig.~\ref{Fig.sector.KC}) directly reflects its effective Poisson statistics resulting from the superposition of uncorrelated each sectors. Conversely, the emergence of Wigner-Dyson level repulsion in each resolved sector restores the prominent peak feature. Crucially, this behavior is only clearly revealed after applying spectral unfolding (right column of Fig.~\ref{Fig.sector.KC}).
%

\subsection{Heisenberg XXZ spin chain}

As our next example, we consider an interacting lattice system: a one-dimensional spin-$1/2$ Heisenberg XXZ chain with a next-to-nearest-neighbor (NNN) perturbation~\cite{Samaj:2013yva,Gubin_2012,Rabinovici:2022beu}. The Hamiltonian is
\begin{align}
H
&=
\sum_{i=1}^{N-1}
\left[
J\left(S_i^xS_{i+1}^x+S_i^yS_{i+1}^y\right)
+
J_{zz}S_i^zS_{i+1}^z
\right]
\nonumber\\
&\quad+
\sum_{i=1}^{N-2}
J_c S_i^zS_{i+2}^z\,,
\end{align}
where $S_i^\alpha$ ($\alpha=x,y,z$) are spin-$1/2$ operators. For $J_c=0$, the model reduces to the integrable XXZ chain, whereas the NNN interaction breaks integrability and produces a chaotic spectrum for the parameters considered below.

Specifically, we fix $(J,J_{zz})=(1,0.5)$ and compare $J_c=0$ with $J_c=1$. Since the Hamiltonian preserves parity and the total magnetization
\begin{align}
M_z=\sum_{i=1}^N S_i^z\,,
\end{align}
the spectral analysis must be performed within a fixed symmetry sector. Following Ref.~\cite{Camargo:2024deu}, we take $N=15$, $M_z=-5/2$, and a fixed parity block.
\begin{figure}[t!]
 \centering
     \vspace{1.2mm}
     {\includegraphics[width=4.1cm]{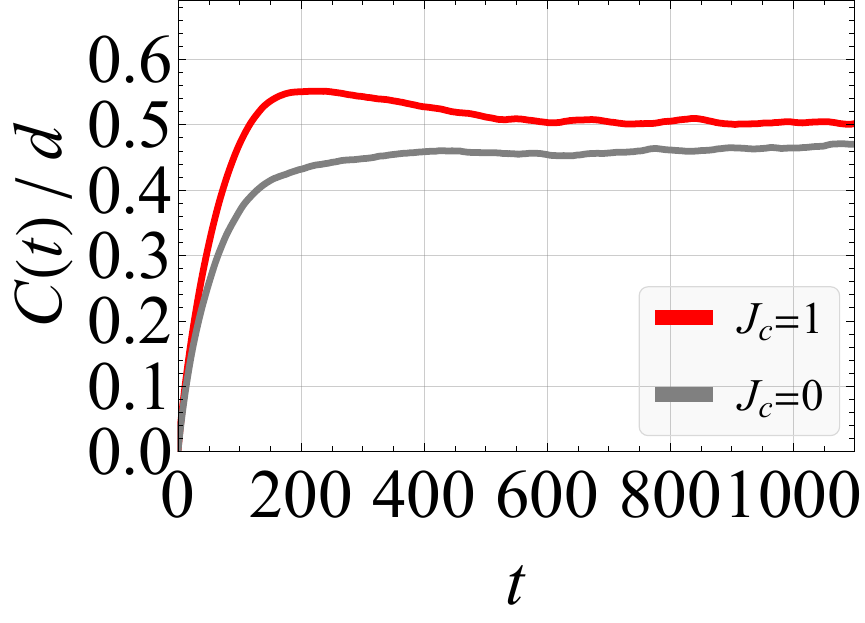}}
     \vspace{1.2mm}
     {\includegraphics[width=4.1cm]{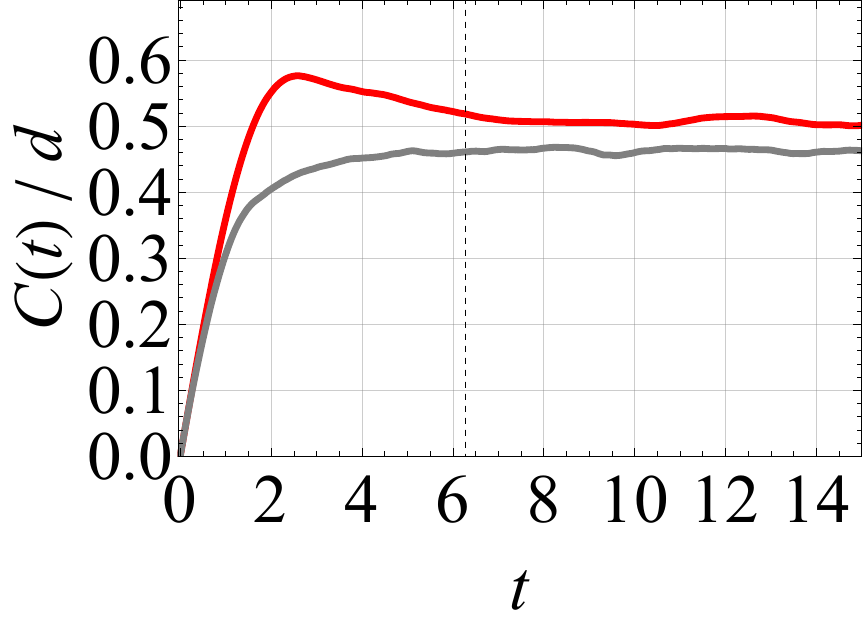}}
     \vspace{1.2mm}
     {\includegraphics[width=4.1cm]{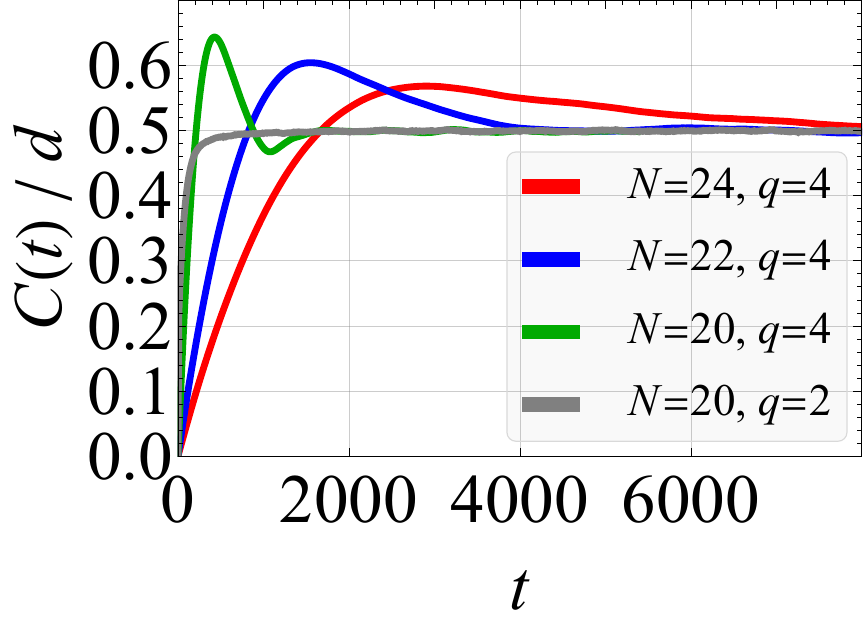}}
     \vspace{1.2mm}
     {\includegraphics[width=4.1cm]{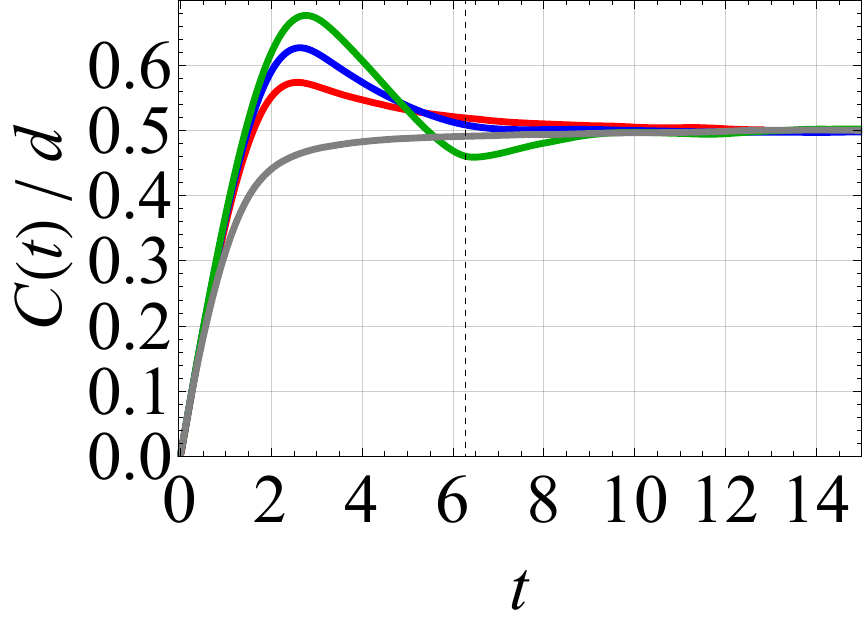}}
\vspace{-5mm}
\caption{Spread complexity in other representative physical systems. The left and right columns show the raw and unfolded dynamics, respectively, for the XXZ spin chain (top), and the SYK model (bottom). In each case, unfolding preserves the distinction between chaotic and integrable behavior while normalizing the saturation timescale.}
\label{Fig.generic.KC}
\end{figure}

The top row of Fig.~\ref{Fig.generic.KC} again shows that unfolding preserves the chaotic-integrable distinction while bringing the saturation scale to approximately $t_{\rm H}=2\pi$~\footnote{While the normalized saturation value $1/2$ is accurately recovered in the chaotic XXZ chain, the integrable case exhibits small fluctuations around the plateau due to spectral degeneracies~\cite{Camargo:2024deu}.}. This rescaling affects the Krylov dynamics already at early times. At $\beta=0$, the first Lanczos coefficient satisfies $b_1^2=\langle H^2\rangle-\langle H\rangle^2$, so the leading short-time growth is controlled by the spectral variance. By fixing the mean level spacing, unfolding removes the model-dependent overall energy scale and correspondingly places the early-time dynamics on a common scale. The same unit-spacing convention fixes the Heisenberg time to $t_{\rm H}=2\pi$, explaining the common late-time spectral scale.

A similar unfolding-induced peak of spread complexity, together with its alignment with the SFF ramp, was recently reported in weakly chaotic long-range spin chains derived from the $\mathcal{N}=4$ super Yang-Mills theory~\cite{Caputa:2026hvh}, providing an independent example consistent with the general mechanism established in this manuscript.

\subsection{Sachdev--Ye--Kitaev model}

As a final example, we consider the SYK model~\cite{Sachdev:1992fk,RevModPhys.94.035004}, which provides a stringent many-body test because its spectral scale depends strongly on both system size and interaction order. For general $q$-body interactions, the corresponding Hamiltonian is given by
\begin{align}\label{SYK}
H =
\frac{i^{q/2}}{q!}
\sum_{i_1,\ldots,i_q=1}^{N}
J_{i_1\cdots i_q}\,
\chi_{i_1}\cdots\chi_{i_q}\,,
\end{align}
where the Majorana fermions satisfy ${\chi_i,\chi_j}=\delta_{ij}$. The couplings $J_{i_1\cdots i_q}$ are Gaussian random variables with zero mean, $\langle J_{i_1\cdots i_q} \rangle=0$, and variance $\langle J_{i_1\cdots i_q}^2\rangle=\frac{(q-1)!\mathcal J^2}{N^{q-1}}$.

Spread complexity in SYK has been studied extensively in Refs.~\cite{Balasubramanian:2022tpr,Erdmenger:2023wjg,Baggioli:2024wbz,Huh:2024ytz}. For the standard $q=4$ SYK model, the RMT symmetry class depends on $N$ modulo $8$~\cite{Cotler:2016fpe,Kanazawa:2017dpd,Garcia-Garcia:2016mno}. Here, we consider $N=20,22,$ and $24$, corresponding respectively to GSE, GUE, and GOE statistics, and contrast them with the integrable $q=2$ model exhibiting Poisson statistics.

The bottom row of Fig.~\ref{Fig.generic.KC} makes the effect of unfolding transparent. In the raw spectrum, changing $N$ or $q$ substantially changes the spectral variance and hence the characteristic timescale of the Krylov dynamics. After unfolding, these model-dependent scales are removed: the RMT hierarchy and distinction from the integrable case are preserved, while the onset of saturation aligns with the Heisenberg scale $t_{\rm H}=2\pi$. The SYK example therefore clearly illustrates that unfolding normalizes the macroscopic spectral scale without erasing the microscopic correlations responsible for RMT universality.

Taken together, the RMT ensembles and physical examples considered above establish the control case for our proposal: whenever conventional spectral diagnostics reliably distinguish chaotic from integrable dynamics, unfolding preserves the corresponding spread complexity signatures while removing non-universal spectral scales. In the next section, we turn to the more challenging question of whether the same procedure can eliminate the false-positive peaks of non-generic systems.

\section{Unfolded Krylov complexity in non-generic systems}\label{sec4}
\setcounter{paragraph}{0}

Having established that unfolding preserves the characteristic spread complexity signatures of generic chaotic systems, we now turn to non-generic cases where conventional diagnostics can yield false positives. Our goal is to test whether these apparent signatures survive once the smooth density of states is removed.

We consider three representative cases. We begin with a logarithmic toy spectrum, which isolates the effect of a strongly varying density of states in the absence of RMT correlations. We then turn to two prototypical saddle-dominated systems: the inverted harmonic oscillator and the Lipkin-Meshkov-Glick model.

\subsection{Logarithmic spectrum}

To isolate the spectral origin of the false-positive signal, we first consider the logarithmic spectrum
\begin{align}\label{toy}
    E_k=\log(k+1)\,,\qquad k=0,1,\ldots,N-1\,.
\end{align}
This toy model was motivated in Ref.~\cite{Das:2023yfj}, where logarithmically spaced levels arise in connection with stretched-horizon modes of black holes and lead to a partition function related to the Riemann zeta function. Notably, the raw logarithmic spectrum generates a ramp-like structure in the SFF reminiscent of chaotic spectra, despite the absence of RMT level correlations~\cite{Das:2023yfj}.

For our purposes, this model is especially useful because it cleanly separates a strongly non-uniform smooth density of states from microscopic spectral correlations. Indeed, the logarithmic spacing produces an exponentially growing density of states, allowing us to test directly whether this macroscopic spectral structure is responsible for the apparent chaotic signatures.

\begin{figure}[t!]
 \centering
    \vspace{1.2mm}
     {\includegraphics[width=4.2cm]{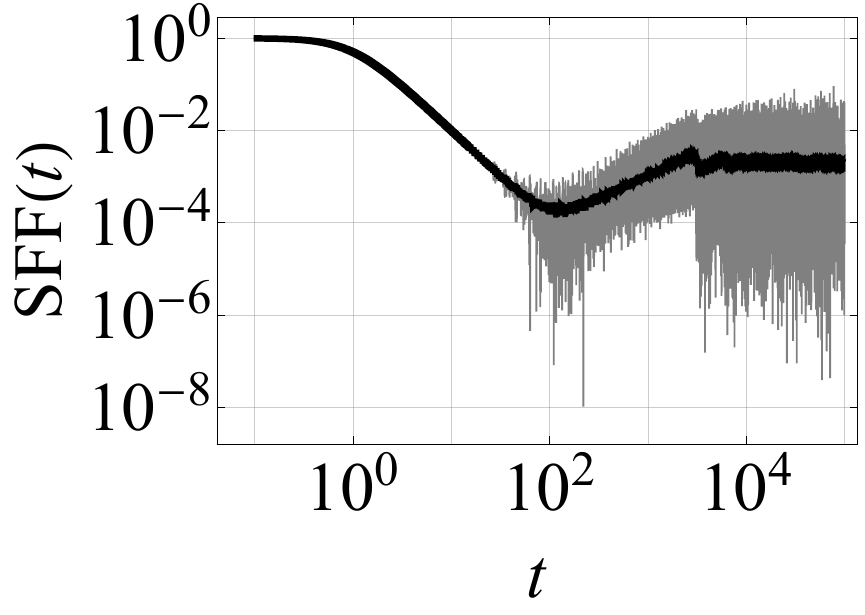}}
    \vspace{1.2mm}
     {\includegraphics[width=4.2cm]{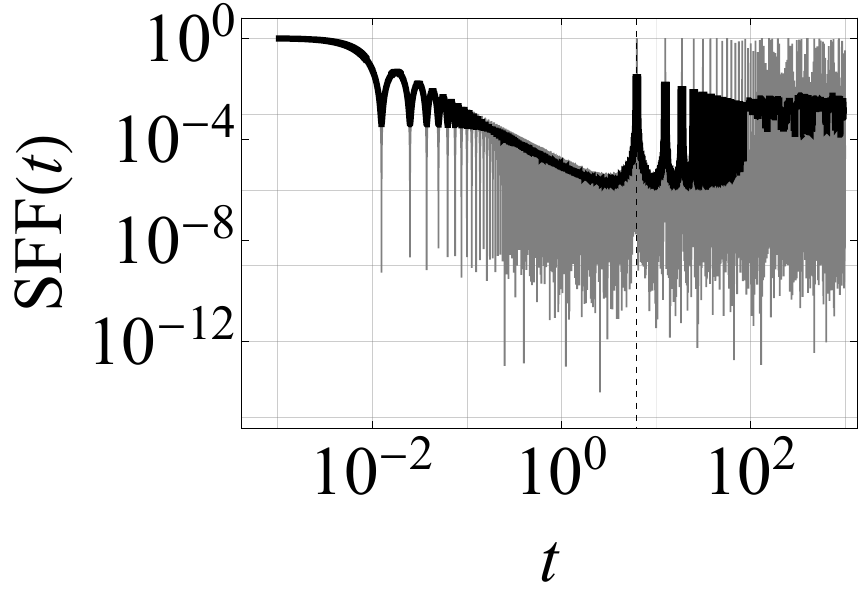}}
\vspace{-5mm}
\caption{Spectral form factor for the logarithmic toy spectrum. The left panel shows the raw spectrum, which exhibits a spurious ramp, while the right panel shows the unfolded result, where this feature disappears. The black curve denotes a moving average used to reduce numerical fluctuations. This shows that the ramp originates from the non-uniform density of states rather than from RMT-type spectral correlations.}
\label{Fig.non-generic.SFF}
\end{figure}

Fig.~\ref{Fig.non-generic.SFF} shows the SFF for this model. As advertised, the raw spectrum leads to a pronounced ramp even though it does not contain RMT correlations. After unfolding, this spurious structure disappears. The spurious ramp therefore originates from the non-uniform density of states rather than from level repulsion or spectral rigidity.

\begin{figure}[t!]
 \centering
     \vspace{1.2mm}
     {\includegraphics[width=4.1cm]{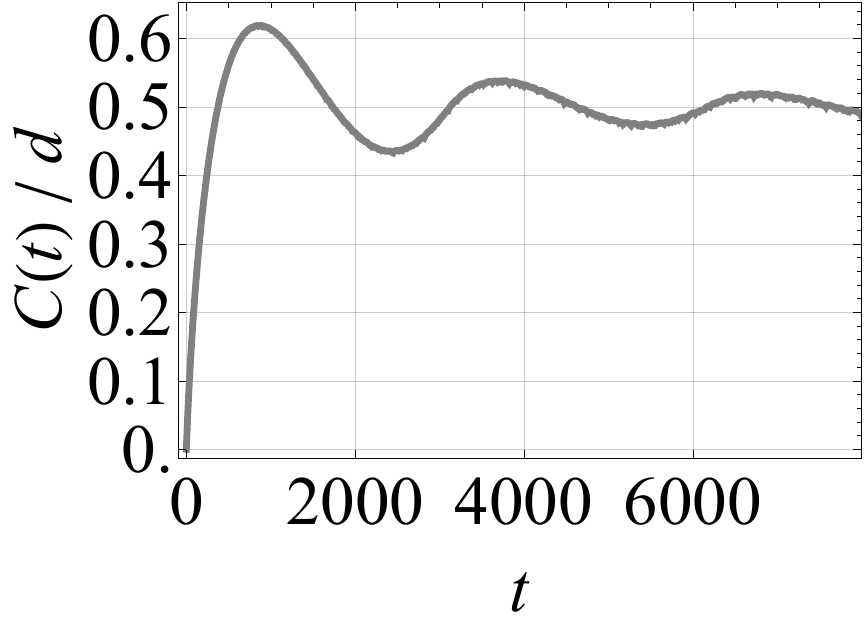}}
    \vspace{1.2mm}
     {\includegraphics[width=4.1cm]{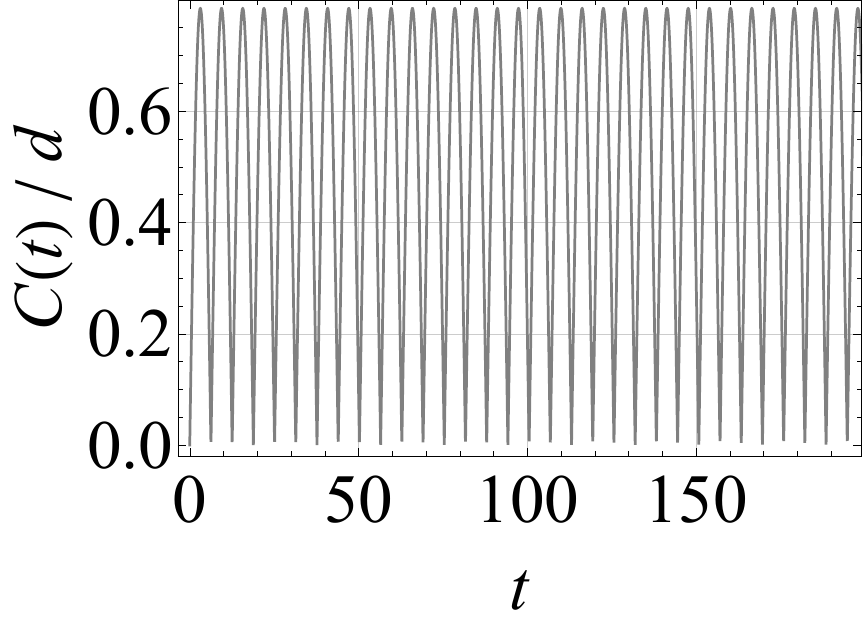}}
    \vspace{1.2mm}
     {\includegraphics[width=4.1cm]{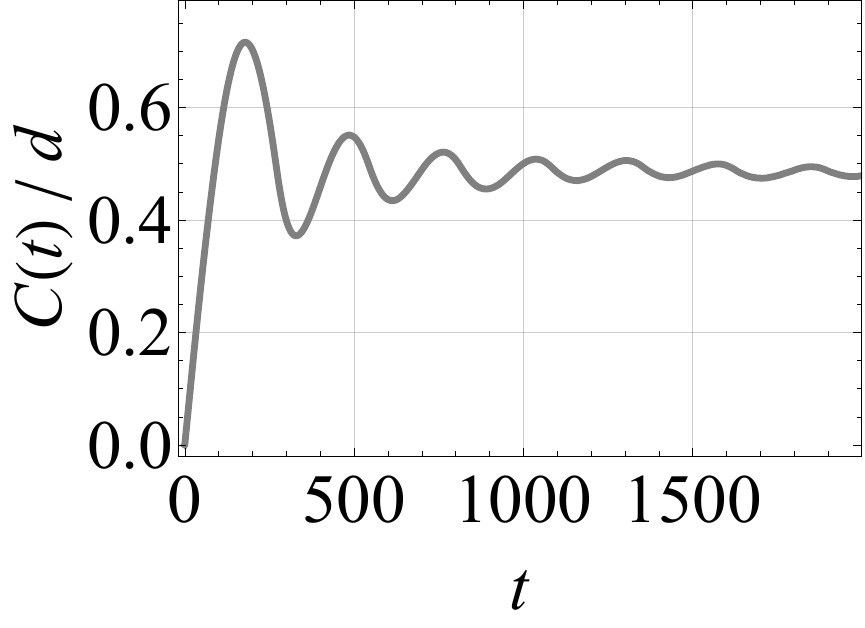}}
    \vspace{1.2mm}
     {\includegraphics[width=4.1cm]{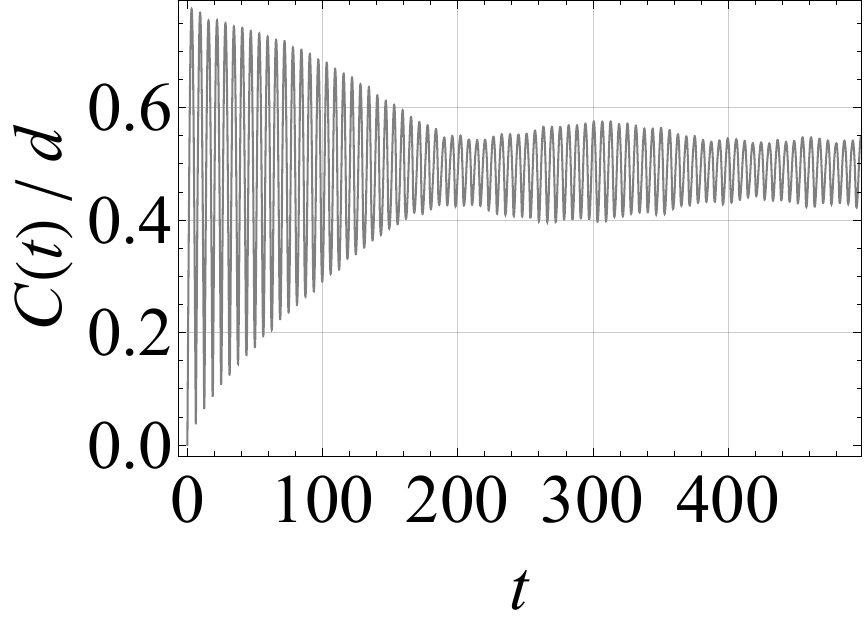}}
    \vspace{1.2mm}
     {\includegraphics[width=4.1cm]{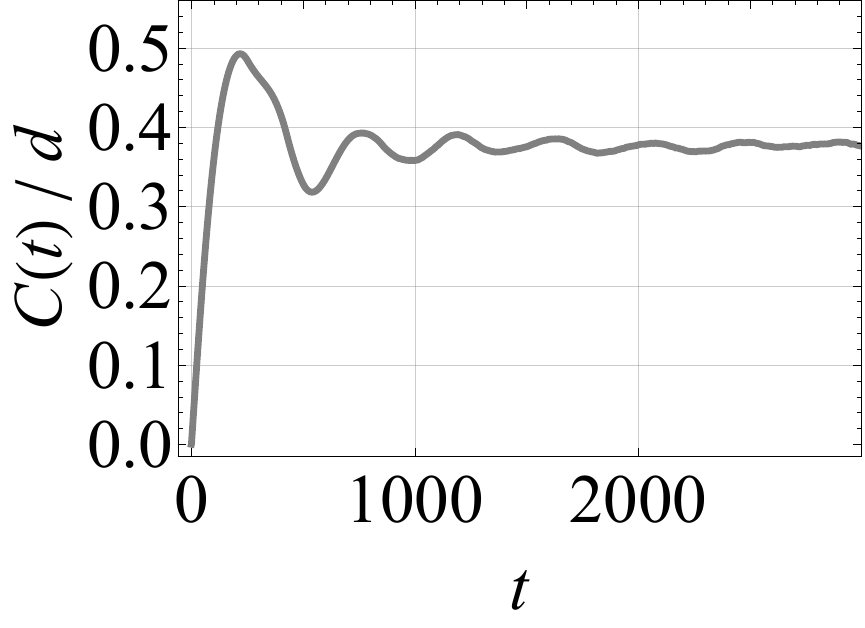}}
    \vspace{1.2mm}
     {\includegraphics[width=4.1cm]{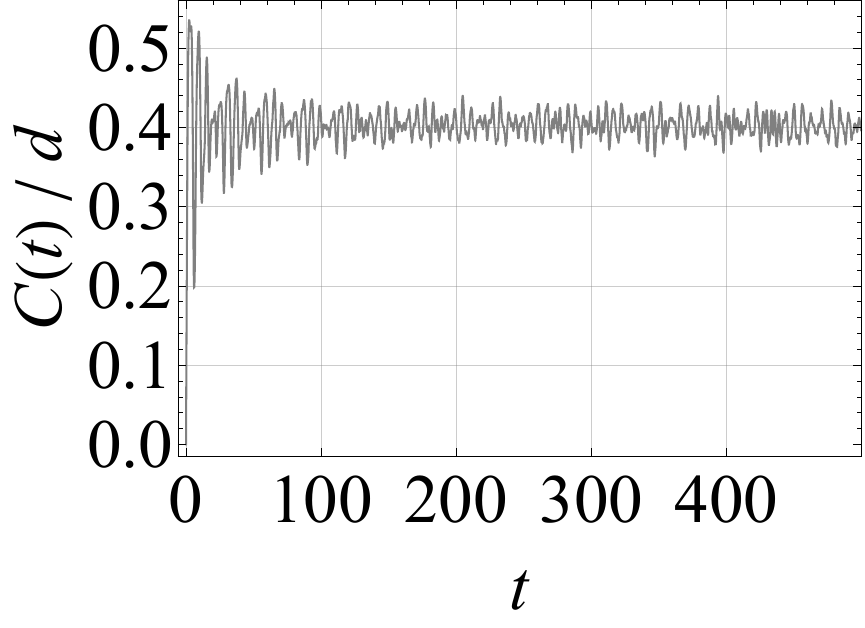}}
\vspace{-5mm}
\caption{Spread complexity in non-generic systems. The left and right columns show the raw and unfolded dynamics, respectively. From top to bottom, the rows correspond to the logarithmic toy spectrum, the IHO, and the LMG model. In each case, unfolding removes the false-positive peak and reveals oscillatory dynamics, indicating across all three examples that the original peak is tied to the non-uniform spectral structure rather than to RMT-type spectral correlations.}
\label{Fig.non-generic.KC}
\end{figure}

A parallel pattern emerges in spread complexity. Indeed, as shown in the top row of Fig.~\ref{Fig.non-generic.KC}, the raw spectrum develops a pronounced peak resembling that of chaotic systems, which gives way to persistent oscillations upon unfolding. Since unfolding maps the logarithmic spectrum onto a uniform lattice, this behavior identifies the smooth spectral profile as the origin of the original peak, rather than microscopic RMT correlations.

%

\subsection{Inverted harmonic oscillator}

We next consider a physical realization of the same mechanism in a saddle-dominated system. The IHO model~\cite{Hashimoto:2020xfr} is described by the following Hamiltonian
\begin{align}
    H=p^2+V(x)\,,
\end{align}
with
\begin{align}\label{IHO_Ham2}
    V(x)
    =
    -\frac{1}{4}\lambda^2x^2
    +gx^4
    +\frac{\lambda^4}{64g}\,.
\end{align}
Here, $\lambda$ sets the scale of the local instability, while $g$ controls the quartic stabilization.

We take $\lambda=2\sqrt{5}$ and $g=1/10$, and truncate the Hilbert space to $N=200$ for the numerical analysis. This system is an example of saddle-dominated scrambling: despite being integrable, the unstable region near the barrier can produce chaos-like signatures in diagnostics such as OTOCs and spread complexity~\cite{Hashimoto:2020xfr,Huh:2023jxt}.

The middle row of Fig.~\ref{Fig.non-generic.KC} shows the effect of unfolding on spread complexity. For the raw spectrum, this observable develops a pronounced peak, providing a clear false positive despite the system being integrable~\cite{Huh:2023jxt}. The saddle produces a strong accumulation of levels and hence a highly non-uniform density of states, which feeds directly into the Krylov dynamics. After unfolding, the peak disappears and is replaced by oscillations whose amplitude gradually settles at late times, indicating that the original peak is tied to the saddle-induced smooth spectral structure rather than to RMT level correlations. In contrast with the logarithmic toy model, however, these residual oscillations exhibit nontrivial dephasing and therefore retain information about the microscopic spectrum.

\subsection{Lipkin-Meshkov-Glick model}

As a final example, we examine the LMG model~\cite{Lipkin:1964yk, Xu:2019lhc}, a spin system that provides a second realization of saddle-dominated scrambling. Its Hamiltonian is
\begin{align}\label{LMG_Ham}
    H=\hat{x}+2\hat{z}^2\,,
\end{align}
where
\begin{align}
\{\hat{x},\hat{y},\hat{z}\}
\equiv
\left\{
\frac{\hat S_x}{S},
\frac{\hat S_y}{S},
\frac{\hat S_z}{S}
\right\}.
\end{align}
These collective variables satisfy
\begin{align}
    [\hat{x},\hat{y}]
    =
    i\hbar_{\rm eff}\hat{z}\,,
    \qquad
    \hbar_{\rm eff}=\frac{1}{S}\,.
\end{align}
In the semiclassical limit $S\rightarrow\infty$, the dynamics is described on the unit sphere,
\begin{align}
x^2+y^2+z^2=1 \,,
\end{align}
with the corresponding $\mathrm{SU}(2)$ Poisson algebra. The classical phase space contains an unstable saddle that generates local exponential sensitivity and produces false-positive signatures in both OTOCs and Krylov dynamics despite its integrability~\cite{Bhattacharjee:2022vlt,Huh:2023jxt}.

The bottom row of Fig.~\ref{Fig.non-generic.KC} shows that the raw spread complexity again develops a pronounced chaos-like peak. As in the IHO, this behavior is associated with the accumulation of spectral weight induced by the saddle. Consistent with this picture, after unfolding, the peak disappears and is replaced by oscillations whose amplitude decays and stabilizes through dephasing.

\vspace{2mm}

The agreement among these three examples points to a common mechanism. In the logarithmic toy spectrum, the effect can be isolated directly, while the IHO and LMG model provide physical saddle-dominated realizations. In each case, the false-positive peak originates from the smooth spectral structure and is removed by unfolding, whereas the remaining oscillatory dynamics reflects the underlying microscopic spectrum. This provides the non-generic counterpart to the numerical results of Sec.~\ref{sec3}: unfolding preserves RMT-universal signatures when present, but suppresses structures generated solely by the non-universal density of states.

\section{Analytic framework}\label{sec5}
\setcounter{paragraph}{0}

The results of Secs.~\ref{sec3} and~\ref{sec4} show that spread complexity is sensitive both to universal spectral correlations and to the non-universal smooth density of states. To make this dependence explicit, we now reformulate the Lanczos construction in terms of orthogonal polynomials. This provides an exact representation of spread complexity for arbitrary finite discrete spectra and clarifies how spectral information enters the Krylov dynamics.

\subsection{Orthogonal-polynomial representation of the Krylov basis}

The Lanczos recursion algorithm admits an exact reformulation in terms of orthogonal polynomials associated with the spectral measure of the Hamiltonian and initial state~\cite{Muck:2022xfc,Balasubramanian:2026azk,Caputa:2025ozd,Balasubramanian:2025xkj}. More explicitly, to set up this framework, consider a Hamiltonian with eigenstates $\{|E_k\rangle\}$. The initial state can then be expanded as
\begin{align}
|\psi(0)\rangle
=
\sum_{k=0}^{N-1}
\sqrt{w_k}\,|E_k\rangle\,,
\label{eq:psi0_energy_basis}
\end{align}
where
\begin{align}
w_k
\equiv
|\langle E_k|\psi(0)\rangle|^2\,,
\qquad
\sum_{k=0}^{N-1}w_k=1\,.
\end{align}
For the infinite-temperature TFD state considered in this work, the spectral weights are uniform, $w_k=1/N$.

The pair $\{E_k,w_k\}$ defines the discrete spectral measure
\begin{align}
\mathrm{d}\mu(E)
=
\sum_{k=0}^{N-1}
w_k\,\delta(E-E_k)\,\mathrm{d}E\,.
\end{align}
By Favard's theorem~\cite{favard1935polynomes, shohat1938polynomes}, the Lanczos recursion is equivalent to the three-term recurrence of a family of orthonormal polynomials $\{h_n(E)\}$ with respect to this measure. The Krylov states can therefore be written as
\begin{align}
|K_n\rangle
=
h_n(H)|\psi(0)\rangle
=
\sum_{k=0}^{N-1}
\sqrt{w_k}\,
h_n(E_k)\,
|E_k\rangle\,.
\label{eq:Krylov_state_polynomial}
\end{align}
Projecting the Lanczos recurrence onto the energy eigenbasis gives
\begin{align}
E_k h_n(E_k)
=
a_n h_n(E_k)
+
b_{n+1}h_{n+1}(E_k)
+
b_n h_{n-1}(E_k)\,,
\label{eq:poly_three_term}
\end{align}
with $h_{-1}(E)=0$ and $h_0(E)=1$. Likewise, the orthonormality of the Krylov basis implies
\begin{align}
\sum_{k=0}^{N-1}
w_k\,h_n(E_k)h_m(E_k)
=
\delta_{nm}\,.
\label{eq:poly_orthonormality}
\end{align}
Thus, the Lanczos coefficients are the recurrence coefficients of the orthogonal polynomials associated with the spectral measure. This equivalence makes the dependence of Krylov dynamics on the spectrum explicit.

\subsection{Krylov amplitudes and complexity kernel}

Under unitary evolution,
\begin{align}
|\psi(t)\rangle
=
\sum_{k=0}^{N-1}
\sqrt{w_k}\,
e^{-iE_kt}|E_k\rangle\,,
\end{align}
the Krylov amplitudes $\psi_n(t)\equiv\langle K_n|\psi(t)\rangle$ become
\begin{align}
\psi_n(t)
=
\sum_{k=0}^{N-1}
w_k\,e^{-iE_kt}\,h_n(E_k)\,.
\label{eq:general_psi_n}
\end{align}
Substituting this expression into the definition of spread complexity yields a double sum over energy levels,
\begin{align}
C(t)
=
\sum_{k,k'=0}^{N-1}
w_kw_{k'}\,
e^{-i(E_k-E_{k'})t}\,
\mathcal{K}(E_k,E_{k'})\,,
\label{eq:general_C_kernel}
\end{align}
where
\begin{align}
\mathcal{K}(E_k,E_{k'})
\equiv
\sum_{n=0}^{N-1}
n\,h_n(E_k)h_n(E_{k'}) \,,
\label{eq:kernel_def}
\end{align}
defines the \emph{complexity kernel}. This representation cleanly separates the explicit time dependence, carried by energy differences, from the spectral information encoded in the polynomial basis of the Krylov dynamics.

To evaluate the kernel, we introduce the partial sum
\begin{align}
F_n(E_k,E_{k'})
\equiv
\sum_{m=0}^{n}
h_m(E_k)h_m(E_{k'})\,.
\end{align}
Applying Abel's summation formula~\cite{apostol1976introduction} directly to Eq.~\eqref{eq:kernel_def} then yields
\begin{align}
\mathcal{K}(E_k,E_{k'})
=
(N-1)F_{N-1}(E_k,E_{k'})
-
\sum_{n=0}^{N-2}
F_n(E_k,E_{k'})\,.
\end{align}
Completeness of the $N$-point polynomial basis implies
\begin{align}
F_{N-1}(E_k,E_{k'})
=
\frac{\delta_{kk'}}{w_k}\,.
\end{align}
For $k\neq k'$, the remaining partial sums can be evaluated in a compact form using the Christoffel--Darboux identity~\cite{christoffel1858ueber,darboux1878memoire,szego1939orthogonal} associated with Eq.~\eqref{eq:poly_three_term},
\begin{align}
F_n(E_k,E_{k'})
&=\nonumber\\
b_{n+1}
&\frac{
h_{n+1}(E_k)h_n(E_{k'})
-
h_n(E_k)h_{n+1}(E_{k'})
}{
E_k-E_{k'}
}\,.
\end{align}
Shifting $n\rightarrow n-1$ yields the exact off-diagonal kernel,
\begin{align}
\mathcal{K}(E_k,E_{k'})
&=\nonumber\\
-\sum_{n=1}^{N-1}
&b_n
\frac{
h_n(E_k)h_{n-1}(E_{k'})
-
h_{n-1}(E_k)h_n(E_{k'})
}{
E_k-E_{k'}
}\,,
\label{eq:offdiag_kernel}
\end{align}
for $k\neq k'$. Finally, separating Eq.~\eqref{eq:general_C_kernel} into diagonal and off-diagonal contributions gives
\begin{align}
C(t)
&=
C_0
-
2\sum_{n=1}^{N-1}b_n
\sum_{0\leq k<k'\leq N-1}
w_kw_{k'}
\frac{\cos[(E_k-E_{k'})t]}{E_k-E_{k'}}
\nonumber\\
&\qquad\times
\Big[
h_n(E_k)h_{n-1}(E_{k'})
-
h_{n-1}(E_k)h_n(E_{k'})
\Big]\,,
\label{eq:most_general_spread_complexity}
\end{align}
where
\begin{align}
C_0 \equiv
\sum_{k=0}^{N-1}
w_k^2\,\mathcal{K}(E_k,E_k) = \sum_{n=0}^{N-1}
n\sum_{k=0}^{N-1}
w_k^2[h_n(E_k)]^2\,.
\end{align}
For a nondegenerate spectrum, $C_0$ gives the infinite-time average and hence the dephased late-time baseline. Eq.~\eqref{eq:most_general_spread_complexity} thus provides an exact expression for the spread complexity of any finite discrete spectral measure.

\begin{figure}[t!]
\renewcommand{\thempfootnote}{\arabic{mpfootnote}}
 \centering
     \vspace{1.2mm}
     {\includegraphics[width=4.1cm]{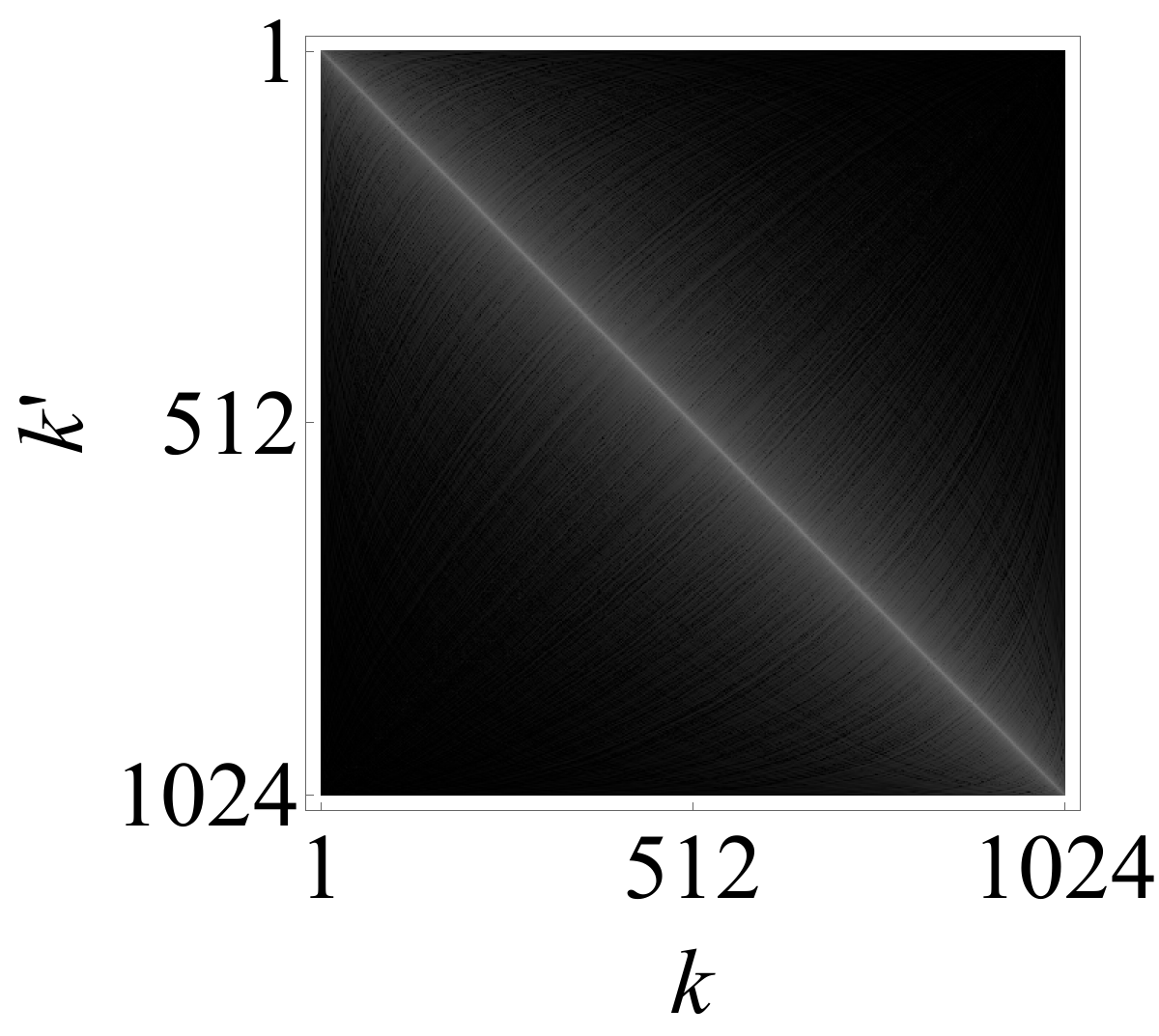}}
     \hspace{1.2mm}
     {\includegraphics[width=4.1cm]{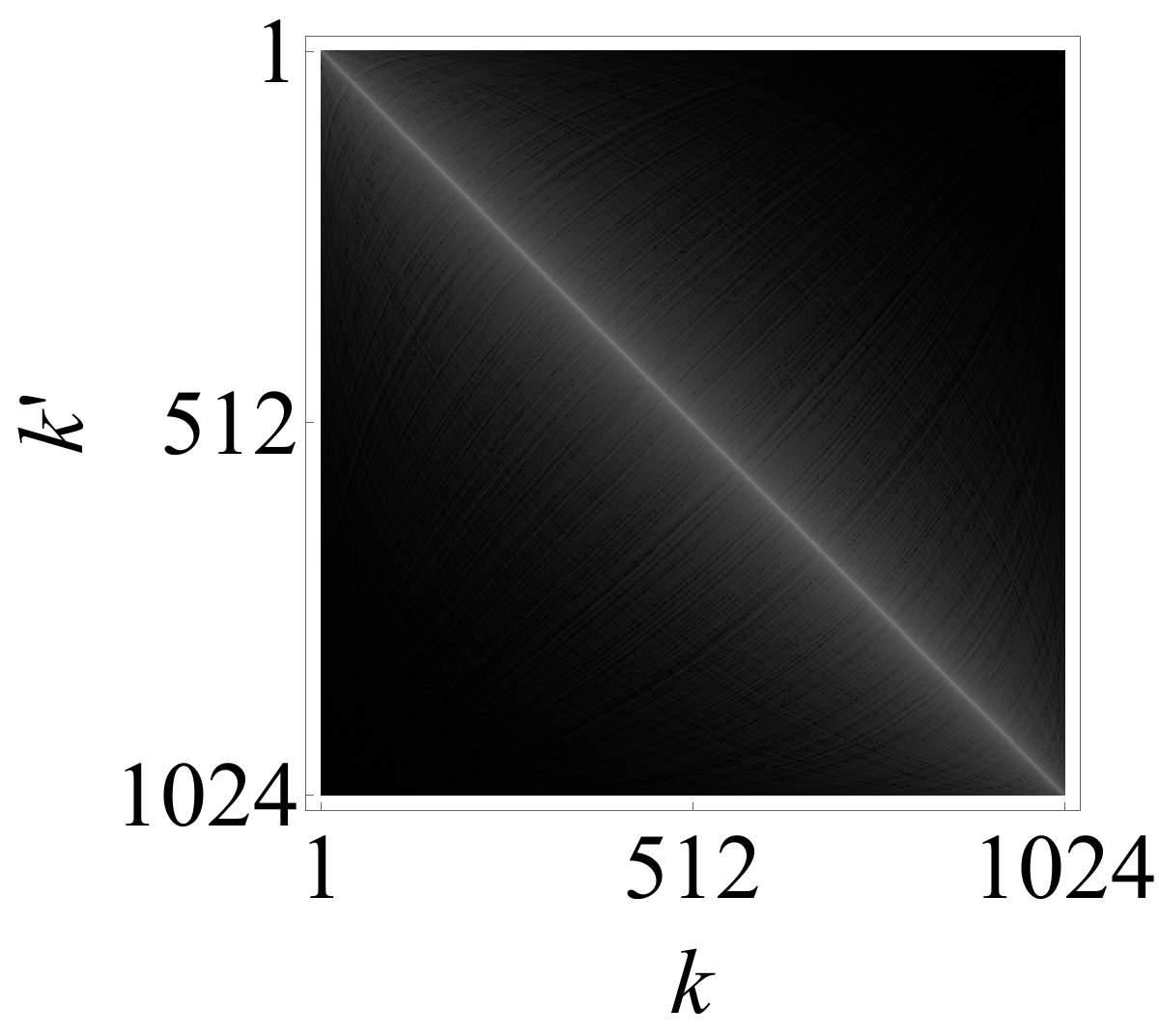}} \\
     \vspace{1.2mm}
     {\includegraphics[width=4.1cm]{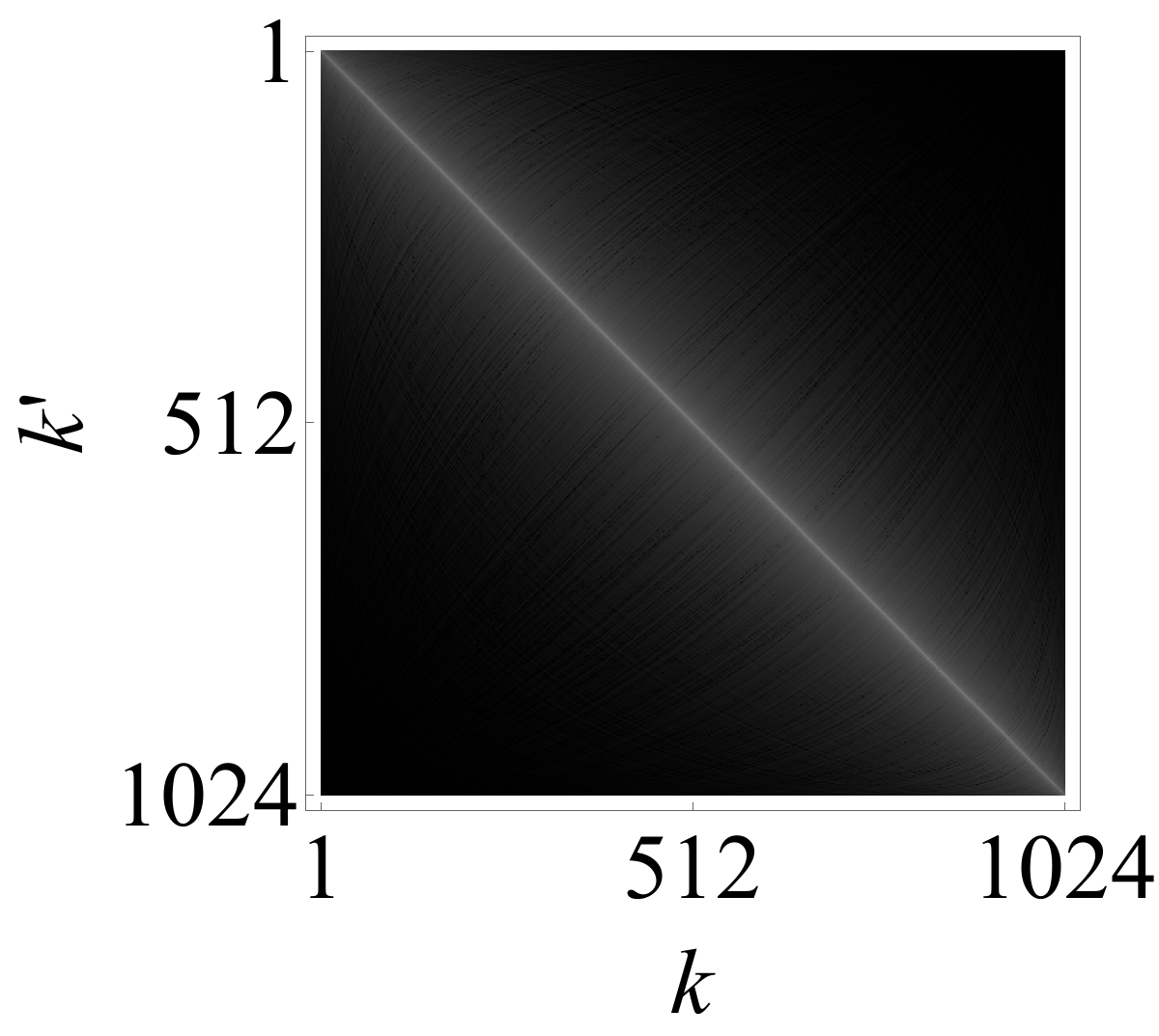}}
     \hspace{1.2mm}
     {\includegraphics[width=4.1cm]{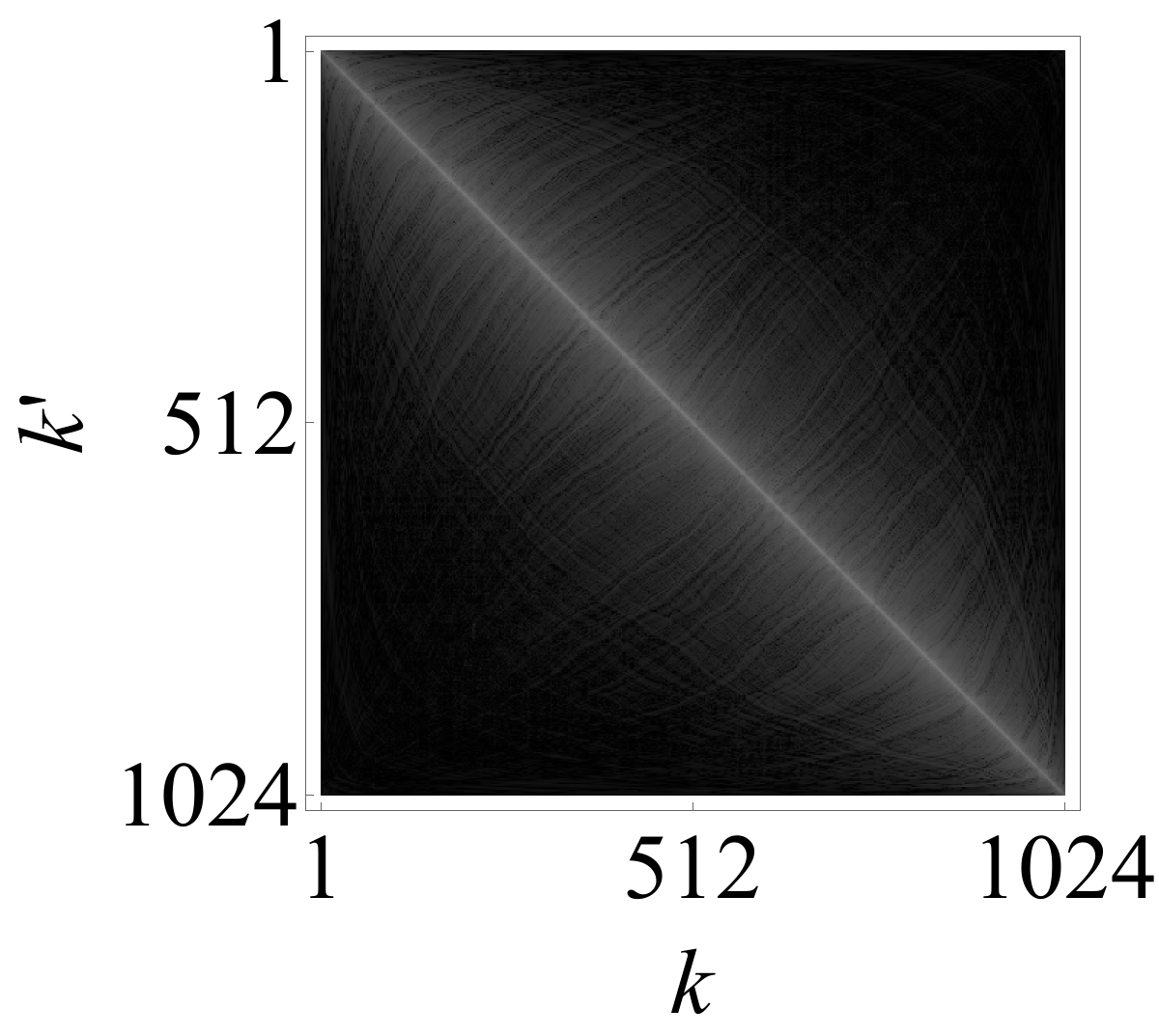}}
\vspace{-3mm}
\caption{Complexity kernel $\mathcal{K}(E_k,E_{k'})$ for GOE, GUE, GSE, and Poisson spectra (top-left, top-right, bottom-left and bottom-right, respectively), with $N=1024$. All four cases exhibit an approximately banded structure around the diagonal, showing that bandedness alone is not a signature of RMT universality. The distinction between RMT and Poisson statistics instead appears in the off-diagonal structure, which encodes the different microscopic spectral correlations.}
\label{Fig.Un_Kernel.KC}
\end{figure}

Fig.~\ref{Fig.Un_Kernel.KC} illustrates the complexity kernel for the GOE, GUE, GSE, and Poisson spectra. In all cases, the kernel exhibits an approximately banded structure, with its largest entries concentrated near the main diagonal. Specifically the $k$-th off-diagonal band weight, defined via the complexity kernel $\mathcal{K}$ given in  Eq.~\eqref{eq:kernel_def}, 
\begin{align}
W_k = \sum_{i=0}^{N-k-1} \mathcal{K}(E_i, E_{i+k})\,,
\end{align}
decays algebraically as $|W_k| \sim k^{-2}$ away from the diagonal. Bandedness is therefore not, by itself, a signature of RMT universality. Instead, the distinction lies in the finer off-diagonal behavior: RMT spectra display a characteristic band profile associated with spectral rigidity and level repulsion, whereas the Poisson spectrum exhibits pronounced fluctuations across the bands. The complexity kernel structure given in Eq.~\eqref{eq:general_C_kernel} and its banded structure align with the independent derivation of Ref.~\cite{Basu:2026gvl}. In addition, here  we explicitly express the spread complexity, Eq. \eqref{eq:most_general_spread_complexity}, in terms of the Lanczos coefficients $b_n$ and orthonormal polynomials $h_n(E)$.

Crucially, the analytic formulation in Eq.~\eqref{eq:most_general_spread_complexity} reveals the spectral ingredients governing spread com\-plexity. In deriving the off-diagonal kernel in Eq.~\eqref{eq:offdiag_kernel} from Eq.~\eqref{eq:kernel_def}, the diagonal Lanczos coefficients $a_n$ cancel explicitly owing to the underlying three-term recurrence relation. The resulting expression depends explicitly on the off-diagonal hopping amplitudes $b_n$, the energy differences $E_k-E_{k'}$, and the adjacent polynomial combinations $h_n(E_k)h_{n-1}(E_{k'})-h_{n-1}(E_k)h_n(E_{k'})$. This representation shows how the interplay between hopping along the Krylov chain and spectral correlations governs the full time dependence of spread complexity.

More generally, the spectral measure $\mu(E)$ determines the orthogonal-polynomial family $h_n(E)$ and the Lanczos coefficients~\cite{Muck:2022xfc,Balasubramanian:2026azk}; see also~\cite{Qu:2025lgo,Pedraza:2026zji,Qu:2026dmv}. Representative examples illustrate this correspondence: in the continuum, a uniform measure on a bounded interval leads to Legendre polynomials, whereas a Gaussian measure produces Hermite polynomials with $b_n\propto\sqrt{n}$. An exponentially decaying measure on the half-line leads to Laguerre polynomials with $b_n\propto n$, while the Wigner semicircle measure yields Chebyshev polynomials of the second kind and asymptotically constant $b_n$. Finally, a finite, uniformly spaced spectrum yields discrete Chebyshev polynomials.

This finite-lattice case is especially relevant for spectral unfolding. As shown in the next section, mapping the logarithmic spectrum exactly onto a uniform lattice allows its unfolded Krylov dynamics to be solved analy\-tically in terms of discrete Chebyshev polynomials.


\section{Unfolded Krylov complexity from orthogonal polynomials}\label{sec6}
\setcounter{paragraph}{0}

\subsection{Unfolded spectral measure and discrete Chebyshev polynomials}
\label{subsec:unfolding_lanczos}

We now apply the orthogonal-polynomial framework of Sec.~\ref{sec5} directly to unfolded spectra. Under the unfolding map $E_k\rightarrow\xi_k$, the eigenstates remain unchanged, $|\xi_k\rangle\equiv|E_k\rangle$, and so do the spectral weights of the initial state. 

In this setting, the logarithmic spectrum analyzed in Sec.~\ref{sec4} is particularly useful because the unfolding can be performed exactly, yielding a uniform discrete lattice and hence an analytically solvable Krylov problem. Specifically, consider the logarithmic spectrum of Eq.~\eqref{toy}. Its smooth density of states grows exponentially,
\begin{align}
\bar{\rho}(E)
=
\frac{\mathrm{d}k}{\mathrm{d}E}
=
e^E\,,
\qquad
E\in[0,\log N]\,.
\end{align}
The cumulative unfolding map is therefore exact,
\begin{align}
\xi(E)
=
\int_0^E
\bar{\rho}(E')\,\mathrm{d}E'
=
e^E-1\,,
\label{eq:toy_unfolding_map}
\end{align}
and maps the original levels to the uniform lattice with
\begin{align}
\xi_k=\xi(E_k)=k\,,
\qquad
k=0,1,\ldots,N-1\,.
\end{align}

For the infinite-temperature TFD state, the spectral weights are uniform, $w_k=1/N$, so the measure becomes
\begin{align}
\mathrm{d}\mu_{\rm U}(\xi)
=
\frac{1}{N}
\sum_{k=0}^{N-1}
\delta(\xi-k)\,\mathrm{d}\xi\,.
\label{eq:toy_unfolded_measure}
\end{align}
The combination of uniform weights and an equidistant finite lattice selects the discrete Chebyshev polynomials $T_n(\xi)$ (also known as Gram polynomials) as the relevant orthogonal-polynomial family $h_n(\xi)$. For completeness, some of their properties and important recurrence relations are summarized in Appendix~\ref{appA}. In the next subsection, we analyze the resulting exact Krylov dynamics and contrast it with our previous numerical results.

\subsection{Exact solution for the logarithmic spectrum}

\begin{figure}[t!]
 \centering
    \vspace{1.2mm}
     {\includegraphics[width=7.1cm]{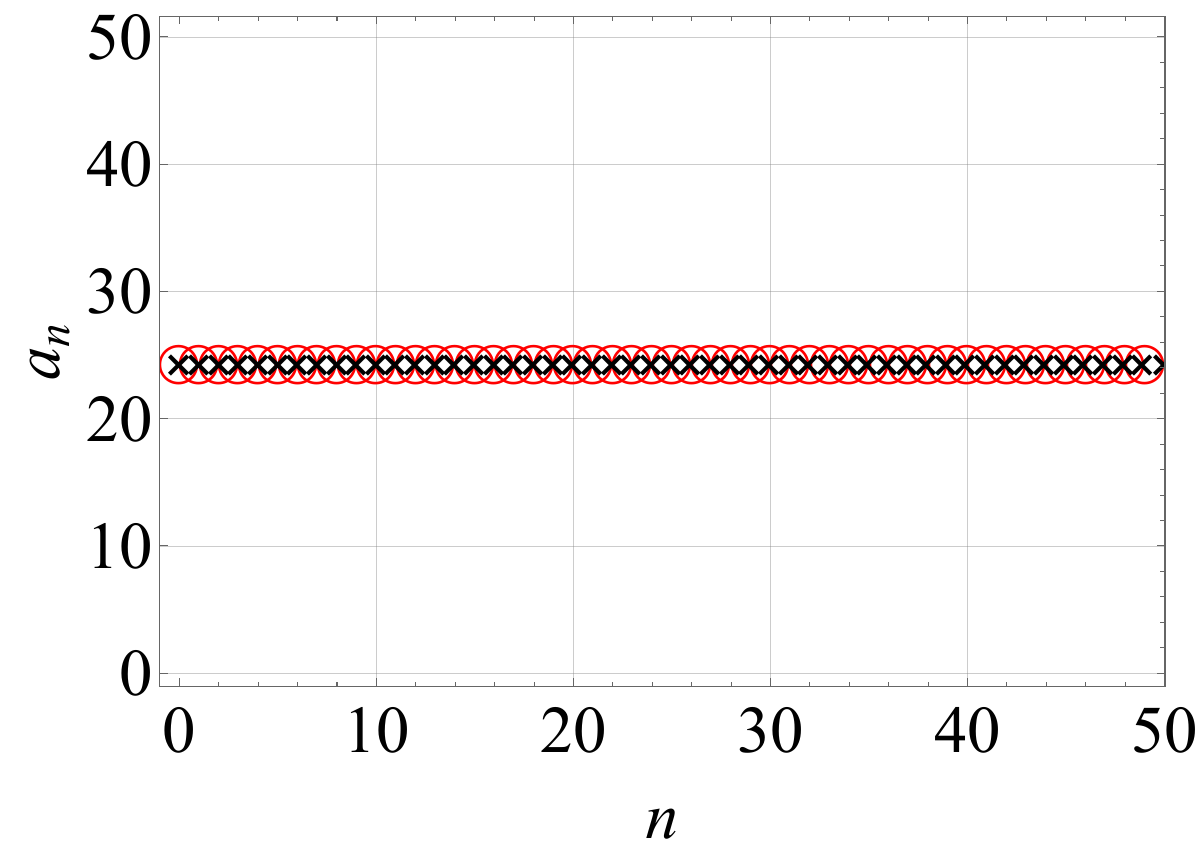}}
     {\includegraphics[width=7.1cm]{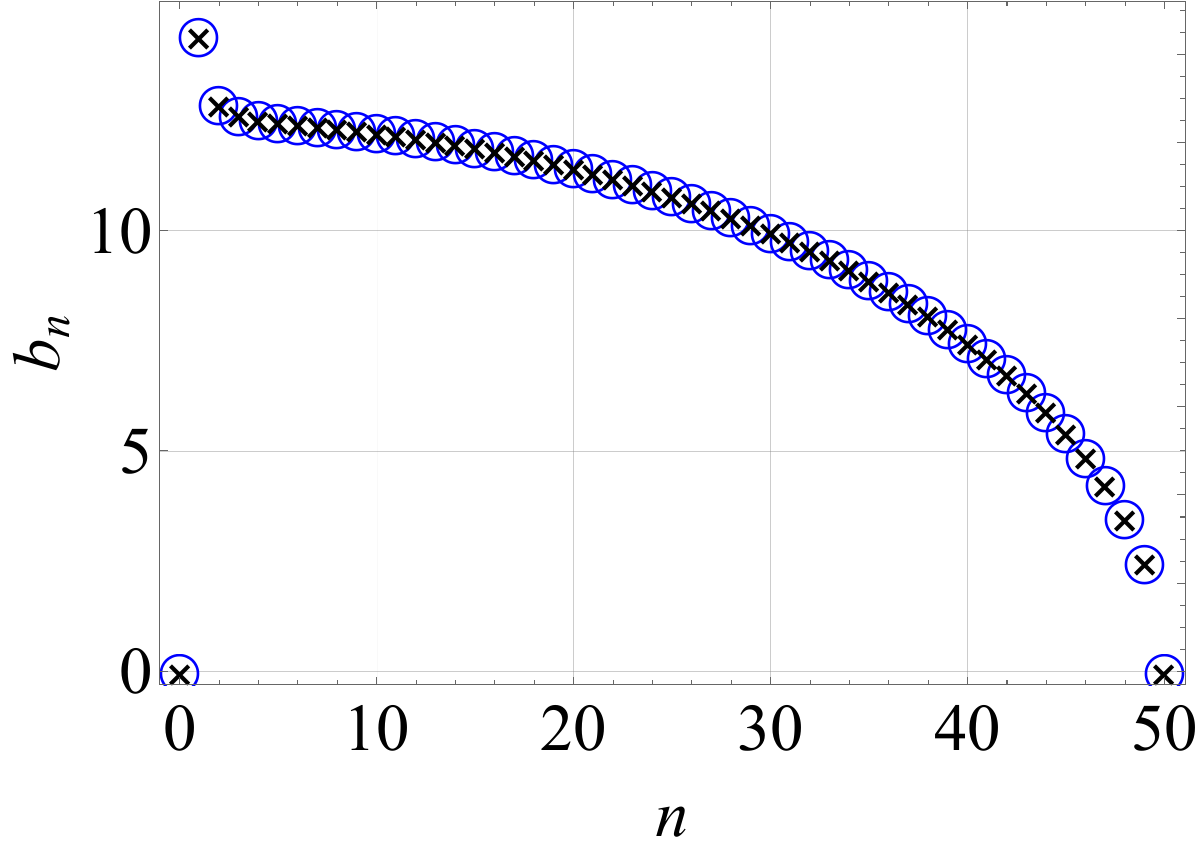}}
\vspace{-3mm}
\caption{Lanczos coefficients for the unfolded logarithmic spectrum with $N=50$. The top and bottom panels show the diagonal $a_n$ and off-diagonal $b_n$ coefficients, respectively. Numerical results (red and blue circles) agree with the exact discrete-Chebyshev predictions (black crosses; Eq.~\eqref{eq:Lanczos_a}), confirming the analytic finite-lattice Krylov construction.}
\label{Fig.Log.Lanc}
\end{figure}

Using the recurrence relations of the discrete Chebyshev polynomials, we can obtain the exact Lanczos coefficients for the unfolded logarithmic spectrum. The derivation is presented in Appendix~\ref{appB}, and yields
\begin{align}\label{eq:Lanczos_a}
\begin{split}
a_n
&=
\frac{N-1}{2}\,,\\
b_n
&=
\frac{n}{2}
\sqrt{
\frac{N^2-n^2}{4n^2-1}
}\,,
\qquad
n=1,\ldots,N-1\,,
\end{split}
\end{align}
together with $b_0=b_N=0$, ensuring termination of the Krylov chain at $n=N$. As shown in Fig.~\ref{Fig.Log.Lanc}, the numerical Lanczos coefficients agree excellently with Eq.~\eqref{eq:Lanczos_a}.

For the uniform measure~\eqref{eq:toy_unfolded_measure}, the orthonormal polynomials $h_n(\xi)$ are proportional to the normalized discrete Chebyshev polynomials $T_n(\xi)$ introduced in Appendix~\ref{appA}. With the normalization adopted there,
\begin{align}
h_n(\xi)=\sqrt{N}\,T_n(\xi)\,.
\end{align}
Using these polynomials together with the Lanczos coefficients~\eqref{eq:Lanczos_a}, the general expression~\eqref{eq:most_general_spread_complexity} becomes
\begin{align}
C(t)
=
\frac{N-1}{2}
-
\frac{2}{N}
\sum_{n=1}^{N-1}
b_n
\sum_{0\leq x<y\leq N-1}
\frac{\cos[(x-y)t]}{x-y}
\nonumber\\
\times
\Big[
T_n(x)T_{n-1}(y)
-
T_{n-1}(x)T_n(y)
\Big]\,.
\label{eq:exact_chebyshev_complexity}
\end{align}
This is an exact analytic expression for the unfolded spread complexity of the logarithmic model.

\begin{figure}[t!]
    \centering
    \vspace{1.2mm}
    {\includegraphics[width=7.1cm]{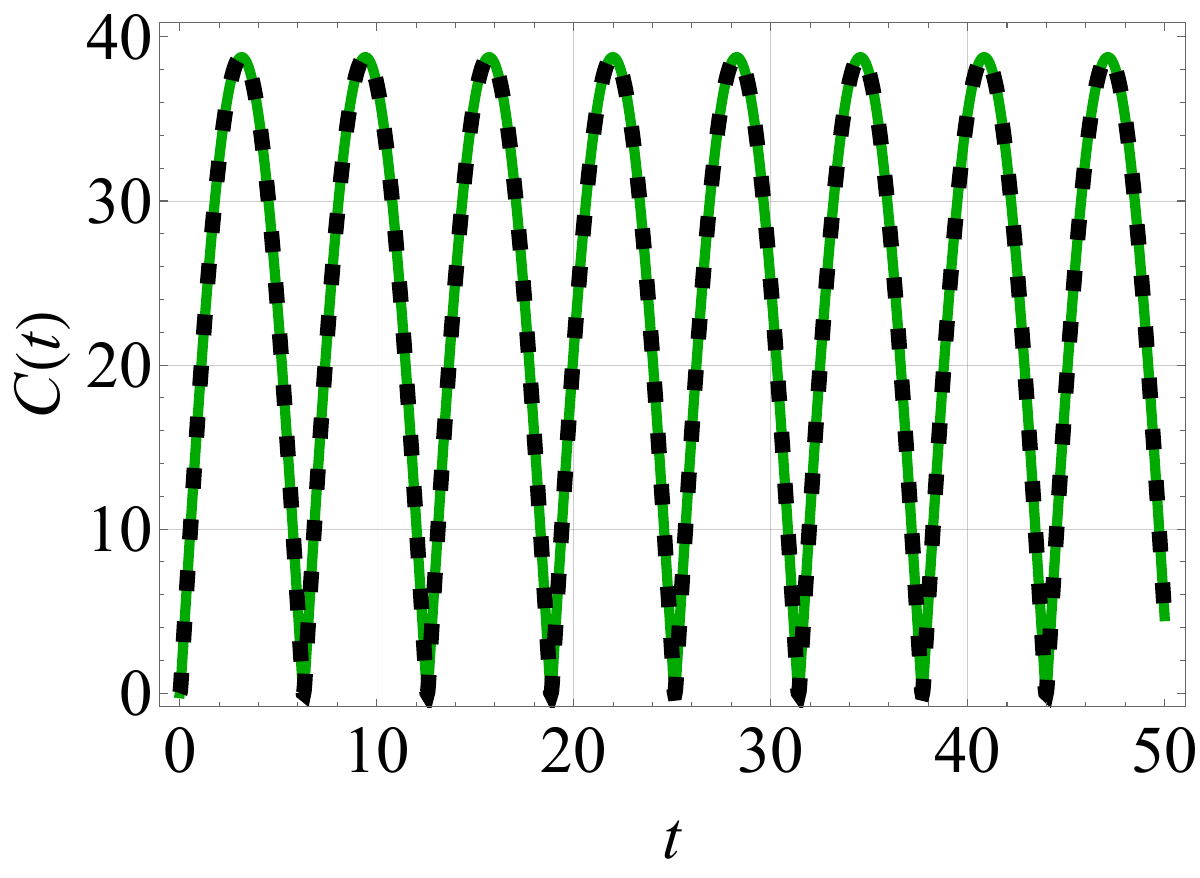}}
    \vspace{-3mm}
    \caption{Unfolded spread complexity for the logarithmic spectrum with $N=50$. The numerical evolution (solid green curve) agrees with the analytical result in Eq.~\eqref{eq:exact_chebyshev_complexity} (black dashed curve). The false-positive peak of the raw spectrum is absent after unfolding, confirming that it originates from the smooth spectrum rather than from RMT-type correlations.}
    \label{Fig.Log.KC}
\end{figure}
Fig.~\ref{Fig.Log.KC} shows exact agreement between Eq.~\eqref{eq:exact_chebyshev_complexity} and the numerical Krylov evolution. Crucially, the analytic result contains no false-positive peak and instead reproduces the oscillatory behavior found numerically in Sec.~\ref{sec4}, more closely resembling integrable dynamics. The logarithmic model thus provides an explicit analytical realization of the mechanism underlying our proposal: removing the exponentially varying smooth density maps the spectrum to a uniform lattice, leaving the unfolded spectral structure to govern the Krylov dynamics.

\section{Discussion}\label{sec7}
\setcounter{paragraph}{0}
In this work, we investigated whether spectral unfolding can sharpen spread complexity (Krylov state complexity) as a diagnostic of quantum chaos. Across RMT ensembles and representative physical systems, we found that unfolding preserves the characteristic signatures of RMT-universal dynamics, including the peak-relaxation structure and the hierarchy among Dyson symmetry classes. The triangular billiards further show that these distinctions persist even in systems without classical exponential instability, separating integrable, pseudo-integrable, and non-integrable regimes. At the same time, it removes the false-positive peaks found in the logarithmic spectrum and in the saddle-dominated IHO and LMG models. These results support unfolded Krylov complexity as a cleaner probe of microscopic spectral correlations than the raw spread complexity.

The underlying mechanism is simple but physically important. Spread complexity constructed from the raw spectrum is sensitive not only to level correlations, but also to the smooth density of states through the moments that determine the Lanczos coefficients. Consequently, a strongly varying spectral density, for instance near a saddle-induced accumulation of levels, can generate nontrivial Krylov dynamics even in the absence of RMT correlations. Unfolding filters out this macroscopic contribution while retaining the local spectral fluctuations. From this perspective, the false-positive peak is not intrinsically a signature of scrambling, but can instead reflect non-universal spectral structure. This mechanism is seen most cleanly in the logarithmic toy spectrum, and persists in the physical saddle-dominated IHO and collective LMG models, suggesting that it is not tied to a particular realization but to the underlying spectral structure.

Our orthogonal-polynomial formulation provides an analytical perspective on this separation. For arbitrary finite discrete spectra, the Lanczos recursion can be recast in terms of orthogonal polynomials, leading to an exact representation of spread complexity through a two-energy kernel. Interestingly, the approximately banded structure of this kernel is common to both chaotic and integrable spectra; the distinction between them is instead encoded in its finer off-diagonal features. For the logarithmic model, unfolding maps the spectrum exactly onto a uniform lattice governed by discrete Chebyshev polynomials, allowing the unfolded Lanczos coefficients and complexity to be obtained analytically. The absence of a false-positive peak in this example gives an explicit realization of the mechanism identified numerically.

Unfolding also places otherwise disparate spectra on a common spectral scale. In RMT-universal systems, we found that the onset of late-time saturation of spread complexity aligns, in unfolded units, with the Heisenberg scale $t\simeq 2\pi$ of the spectral form factor. This correspondence supports the view that the late-time structure of spread complexity is controlled by microscopic spectral correlations rather than by the overall density of states.

Several questions remain open. Our analysis focused primarily on the infinite-temperature TFD state, for which the spectral weights are uniform. It would be interesting to determine how the picture changes at finite temperature or for more general initial states, where nontrivial spectral weights provide an additional source of structure. The dependence on the unfolding prescription, particularly in finite systems or near spectral singularities, also deserves further study. More broadly, extending these ideas to operator Krylov complexity would require identifying the appropriate unfolding of the Liouvillian spectral measure and could clarify whether saddle-dominated exponential growth can be filtered in an ana\-logous way. Moreover, applying unfolded Krylov complexity across chaos--integrability transitions, 
such as in 
\cite{Aguilar-Gutierrez:2026jjv},
may help determine which features of Krylov dynamics are genuinely universal and which retain model-dependent information. Finally, it will also be of interest to analyze unfolded Krylov complexity in open quantum systems, extending \cite{Bhattacharya:2023yec,Nandy:2024mml}.

\vspace{2mm}
\textit{Acknowledgments.}
JE acknowledges financial support  by the Deutsche Forschungsgemeinschaft (DFG, German Research Foundation) by the German-Korean exchange grant ChaosAdS\-CFT (project-id 566966698), as well as under Germany's Excellence Strategy through the W\"{u}rzburg-Dresden Cluster of Excellence on Complexity and Topology in Quantum Matter - ctd.qmat (EXC 2147, project-id 390858490).
KBH acknowledges the support of the Foreign Young Scholars Research Fund Project (Grant No.22Z033100604). 
HSJ is supported by an appointment to the JRG Program at the APCTP through the Science and Technology Promotion Fund and Lottery Fund of the Korean Government. This is also supported by the Korean Local Governments -- Gyeongsangbuk-do Province and Pohang City.
JFP is supported by the ‘Atracción de Talento’ program of the Comunidad de Madrid under grant 2020-T1/TIC-20495, the Spanish Agencia Estatal de Investigación through grants CEX2025-001574-S and PID2024-156043NB-I00, funded by MCIN/AEI/10.13039/501100011033 (the research presented in this publication falls within the research line Strings and Quantum Gravity).
We would also like to thank the APCTP Focus Program, ``Holography 2026: Correlation and Entanglement in Quantum Matter'' in Pohang, Korea, held in Pohang, Korea, where part of this work was completed.

\appendix

\section{Discrete Chebyshev polynomials}\label{appA}

For the uniform finite lattice
\begin{align}
x\in\{0,1,\ldots,N-1\}\,,
\end{align}
the relevant orthogonal basis is provided by the discrete Chebyshev polynomials, also known as Gram polynomials~\cite{Muck:2022xfc,Balasubramanian:2026azk}. We collect here the recurrence and normalization formulas needed in our analysis.

The unnormalized polynomials $t_n(x)$ satisfy the three-term recurrence
\begin{align}\label{eq:discrete_chebyshev_recurrence}
(n+1)t_{n+1}(x)
&=
(2n+1)(2x-N+1)t_n(x)
\nonumber\\
&\quad
-n(N^2-n^2)t_{n-1}(x)\,,
\end{align}
for $1\leq n\leq N-2$, with
\begin{align}
t_0(x)=1\,,
\qquad
t_1(x)=2x-N+1\,.
\end{align}
Their discrete orthogonality relation is
\begin{align}\label{eq:unnormalized_orthogonality}
\sum_{x=0}^{N-1}
t_n(x)t_m(x)
=
H_n^2\,\delta_{nm}\,,
\end{align}
where
\begin{align}\label{eq:squared_norm}
H_n^2
&=
\frac{
N\prod_{j=1}^{n}(N^2-j^2)
}{
2n+1
}
\nonumber\\
&=
\frac{2n-1}{2n+1}
(N^2-n^2)H_{n-1}^2\,,
\end{align}
with $H_0^2=N$. Defining the normalized polynomials
\begin{align}
T_n(x)\equiv\frac{t_n(x)}{H_n}\,,
\end{align}
we obtain
\begin{align}\label{eq:normalized_orthogonality}
\sum_{x=0}^{N-1}
T_n(x)T_m(x)
=
\delta_{nm}\,.
\end{align}

Substituting $t_n(x)=H_nT_n(x)$ into Eq.~\eqref{eq:discrete_chebyshev_recurrence} gives the normalized recurrence
\begin{align}\label{eq:orthonormal_recurrence}
T_{n+1}(x)
=
\alpha_n(2x-N+1)T_n(x)
-
\beta_nT_{n-1}(x)\,,
\end{align}
where
\begin{align}
\alpha_n
&=
\frac{\sqrt{(2n+1)(2n+3)}}{n+1}
\frac{1}{\sqrt{N^2-(n+1)^2}}\,,
\label{eq:alpha_n}
\\[0.5em]
\beta_n
&=
\frac{n}{n+1}
\sqrt{\frac{2n+3}{2n-1}}
\sqrt{\frac{N^2-n^2}{N^2-(n+1)^2}}\,.
\label{eq:beta_n}
\end{align}
These coefficients provide the starting point for the exact Lanczos coefficients derived below in Appendix~\ref{appB}.

\smallskip


\section{Lanczos coefficients from discrete Chebyshev polynomials}\label{appB}

We now extract the Lanczos coefficients associated with the uniform finite
lattice. As discussed in Sec.~\ref{sec6}, the orthonormal polynomials with
respect to the measure~\eqref{eq:toy_unfolded_measure} are
$h_n(x)=\sqrt{N}\,T_n(x)$. Since this overall normalization is independent
of $n$, the polynomials $h_n$ and $T_n$ obey the same three-term recurrence
coefficients.

Starting from the normalized recurrence relation~\eqref{eq:orthonormal_recurrence},
\begin{align}
\alpha_n(2x-N+1)T_n(x)
=
T_{n+1}(x)
+
\beta_nT_{n-1}(x)\,,
\end{align}
we solve for the action of the multiplication operator $x$:
\begin{align}
xT_n(x)
&=
\frac{1}{2\alpha_n}T_{n+1}(x)
+
\frac{N-1}{2}T_n(x)
+
\frac{\beta_n}{2\alpha_n}T_{n-1}(x)\,.
\label{eq:chebyshev_lanczos_form}
\end{align}
The recursion is initialized by~\cite{denBrinker2020Controlled}
\begin{align}
T_0(x)
&=
\frac{1}{\sqrt{N}}\,,
\\[0.5em]
T_1(x)
&=
\frac{\sqrt{3}(2x-N+1)}
{\sqrt{N(N^2-1)}}\,.
\end{align}

Comparing Eq.~\eqref{eq:chebyshev_lanczos_form} with the polynomial form
of the Lanczos recurrence,
\begin{align}
x h_n(x)
=
b_{n+1}h_{n+1}(x)
+
a_nh_n(x)
+
b_nh_{n-1}(x)\,,
\end{align}
we identify
\begin{align}
a_n
&=
\frac{N-1}{2}\,,
\\
b_{n+1}
&=
\frac{1}{2\alpha_n}\,.
\end{align}
Using Eq.~\eqref{eq:alpha_n} and shifting $n\rightarrow n-1$ gives
\begin{align}
b_n
&=
\frac{n}{2}
\frac{\sqrt{N^2-n^2}}
{\sqrt{(2n-1)(2n+1)}}
\nonumber\\
&=
\frac{n}{2}
\sqrt{\frac{N^2-n^2}{4n^2-1}}\,,
\qquad
n=1,2,\ldots,N-1\,.
\label{eq:lanczos_bn_exact}
\end{align}
Together with $b_0=b_N=0$, these coefficients reproduce
Eq.~\eqref{eq:Lanczos_a} reported in
Sec.~\ref{sec6}.

\bibliographystyle{apsrev4-1}
\bibliography{Ref}

\begin{thebibliography}{95}%
\makeatletter
\providecommand \@ifxundefined [1]{%
 \@ifx{#1\undefined}
}%
\providecommand \@ifnum [1]{%
 \ifnum #1\expandafter \@firstoftwo
 \else \expandafter \@secondoftwo
 \fi
}%
\providecommand \@ifx [1]{%
 \ifx #1\expandafter \@firstoftwo
 \else \expandafter \@secondoftwo
 \fi
}%
\providecommand \natexlab [1]{#1}%
\providecommand \enquote  [1]{``#1''}%
\providecommand \bibnamefont  [1]{#1}%
\providecommand \bibfnamefont [1]{#1}%
\providecommand \citenamefont [1]{#1}%
\providecommand \href@noop [0]{\@secondoftwo}%
\providecommand \href [0]{\begingroup \@sanitize@url \@href}%
\providecommand \@href[1]{\@@startlink{#1}\@@href}%
\providecommand \@@href[1]{\endgroup#1\@@endlink}%
\providecommand \@sanitize@url [0]{\catcode `\\12\catcode `\$12\catcode
  `\&12\catcode `\#12\catcode `\^12\catcode `\_12\catcode `\%12\relax}%
\providecommand \@@startlink[1]{}%
\providecommand \@@endlink[0]{}%
\providecommand \url  [0]{\begingroup\@sanitize@url \@url }%
\providecommand \@url [1]{\endgroup\@href {#1}{\urlprefix }}%
\providecommand \urlprefix  [0]{URL }%
\providecommand \Eprint [0]{\href }%
\providecommand \doibase [0]{http://dx.doi.org/}%
\providecommand \selectlanguage [0]{\@gobble}%
\providecommand \bibinfo  [0]{\@secondoftwo}%
\providecommand \bibfield  [0]{\@secondoftwo}%
\providecommand \translation [1]{[#1]}%
\providecommand \BibitemOpen [0]{}%
\providecommand \bibitemStop [0]{}%
\providecommand \bibitemNoStop [0]{.\EOS\space}%
\providecommand \EOS [0]{\spacefactor3000\relax}%
\providecommand \BibitemShut  [1]{\csname bibitem#1\endcsname}%
\let\auto@bib@innerbib\@empty
\bibitem [{\citenamefont {Shenker}\ and\ \citenamefont
  {Stanford}(2014)}]{Shenker:2013pqa}%
  \BibitemOpen
  \bibfield  {author} {\bibinfo {author} {\bibfnamefont {S.~H.}\ \bibnamefont
  {Shenker}}\ and\ \bibinfo {author} {\bibfnamefont {D.}~\bibnamefont
  {Stanford}},\ }\href {\doibase 10.1007/JHEP03(2014)067} {\bibfield  {journal}
  {\bibinfo  {journal} {JHEP}\ }\textbf {\bibinfo {volume} {03}},\ \bibinfo
  {pages} {067} (\bibinfo {year} {2014})},\ \Eprint
  {http://arxiv.org/abs/1306.0622} {arXiv:1306.0622 [hep-th]} \BibitemShut
  {NoStop}%
\bibitem [{\citenamefont {Shenker}\ and\ \citenamefont
  {Stanford}(2015)}]{Shenker:2014cwa}%
  \BibitemOpen
  \bibfield  {author} {\bibinfo {author} {\bibfnamefont {S.~H.}\ \bibnamefont
  {Shenker}}\ and\ \bibinfo {author} {\bibfnamefont {D.}~\bibnamefont
  {Stanford}},\ }\href {\doibase 10.1007/JHEP05(2015)132} {\bibfield  {journal}
  {\bibinfo  {journal} {JHEP}\ }\textbf {\bibinfo {volume} {05}},\ \bibinfo
  {pages} {132} (\bibinfo {year} {2015})},\ \Eprint
  {http://arxiv.org/abs/1412.6087} {arXiv:1412.6087 [hep-th]} \BibitemShut
  {NoStop}%
\bibitem [{\citenamefont {Garc\'{\i}a-Garc\'{\i}a}\ and\ \citenamefont
  {Verbaarschot}(2016)}]{PhysRevD.94.126010}%
  \BibitemOpen
  \bibfield  {author} {\bibinfo {author} {\bibfnamefont {A.~M.}\ \bibnamefont
  {Garc\'{\i}a-Garc\'{\i}a}}\ and\ \bibinfo {author} {\bibfnamefont {J.~J.~M.}\
  \bibnamefont {Verbaarschot}},\ }\href {\doibase 10.1103/PhysRevD.94.126010}
  {\bibfield  {journal} {\bibinfo  {journal} {Phys. Rev. D}\ }\textbf {\bibinfo
  {volume} {94}},\ \bibinfo {pages} {126010} (\bibinfo {year}
  {2016})}\BibitemShut {NoStop}%
\bibitem [{\citenamefont {Cotler}\ \emph {et~al.}(2017)\citenamefont {Cotler},
  \citenamefont {Gur-Ari}, \citenamefont {Hanada}, \citenamefont {Polchinski},
  \citenamefont {Saad}, \citenamefont {Shenker}, \citenamefont {Stanford},
  \citenamefont {Streicher},\ and\ \citenamefont {Tezuka}}]{Cotler:2016fpe}%
  \BibitemOpen
  \bibfield  {author} {\bibinfo {author} {\bibfnamefont {J.~S.}\ \bibnamefont
  {Cotler}}, \bibinfo {author} {\bibfnamefont {G.}~\bibnamefont {Gur-Ari}},
  \bibinfo {author} {\bibfnamefont {M.}~\bibnamefont {Hanada}}, \bibinfo
  {author} {\bibfnamefont {J.}~\bibnamefont {Polchinski}}, \bibinfo {author}
  {\bibfnamefont {P.}~\bibnamefont {Saad}}, \bibinfo {author} {\bibfnamefont
  {S.~H.}\ \bibnamefont {Shenker}}, \bibinfo {author} {\bibfnamefont
  {D.}~\bibnamefont {Stanford}}, \bibinfo {author} {\bibfnamefont
  {A.}~\bibnamefont {Streicher}}, \ and\ \bibinfo {author} {\bibfnamefont
  {M.}~\bibnamefont {Tezuka}},\ }\href {\doibase 10.1007/JHEP05(2017)118}
  {\bibfield  {journal} {\bibinfo  {journal} {JHEP}\ }\textbf {\bibinfo
  {volume} {05}},\ \bibinfo {pages} {118} (\bibinfo {year} {2017})},\ \bibinfo
  {note} {[Erratum: JHEP 09, 002 (2018)]},\ \Eprint
  {http://arxiv.org/abs/1611.04650} {arXiv:1611.04650 [hep-th]} \BibitemShut
  {NoStop}%
\bibitem [{\citenamefont {de~Boer}\ \emph {et~al.}(2018)\citenamefont
  {de~Boer}, \citenamefont {Llabr\'es}, \citenamefont {Pedraza},\ and\
  \citenamefont {Vegh}}]{deBoer:2017xdk}%
  \BibitemOpen
  \bibfield  {author} {\bibinfo {author} {\bibfnamefont {J.}~\bibnamefont
  {de~Boer}}, \bibinfo {author} {\bibfnamefont {E.}~\bibnamefont {Llabr\'es}},
  \bibinfo {author} {\bibfnamefont {J.~F.}\ \bibnamefont {Pedraza}}, \ and\
  \bibinfo {author} {\bibfnamefont {D.}~\bibnamefont {Vegh}},\ }\href {\doibase
  10.1103/PhysRevLett.120.201604} {\bibfield  {journal} {\bibinfo  {journal}
  {Phys. Rev. Lett.}\ }\textbf {\bibinfo {volume} {120}},\ \bibinfo {pages}
  {201604} (\bibinfo {year} {2018})},\ \Eprint
  {http://arxiv.org/abs/1709.01052} {arXiv:1709.01052 [hep-th]} \BibitemShut
  {NoStop}%
\bibitem [{\citenamefont {Stanford}\ and\ \citenamefont
  {Witten}(2020)}]{Stanford:2019vob}%
  \BibitemOpen
  \bibfield  {author} {\bibinfo {author} {\bibfnamefont {D.}~\bibnamefont
  {Stanford}}\ and\ \bibinfo {author} {\bibfnamefont {E.}~\bibnamefont
  {Witten}},\ }\href {\doibase 10.4310/ATMP.2020.v24.n6.a4} {\bibfield
  {journal} {\bibinfo  {journal} {Adv. Theor. Math. Phys.}\ }\textbf {\bibinfo
  {volume} {24}},\ \bibinfo {pages} {1475} (\bibinfo {year} {2020})},\ \Eprint
  {http://arxiv.org/abs/1907.03363} {arXiv:1907.03363 [hep-th]} \BibitemShut
  {NoStop}%
\bibitem [{\citenamefont {Altland}\ and\ \citenamefont
  {Sonner}(2026)}]{Altland:2026tog}%
  \BibitemOpen
  \bibfield  {author} {\bibinfo {author} {\bibfnamefont {A.}~\bibnamefont
  {Altland}}\ and\ \bibinfo {author} {\bibfnamefont {J.}~\bibnamefont
  {Sonner}},\ }\href@noop {} {\  (\bibinfo {year} {2026})},\ \Eprint
  {http://arxiv.org/abs/2604.12784} {arXiv:2604.12784 [quant-ph]} \BibitemShut
  {NoStop}%
\bibitem [{\citenamefont {Sekino}\ and\ \citenamefont
  {Susskind}(2008)}]{Sekino:2008he}%
  \BibitemOpen
  \bibfield  {author} {\bibinfo {author} {\bibfnamefont {Y.}~\bibnamefont
  {Sekino}}\ and\ \bibinfo {author} {\bibfnamefont {L.}~\bibnamefont
  {Susskind}},\ }\href {\doibase 10.1088/1126-6708/2008/10/065} {\bibfield
  {journal} {\bibinfo  {journal} {JHEP}\ }\textbf {\bibinfo {volume} {10}},\
  \bibinfo {pages} {065} (\bibinfo {year} {2008})},\ \Eprint
  {http://arxiv.org/abs/0808.2096} {arXiv:0808.2096 [hep-th]} \BibitemShut
  {NoStop}%
\bibitem [{\citenamefont {Larkin}\ and\ \citenamefont
  {Ovchinnikov}(1969)}]{larkin1969quasiclassical}%
  \BibitemOpen
  \bibfield  {author} {\bibinfo {author} {\bibfnamefont {A.~I.}\ \bibnamefont
  {Larkin}}\ and\ \bibinfo {author} {\bibfnamefont {Y.~N.}\ \bibnamefont
  {Ovchinnikov}},\ }\href@noop {} {\bibfield  {journal} {\bibinfo  {journal}
  {Sov Phys JETP}\ }\textbf {\bibinfo {volume} {28}},\ \bibinfo {pages} {1200}
  (\bibinfo {year} {1969})}\BibitemShut {NoStop}%
\bibitem [{\citenamefont {Berman}\ and\ \citenamefont
  {Zaslavsky}(1978)}]{berman1978condition}%
  \BibitemOpen
  \bibfield  {author} {\bibinfo {author} {\bibfnamefont {G.~P.}\ \bibnamefont
  {Berman}}\ and\ \bibinfo {author} {\bibfnamefont {G.~M.}\ \bibnamefont
  {Zaslavsky}},\ }\href@noop {} {\bibfield  {journal} {\bibinfo  {journal}
  {Physica A: Statistical Mechanics and its Applications}\ }\textbf {\bibinfo
  {volume} {91}},\ \bibinfo {pages} {450} (\bibinfo {year} {1978})}\BibitemShut
  {NoStop}%
\bibitem [{\citenamefont {Maldacena}\ \emph {et~al.}(2016)\citenamefont
  {Maldacena}, \citenamefont {Shenker},\ and\ \citenamefont
  {Stanford}}]{Maldacena:2015waa}%
  \BibitemOpen
  \bibfield  {author} {\bibinfo {author} {\bibfnamefont {J.}~\bibnamefont
  {Maldacena}}, \bibinfo {author} {\bibfnamefont {S.~H.}\ \bibnamefont
  {Shenker}}, \ and\ \bibinfo {author} {\bibfnamefont {D.}~\bibnamefont
  {Stanford}},\ }\href {\doibase 10.1007/JHEP08(2016)106} {\bibfield  {journal}
  {\bibinfo  {journal} {JHEP}\ }\textbf {\bibinfo {volume} {08}},\ \bibinfo
  {pages} {106} (\bibinfo {year} {2016})},\ \Eprint
  {http://arxiv.org/abs/1503.01409} {arXiv:1503.01409 [hep-th]} \BibitemShut
  {NoStop}%
\bibitem [{\citenamefont {Lipkin}\ \emph {et~al.}(1965)\citenamefont {Lipkin},
  \citenamefont {Meshkov},\ and\ \citenamefont {Glick}}]{Lipkin:1964yk}%
  \BibitemOpen
  \bibfield  {author} {\bibinfo {author} {\bibfnamefont {H.~J.}\ \bibnamefont
  {Lipkin}}, \bibinfo {author} {\bibfnamefont {N.}~\bibnamefont {Meshkov}}, \
  and\ \bibinfo {author} {\bibfnamefont {A.~J.}\ \bibnamefont {Glick}},\ }\href
  {\doibase 10.1016/0029-5582(65)90862-X} {\bibfield  {journal} {\bibinfo
  {journal} {Nucl. Phys.}\ }\textbf {\bibinfo {volume} {62}},\ \bibinfo {pages}
  {188} (\bibinfo {year} {1965})}\BibitemShut {NoStop}%
\bibitem [{\citenamefont {Xu}\ \emph {et~al.}(2020)\citenamefont {Xu},
  \citenamefont {Scaffidi},\ and\ \citenamefont {Cao}}]{Xu:2019lhc}%
  \BibitemOpen
  \bibfield  {author} {\bibinfo {author} {\bibfnamefont {T.}~\bibnamefont
  {Xu}}, \bibinfo {author} {\bibfnamefont {T.}~\bibnamefont {Scaffidi}}, \ and\
  \bibinfo {author} {\bibfnamefont {X.}~\bibnamefont {Cao}},\ }\href {\doibase
  10.1103/PhysRevLett.124.140602} {\bibfield  {journal} {\bibinfo  {journal}
  {Phys. Rev. Lett.}\ }\textbf {\bibinfo {volume} {124}},\ \bibinfo {pages}
  {140602} (\bibinfo {year} {2020})},\ \Eprint
  {http://arxiv.org/abs/1912.11063} {arXiv:1912.11063 [cond-mat.stat-mech]}
  \BibitemShut {NoStop}%
\bibitem [{\citenamefont {Hashimoto}\ \emph {et~al.}(2020)\citenamefont
  {Hashimoto}, \citenamefont {Huh}, \citenamefont {Kim},\ and\ \citenamefont
  {Watanabe}}]{Hashimoto:2020xfr}%
  \BibitemOpen
  \bibfield  {author} {\bibinfo {author} {\bibfnamefont {K.}~\bibnamefont
  {Hashimoto}}, \bibinfo {author} {\bibfnamefont {K.-B.}\ \bibnamefont {Huh}},
  \bibinfo {author} {\bibfnamefont {K.-Y.}\ \bibnamefont {Kim}}, \ and\
  \bibinfo {author} {\bibfnamefont {R.}~\bibnamefont {Watanabe}},\ }\href
  {\doibase 10.1007/JHEP11(2020)068} {\bibfield  {journal} {\bibinfo  {journal}
  {JHEP}\ }\textbf {\bibinfo {volume} {11}},\ \bibinfo {pages} {068} (\bibinfo
  {year} {2020})},\ \Eprint {http://arxiv.org/abs/2007.04746} {arXiv:2007.04746
  [hep-th]} \BibitemShut {NoStop}%
\bibitem [{\citenamefont {Mehta}(2004)}]{mehta2004random}%
  \BibitemOpen
  \bibfield  {author} {\bibinfo {author} {\bibfnamefont {M.~L.}\ \bibnamefont
  {Mehta}},\ }\href@noop {} {\emph {\bibinfo {title} {Random matrices}}},\
  Vol.\ \bibinfo {volume} {142}\ (\bibinfo  {publisher} {Elsevier},\ \bibinfo
  {address} {Amsterdam},\ \bibinfo {year} {2004})\BibitemShut {NoStop}%
\bibitem [{\citenamefont {Bohigas}\ \emph {et~al.}(1984)\citenamefont
  {Bohigas}, \citenamefont {Giannoni},\ and\ \citenamefont
  {Schmit}}]{Bohigas:1983er}%
  \BibitemOpen
  \bibfield  {author} {\bibinfo {author} {\bibfnamefont {O.}~\bibnamefont
  {Bohigas}}, \bibinfo {author} {\bibfnamefont {M.~J.}\ \bibnamefont
  {Giannoni}}, \ and\ \bibinfo {author} {\bibfnamefont {C.}~\bibnamefont
  {Schmit}},\ }\href {\doibase 10.1103/PhysRevLett.52.1} {\bibfield  {journal}
  {\bibinfo  {journal} {Phys. Rev. Lett.}\ }\textbf {\bibinfo {volume} {52}},\
  \bibinfo {pages} {1} (\bibinfo {year} {1984})}\BibitemShut {NoStop}%
\bibitem [{\citenamefont {Bohigas}(1984)}]{Bohigas:1984aa}%
  \BibitemOpen
  \bibfield  {author} {\bibinfo {author} {\bibfnamefont {O.}~\bibnamefont
  {Bohigas}},\ }\href {\doibase 10.1103/PhysRevLett.52.1} {\bibfield  {journal}
  {\bibinfo  {journal} {Physical Review Letters}\ }\textbf {\bibinfo {volume}
  {52}},\ \bibinfo {pages} {1} (\bibinfo {year} {1984})}\BibitemShut {NoStop}%
\bibitem [{\citenamefont {Guhr}\ \emph {et~al.}(1998)\citenamefont {Guhr},
  \citenamefont {Muller-Groeling},\ and\ \citenamefont
  {Weidenmuller}}]{Guhr:1997ve}%
  \BibitemOpen
  \bibfield  {author} {\bibinfo {author} {\bibfnamefont {T.}~\bibnamefont
  {Guhr}}, \bibinfo {author} {\bibfnamefont {A.}~\bibnamefont
  {Muller-Groeling}}, \ and\ \bibinfo {author} {\bibfnamefont {H.~A.}\
  \bibnamefont {Weidenmuller}},\ }\href {\doibase
  10.1016/S0370-1573(97)00088-4} {\bibfield  {journal} {\bibinfo  {journal}
  {Phys. Rept.}\ }\textbf {\bibinfo {volume} {299}},\ \bibinfo {pages} {189}
  (\bibinfo {year} {1998})},\ \Eprint {http://arxiv.org/abs/cond-mat/9707301}
  {arXiv:cond-mat/9707301} \BibitemShut {NoStop}%
\bibitem [{\citenamefont {Bohigas}\ and\ \citenamefont
  {Giannoni}(2005)}]{Bohigas}%
  \BibitemOpen
  \bibfield  {author} {\bibinfo {author} {\bibfnamefont {O.}~\bibnamefont
  {Bohigas}}\ and\ \bibinfo {author} {\bibfnamefont {M.-J.}\ \bibnamefont
  {Giannoni}},\ }\href@noop {} {\emph {\bibinfo {title} {Mathematical and
  Computational Methods in Nuclear Physics}}}\ (\bibinfo  {publisher} {Springer
  Berlin Heidelberg},\ \bibinfo {year} {31 May 2005})\BibitemShut {NoStop}%
\bibitem [{\citenamefont {Berry}\ and\ \citenamefont
  {Tabor}(1977)}]{Berry1977Level}%
  \BibitemOpen
  \bibfield  {author} {\bibinfo {author} {\bibfnamefont {M.~V.}\ \bibnamefont
  {Berry}}\ and\ \bibinfo {author} {\bibfnamefont {M.}~\bibnamefont {Tabor}},\
  }\href {\doibase 10.1098/rspa.1977.0140} {\bibfield  {journal} {\bibinfo
  {journal} {Proc. R. Soc. Lond. A}\ }\textbf {\bibinfo {volume} {356}},\
  \bibinfo {pages} {375} (\bibinfo {year} {1977})}\BibitemShut {NoStop}%
\bibitem [{\citenamefont {Parker}\ \emph {et~al.}(2019)\citenamefont {Parker},
  \citenamefont {Cao}, \citenamefont {Avdoshkin}, \citenamefont {Scaffidi},\
  and\ \citenamefont {Altman}}]{Parker:2018yvk}%
  \BibitemOpen
  \bibfield  {author} {\bibinfo {author} {\bibfnamefont {D.~E.}\ \bibnamefont
  {Parker}}, \bibinfo {author} {\bibfnamefont {X.}~\bibnamefont {Cao}},
  \bibinfo {author} {\bibfnamefont {A.}~\bibnamefont {Avdoshkin}}, \bibinfo
  {author} {\bibfnamefont {T.}~\bibnamefont {Scaffidi}}, \ and\ \bibinfo
  {author} {\bibfnamefont {E.}~\bibnamefont {Altman}},\ }\href {\doibase
  10.1103/PhysRevX.9.041017} {\bibfield  {journal} {\bibinfo  {journal} {Phys.
  Rev. X}\ }\textbf {\bibinfo {volume} {9}},\ \bibinfo {pages} {041017}
  (\bibinfo {year} {2019})},\ \Eprint {http://arxiv.org/abs/1812.08657}
  {arXiv:1812.08657 [cond-mat.stat-mech]} \BibitemShut {NoStop}%
\bibitem [{\citenamefont {Balasubramanian}\ \emph {et~al.}(2022)\citenamefont
  {Balasubramanian}, \citenamefont {Caputa}, \citenamefont {Magan},\ and\
  \citenamefont {Wu}}]{Balasubramanian:2022tpr}%
  \BibitemOpen
  \bibfield  {author} {\bibinfo {author} {\bibfnamefont {V.}~\bibnamefont
  {Balasubramanian}}, \bibinfo {author} {\bibfnamefont {P.}~\bibnamefont
  {Caputa}}, \bibinfo {author} {\bibfnamefont {J.~M.}\ \bibnamefont {Magan}}, \
  and\ \bibinfo {author} {\bibfnamefont {Q.}~\bibnamefont {Wu}},\ }\href
  {\doibase 10.1103/PhysRevD.106.046007} {\bibfield  {journal} {\bibinfo
  {journal} {Phys. Rev. D}\ }\textbf {\bibinfo {volume} {106}},\ \bibinfo
  {pages} {046007} (\bibinfo {year} {2022})},\ \Eprint
  {http://arxiv.org/abs/2202.06957} {arXiv:2202.06957 [hep-th]} \BibitemShut
  {NoStop}%
\bibitem [{\citenamefont {Caputa}\ \emph {et~al.}(2024)\citenamefont {Caputa},
  \citenamefont {Jeong}, \citenamefont {Liu}, \citenamefont {Pedraza},\ and\
  \citenamefont {Qu}}]{Caputa:2024vrn}%
  \BibitemOpen
  \bibfield  {author} {\bibinfo {author} {\bibfnamefont {P.}~\bibnamefont
  {Caputa}}, \bibinfo {author} {\bibfnamefont {H.-S.}\ \bibnamefont {Jeong}},
  \bibinfo {author} {\bibfnamefont {S.}~\bibnamefont {Liu}}, \bibinfo {author}
  {\bibfnamefont {J.~F.}\ \bibnamefont {Pedraza}}, \ and\ \bibinfo {author}
  {\bibfnamefont {L.-C.}\ \bibnamefont {Qu}},\ }\href {\doibase
  10.1007/JHEP05(2024)337} {\bibfield  {journal} {\bibinfo  {journal} {JHEP}\
  }\textbf {\bibinfo {volume} {05}},\ \bibinfo {pages} {337} (\bibinfo {year}
  {2024})},\ \Eprint {http://arxiv.org/abs/2402.09522} {arXiv:2402.09522
  [hep-th]} \BibitemShut {NoStop}%
\bibitem [{\citenamefont {Nandy}\ \emph
  {et~al.}(2025{\natexlab{a}})\citenamefont {Nandy}, \citenamefont
  {Matsoukas-Roubeas}, \citenamefont {Mart{\'\i}nez-Azcona}, \citenamefont
  {Dymarsky},\ and\ \citenamefont {del Campo}}]{Nandy:2024evd}%
  \BibitemOpen
  \bibfield  {author} {\bibinfo {author} {\bibfnamefont {P.}~\bibnamefont
  {Nandy}}, \bibinfo {author} {\bibfnamefont {A.~S.}\ \bibnamefont
  {Matsoukas-Roubeas}}, \bibinfo {author} {\bibfnamefont {P.}~\bibnamefont
  {Mart{\'\i}nez-Azcona}}, \bibinfo {author} {\bibfnamefont {A.}~\bibnamefont
  {Dymarsky}}, \ and\ \bibinfo {author} {\bibfnamefont {A.}~\bibnamefont {del
  Campo}},\ }\href {\doibase 10.1016/j.physrep.2025.05.001} {\bibfield
  {journal} {\bibinfo  {journal} {Phys. Rept.}\ }\textbf {\bibinfo {volume}
  {1125-1128}},\ \bibinfo {pages} {1} (\bibinfo {year} {2025}{\natexlab{a}})},\
  \Eprint {http://arxiv.org/abs/2405.09628} {arXiv:2405.09628 [quant-ph]}
  \BibitemShut {NoStop}%
\bibitem [{\citenamefont {Rabinovici}\ \emph {et~al.}(2025)\citenamefont
  {Rabinovici}, \citenamefont {S{\'a}nchez-Garrido}, \citenamefont {Shir},\
  and\ \citenamefont {Sonner}}]{Rabinovici:2025otw}%
  \BibitemOpen
  \bibfield  {author} {\bibinfo {author} {\bibfnamefont {E.}~\bibnamefont
  {Rabinovici}}, \bibinfo {author} {\bibfnamefont {A.}~\bibnamefont
  {S{\'a}nchez-Garrido}}, \bibinfo {author} {\bibfnamefont {R.}~\bibnamefont
  {Shir}}, \ and\ \bibinfo {author} {\bibfnamefont {J.}~\bibnamefont
  {Sonner}},\ }\href@noop {} {\  (\bibinfo {year} {2025})},\ \Eprint
  {http://arxiv.org/abs/2507.06286} {arXiv:2507.06286 [hep-th]} \BibitemShut
  {NoStop}%
\bibitem [{\citenamefont {Jeong}\ and\ \citenamefont
  {Pedraza}(2026)}]{Jeong:2026gdc}%
  \BibitemOpen
  \bibfield  {author} {\bibinfo {author} {\bibfnamefont {H.-S.}\ \bibnamefont
  {Jeong}}\ and\ \bibinfo {author} {\bibfnamefont {J.~F.}\ \bibnamefont
  {Pedraza}},\ }\href@noop {} {\  (\bibinfo {year} {2026})},\ \Eprint
  {http://arxiv.org/abs/2608.26825} {arXiv:2608.26825 [hep-th]} \BibitemShut
  {NoStop}%
\bibitem [{\citenamefont {Lanczos}(1950)}]{Lanczos:1950zz}%
  \BibitemOpen
  \bibfield  {author} {\bibinfo {author} {\bibfnamefont {C.}~\bibnamefont
  {Lanczos}},\ }\href {\doibase 10.6028/jres.045.026} {\bibfield  {journal}
  {\bibinfo  {journal} {J. Res. Natl. Bur. Stand. B}\ }\textbf {\bibinfo
  {volume} {45}},\ \bibinfo {pages} {255} (\bibinfo {year} {1950})}\BibitemShut
  {NoStop}%
\bibitem [{\citenamefont {Bhattacharjee}\ \emph {et~al.}(2022)\citenamefont
  {Bhattacharjee}, \citenamefont {Cao}, \citenamefont {Nandy},\ and\
  \citenamefont {Pathak}}]{Bhattacharjee:2022vlt}%
  \BibitemOpen
  \bibfield  {author} {\bibinfo {author} {\bibfnamefont {B.}~\bibnamefont
  {Bhattacharjee}}, \bibinfo {author} {\bibfnamefont {X.}~\bibnamefont {Cao}},
  \bibinfo {author} {\bibfnamefont {P.}~\bibnamefont {Nandy}}, \ and\ \bibinfo
  {author} {\bibfnamefont {T.}~\bibnamefont {Pathak}},\ }\href {\doibase
  10.1007/JHEP05(2022)174} {\bibfield  {journal} {\bibinfo  {journal} {JHEP}\
  }\textbf {\bibinfo {volume} {05}},\ \bibinfo {pages} {174} (\bibinfo {year}
  {2022})},\ \Eprint {http://arxiv.org/abs/2203.03534} {arXiv:2203.03534
  [quant-ph]} \BibitemShut {NoStop}%
\bibitem [{\citenamefont {Huh}\ \emph {et~al.}(2024)\citenamefont {Huh},
  \citenamefont {Jeong},\ and\ \citenamefont {Pedraza}}]{Huh:2023jxt}%
  \BibitemOpen
  \bibfield  {author} {\bibinfo {author} {\bibfnamefont {K.-B.}\ \bibnamefont
  {Huh}}, \bibinfo {author} {\bibfnamefont {H.-S.}\ \bibnamefont {Jeong}}, \
  and\ \bibinfo {author} {\bibfnamefont {J.~F.}\ \bibnamefont {Pedraza}},\
  }\href {\doibase 10.1007/JHEP05(2024)137} {\bibfield  {journal} {\bibinfo
  {journal} {JHEP}\ }\textbf {\bibinfo {volume} {05}},\ \bibinfo {pages} {137}
  (\bibinfo {year} {2024})},\ \Eprint {http://arxiv.org/abs/2312.12593}
  {arXiv:2312.12593 [hep-th]} \BibitemShut {NoStop}%
\bibitem [{\citenamefont {Craps}\ \emph {et~al.}(2025)\citenamefont {Craps},
  \citenamefont {Evnin},\ and\ \citenamefont
  {Pascuzzi}}]{PhysRevLett.134.050402}%
  \BibitemOpen
  \bibfield  {author} {\bibinfo {author} {\bibfnamefont {B.}~\bibnamefont
  {Craps}}, \bibinfo {author} {\bibfnamefont {O.}~\bibnamefont {Evnin}}, \ and\
  \bibinfo {author} {\bibfnamefont {G.}~\bibnamefont {Pascuzzi}},\ }\href
  {\doibase 10.1103/PhysRevLett.134.050402} {\bibfield  {journal} {\bibinfo
  {journal} {Phys. Rev. Lett.}\ }\textbf {\bibinfo {volume} {134}},\ \bibinfo
  {pages} {050402} (\bibinfo {year} {2025})}\BibitemShut {NoStop}%
\bibitem [{\citenamefont {Erdmenger}\ \emph {et~al.}(2023)\citenamefont
  {Erdmenger}, \citenamefont {Jian},\ and\ \citenamefont
  {Xian}}]{Erdmenger:2023wjg}%
  \BibitemOpen
  \bibfield  {author} {\bibinfo {author} {\bibfnamefont {J.}~\bibnamefont
  {Erdmenger}}, \bibinfo {author} {\bibfnamefont {S.-K.}\ \bibnamefont {Jian}},
  \ and\ \bibinfo {author} {\bibfnamefont {Z.-Y.}\ \bibnamefont {Xian}},\
  }\href {\doibase 10.1007/JHEP08(2023)176} {\bibfield  {journal} {\bibinfo
  {journal} {JHEP}\ }\textbf {\bibinfo {volume} {08}},\ \bibinfo {pages} {176}
  (\bibinfo {year} {2023})},\ \Eprint {http://arxiv.org/abs/2303.12151}
  {arXiv:2303.12151 [hep-th]} \BibitemShut {NoStop}%
\bibitem [{\citenamefont {Bohigas}(1989)}]{Bohigas1989Random}%
  \BibitemOpen
  \bibfield  {author} {\bibinfo {author} {\bibfnamefont {O.}~\bibnamefont
  {Bohigas}},\ }in\ \href@noop {} {\emph {\bibinfo {booktitle} {Chaos and
  Quantum Physics (Les Houches Session LII)}}},\ \bibinfo {editor} {edited by\
  \bibinfo {editor} {\bibfnamefont {M.-J.}\ \bibnamefont {Giannoni}}, \bibinfo
  {editor} {\bibfnamefont {A.}~\bibnamefont {Voros}}, \ and\ \bibinfo {editor}
  {\bibfnamefont {J.}~\bibnamefont {Zinn-Justin}}}\ (\bibinfo  {publisher}
  {North-Holland},\ \bibinfo {address} {Amsterdam},\ \bibinfo {year} {1989})\
  pp.\ \bibinfo {pages} {87--199}\BibitemShut {NoStop}%
\bibitem [{\citenamefont {Guhr}\ \emph
  {et~al.}(1999{\natexlab{a}})\citenamefont {Guhr}, \citenamefont {Ma},
  \citenamefont {Meyer},\ and\ \citenamefont {Wilke}}]{Guhr1999Random}%
  \BibitemOpen
  \bibfield  {author} {\bibinfo {author} {\bibfnamefont {T.}~\bibnamefont
  {Guhr}}, \bibinfo {author} {\bibfnamefont {J.-Z.}\ \bibnamefont {Ma}},
  \bibinfo {author} {\bibfnamefont {S.}~\bibnamefont {Meyer}}, \ and\ \bibinfo
  {author} {\bibfnamefont {T.}~\bibnamefont {Wilke}},\ }\href {\doibase
  10.1103/PhysRevD.59.054501} {\bibfield  {journal} {\bibinfo  {journal} {Phys.
  Rev. D}\ }\textbf {\bibinfo {volume} {59}},\ \bibinfo {pages} {054501}
  (\bibinfo {year} {1999}{\natexlab{a}})}\BibitemShut {NoStop}%
\bibitem [{\citenamefont {G{\'o}mez}\ \emph
  {et~al.}(2002{\natexlab{a}})\citenamefont {G{\'o}mez}, \citenamefont
  {Molina}, \citenamefont {Rela{\~n}o},\ and\ \citenamefont
  {Retamosa}}]{Gomez2002Spectral}%
  \BibitemOpen
  \bibfield  {author} {\bibinfo {author} {\bibfnamefont {J.~M.~G.}\
  \bibnamefont {G{\'o}mez}}, \bibinfo {author} {\bibfnamefont {R.~A.}\
  \bibnamefont {Molina}}, \bibinfo {author} {\bibfnamefont {A.}~\bibnamefont
  {Rela{\~n}o}}, \ and\ \bibinfo {author} {\bibfnamefont {J.}~\bibnamefont
  {Retamosa}},\ }\href {\doibase 10.1103/PhysRevE.66.036209} {\bibfield
  {journal} {\bibinfo  {journal} {Phys. Rev. E}\ }\textbf {\bibinfo {volume}
  {66}},\ \bibinfo {pages} {036209} (\bibinfo {year}
  {2002}{\natexlab{a}})}\BibitemShut {NoStop}%
\bibitem [{\citenamefont {Morales}\ \emph
  {et~al.}(2011{\natexlab{a}})\citenamefont {Morales}, \citenamefont {Landa},
  \citenamefont {Str{\'a}nsk{\'y}},\ and\ \citenamefont
  {Frank}}]{Morales2011Spectral}%
  \BibitemOpen
  \bibfield  {author} {\bibinfo {author} {\bibfnamefont {I.~O.}\ \bibnamefont
  {Morales}}, \bibinfo {author} {\bibfnamefont {E.}~\bibnamefont {Landa}},
  \bibinfo {author} {\bibfnamefont {P.}~\bibnamefont {Str{\'a}nsk{\'y}}}, \
  and\ \bibinfo {author} {\bibfnamefont {A.}~\bibnamefont {Frank}},\ }\href
  {\doibase 10.1103/PhysRevE.84.016203} {\bibfield  {journal} {\bibinfo
  {journal} {Phys. Rev. E}\ }\textbf {\bibinfo {volume} {84}},\ \bibinfo
  {pages} {016203} (\bibinfo {year} {2011}{\natexlab{a}})}\BibitemShut
  {NoStop}%
\bibitem [{\citenamefont {Das}\ \emph {et~al.}(2024)\citenamefont {Das},
  \citenamefont {Garg}, \citenamefont {Krishnan},\ and\ \citenamefont
  {Kundu}}]{Das:2023yfj}%
  \BibitemOpen
  \bibfield  {author} {\bibinfo {author} {\bibfnamefont {S.}~\bibnamefont
  {Das}}, \bibinfo {author} {\bibfnamefont {S.~K.}\ \bibnamefont {Garg}},
  \bibinfo {author} {\bibfnamefont {C.}~\bibnamefont {Krishnan}}, \ and\
  \bibinfo {author} {\bibfnamefont {A.}~\bibnamefont {Kundu}},\ }\href
  {\doibase 10.1007/JHEP01(2024)172} {\bibfield  {journal} {\bibinfo  {journal}
  {JHEP}\ }\textbf {\bibinfo {volume} {01}},\ \bibinfo {pages} {172} (\bibinfo
  {year} {2024})},\ \Eprint {http://arxiv.org/abs/2308.11704} {arXiv:2308.11704
  [hep-th]} \BibitemShut {NoStop}%
\bibitem [{\citenamefont {Basu}\ \emph {et~al.}(2026)\citenamefont {Basu},
  \citenamefont {Das},\ and\ \citenamefont {Madlala}}]{Basu:2026gvl}%
  \BibitemOpen
  \bibfield  {author} {\bibinfo {author} {\bibfnamefont {P.}~\bibnamefont
  {Basu}}, \bibinfo {author} {\bibfnamefont {S.}~\bibnamefont {Das}}, \ and\
  \bibinfo {author} {\bibfnamefont {B.}~\bibnamefont {Madlala}},\ }\href@noop
  {} {\  (\bibinfo {year} {2026})},\ \Eprint {http://arxiv.org/abs/2608.07412}
  {arXiv:2608.07412 [hep-th]} \BibitemShut {NoStop}%
\bibitem [{\citenamefont {Berry}(1985)}]{Berry1985-mx}%
  \BibitemOpen
  \bibfield  {author} {\bibinfo {author} {\bibfnamefont {M.~V.}\ \bibnamefont
  {Berry}},\ }\href@noop {} {\bibfield  {journal} {\bibinfo  {journal} {Proc.
  R. Soc. Lond.}\ }\textbf {\bibinfo {volume} {400}},\ \bibinfo {pages} {229}
  (\bibinfo {year} {1985})}\BibitemShut {NoStop}%
\bibitem [{\citenamefont {M\"uller}\ \emph {et~al.}(2004)\citenamefont
  {M\"uller}, \citenamefont {Heusler}, \citenamefont {Braun}, \citenamefont
  {Haake},\ and\ \citenamefont {Altland}}]{Muller:2004nb}%
  \BibitemOpen
  \bibfield  {author} {\bibinfo {author} {\bibfnamefont {S.}~\bibnamefont
  {M\"uller}}, \bibinfo {author} {\bibfnamefont {S.}~\bibnamefont {Heusler}},
  \bibinfo {author} {\bibfnamefont {P.}~\bibnamefont {Braun}}, \bibinfo
  {author} {\bibfnamefont {F.}~\bibnamefont {Haake}}, \ and\ \bibinfo {author}
  {\bibfnamefont {A.}~\bibnamefont {Altland}},\ }\href {\doibase
  10.1103/PhysRevLett.93.014103} {\bibfield  {journal} {\bibinfo  {journal}
  {Phys. Rev. Lett.}\ }\textbf {\bibinfo {volume} {93}},\ \bibinfo {pages}
  {014103} (\bibinfo {year} {2004})},\ \Eprint
  {http://arxiv.org/abs/nlin/0401021} {arXiv:nlin/0401021} \BibitemShut
  {NoStop}%
\bibitem [{\citenamefont {Gutzwiller}(1990)}]{Gutzwiller1990}%
  \BibitemOpen
  \bibfield  {author} {\bibinfo {author} {\bibfnamefont {M.~C.}\ \bibnamefont
  {Gutzwiller}},\ }\href@noop {} {\emph {\bibinfo {title} {Chaos in Classical
  and Quantum Mechanics}}}\ (\bibinfo  {publisher} {Springer-Verlag},\ \bibinfo
  {address} {New York},\ \bibinfo {year} {1990})\BibitemShut {NoStop}%
\bibitem [{\citenamefont {Ratcliff}(1971)}]{PhysRevC.3.117}%
  \BibitemOpen
  \bibfield  {author} {\bibinfo {author} {\bibfnamefont {K.~F.}\ \bibnamefont
  {Ratcliff}},\ }\href {\doibase 10.1103/PhysRevC.3.117} {\bibfield  {journal}
  {\bibinfo  {journal} {Phys. Rev. C}\ }\textbf {\bibinfo {volume} {3}},\
  \bibinfo {pages} {117} (\bibinfo {year} {1971})}\BibitemShut {NoStop}%
\bibitem [{\citenamefont {Flores}\ \emph {et~al.}(2001)\citenamefont {Flores},
  \citenamefont {Horoi}, \citenamefont {M\"uller},\ and\ \citenamefont
  {Seligman}}]{PhysRevE.63.026204}%
  \BibitemOpen
  \bibfield  {author} {\bibinfo {author} {\bibfnamefont {J.}~\bibnamefont
  {Flores}}, \bibinfo {author} {\bibfnamefont {M.}~\bibnamefont {Horoi}},
  \bibinfo {author} {\bibfnamefont {M.}~\bibnamefont {M\"uller}}, \ and\
  \bibinfo {author} {\bibfnamefont {T.~H.}\ \bibnamefont {Seligman}},\ }\href
  {\doibase 10.1103/PhysRevE.63.026204} {\bibfield  {journal} {\bibinfo
  {journal} {Phys. Rev. E}\ }\textbf {\bibinfo {volume} {63}},\ \bibinfo
  {pages} {026204} (\bibinfo {year} {2001})}\BibitemShut {NoStop}%
\bibitem [{\citenamefont {Morales}\ \emph
  {et~al.}(2011{\natexlab{b}})\citenamefont {Morales}, \citenamefont {Landa},
  \citenamefont {Str\'ansk\'y},\ and\ \citenamefont
  {Frank}}]{PhysRevE.84.016203}%
  \BibitemOpen
  \bibfield  {author} {\bibinfo {author} {\bibfnamefont {I.~O.}\ \bibnamefont
  {Morales}}, \bibinfo {author} {\bibfnamefont {E.}~\bibnamefont {Landa}},
  \bibinfo {author} {\bibfnamefont {P.}~\bibnamefont {Str\'ansk\'y}}, \ and\
  \bibinfo {author} {\bibfnamefont {A.}~\bibnamefont {Frank}},\ }\href
  {\doibase 10.1103/PhysRevE.84.016203} {\bibfield  {journal} {\bibinfo
  {journal} {Phys. Rev. E}\ }\textbf {\bibinfo {volume} {84}},\ \bibinfo
  {pages} {016203} (\bibinfo {year} {2011}{\natexlab{b}})}\BibitemShut
  {NoStop}%
\bibitem [{Note1()}]{Note1}%
  \BibitemOpen
  \bibinfo {note} {The extracted fluctuations can depend on the smoothing
  prescription, including the polynomial degree, motivating alternative
  unfolding procedures~\cite
  {Guhr1999Semiclassical,Gomez2002Misleading,Morales2011Improved,ABULMAGD2014185,10.1119/1.2198883}.}\BibitemShut
  {Stop}%
\bibitem [{\citenamefont {Viswanath}\ and\ \citenamefont
  {M{\"u}ller}(1994)}]{RecursionBook}%
  \BibitemOpen
  \bibfield  {author} {\bibinfo {author} {\bibfnamefont {V.~S.}\ \bibnamefont
  {Viswanath}}\ and\ \bibinfo {author} {\bibfnamefont {G.}~\bibnamefont
  {M{\"u}ller}},\ }\href@noop {} {\emph {\bibinfo {title} {The Recursion
  Method: Application to Many-Body Dynamics}}}\ (\bibinfo  {publisher}
  {Springer Berlin},\ \bibinfo {address} {Heidelberg, Germany},\ \bibinfo
  {year} {1994})\BibitemShut {NoStop}%
\bibitem [{\citenamefont {Parlett}(1998)}]{parlett1998symmetric}%
  \BibitemOpen
  \bibfield  {author} {\bibinfo {author} {\bibfnamefont {B.~N.}\ \bibnamefont
  {Parlett}},\ }\href {\doibase 10.1137/1.9781611971163} {\emph {\bibinfo
  {title} {The Symmetric Eigenvalue Problem}}},\ Classics in Applied
  Mathematics\ (\bibinfo  {publisher} {Society for Industrial and Applied
  Mathematics},\ \bibinfo {address} {Philadelphia, PA},\ \bibinfo {year}
  {1998})\BibitemShut {NoStop}%
\bibitem [{\citenamefont {Hashimoto}\ \emph {et~al.}(2023)\citenamefont
  {Hashimoto}, \citenamefont {Murata}, \citenamefont {Tanahashi},\ and\
  \citenamefont {Watanabe}}]{Hashimoto:2023swv}%
  \BibitemOpen
  \bibfield  {author} {\bibinfo {author} {\bibfnamefont {K.}~\bibnamefont
  {Hashimoto}}, \bibinfo {author} {\bibfnamefont {K.}~\bibnamefont {Murata}},
  \bibinfo {author} {\bibfnamefont {N.}~\bibnamefont {Tanahashi}}, \ and\
  \bibinfo {author} {\bibfnamefont {R.}~\bibnamefont {Watanabe}},\ }\href
  {\doibase 10.1007/JHEP11(2023)040} {\bibfield  {journal} {\bibinfo  {journal}
  {JHEP}\ }\textbf {\bibinfo {volume} {11}},\ \bibinfo {pages} {040} (\bibinfo
  {year} {2023})},\ \Eprint {http://arxiv.org/abs/2305.16669} {arXiv:2305.16669
  [hep-th]} \BibitemShut {NoStop}%
\bibitem [{\citenamefont {Baggioli}\ \emph {et~al.}(2025)\citenamefont
  {Baggioli}, \citenamefont {Huh}, \citenamefont {Jeong}, \citenamefont {Kim},\
  and\ \citenamefont {Pedraza}}]{Baggioli:2024wbz}%
  \BibitemOpen
  \bibfield  {author} {\bibinfo {author} {\bibfnamefont {M.}~\bibnamefont
  {Baggioli}}, \bibinfo {author} {\bibfnamefont {K.-B.}\ \bibnamefont {Huh}},
  \bibinfo {author} {\bibfnamefont {H.-S.}\ \bibnamefont {Jeong}}, \bibinfo
  {author} {\bibfnamefont {K.-Y.}\ \bibnamefont {Kim}}, \ and\ \bibinfo
  {author} {\bibfnamefont {J.~F.}\ \bibnamefont {Pedraza}},\ }\href {\doibase
  10.1103/PhysRevResearch.7.023028} {\bibfield  {journal} {\bibinfo  {journal}
  {Phys. Rev. Res.}\ }\textbf {\bibinfo {volume} {7}},\ \bibinfo {pages}
  {023028} (\bibinfo {year} {2025})},\ \Eprint
  {http://arxiv.org/abs/2407.17054} {arXiv:2407.17054 [hep-th]} \BibitemShut
  {NoStop}%
\bibitem [{\citenamefont {Caputa}\ \emph {et~al.}(2023)\citenamefont {Caputa},
  \citenamefont {Gupta}, \citenamefont {Haque}, \citenamefont {Liu},
  \citenamefont {Murugan},\ and\ \citenamefont {Van~Zyl}}]{Caputa:2022yju}%
  \BibitemOpen
  \bibfield  {author} {\bibinfo {author} {\bibfnamefont {P.}~\bibnamefont
  {Caputa}}, \bibinfo {author} {\bibfnamefont {N.}~\bibnamefont {Gupta}},
  \bibinfo {author} {\bibfnamefont {S.~S.}\ \bibnamefont {Haque}}, \bibinfo
  {author} {\bibfnamefont {S.}~\bibnamefont {Liu}}, \bibinfo {author}
  {\bibfnamefont {J.}~\bibnamefont {Murugan}}, \ and\ \bibinfo {author}
  {\bibfnamefont {H.~J.~R.}\ \bibnamefont {Van~Zyl}},\ }\href {\doibase
  10.1007/JHEP01(2023)120} {\bibfield  {journal} {\bibinfo  {journal} {JHEP}\
  }\textbf {\bibinfo {volume} {01}},\ \bibinfo {pages} {120} (\bibinfo {year}
  {2023})},\ \Eprint {http://arxiv.org/abs/2208.06311} {arXiv:2208.06311
  [hep-th]} \BibitemShut {NoStop}%
\bibitem [{\citenamefont {Wigner}(1951)}]{Wigner_1951}%
  \BibitemOpen
  \bibfield  {author} {\bibinfo {author} {\bibfnamefont {E.~P.}\ \bibnamefont
  {Wigner}},\ }\href {\doibase 10.1017/s0305004100027237} {\bibfield  {journal}
  {\bibinfo  {journal} {Mathematical Proceedings of the Cambridge Philosophical
  Society}\ }\textbf {\bibinfo {volume} {47}},\ \bibinfo {pages} {790}
  (\bibinfo {year} {1951})}\BibitemShut {NoStop}%
\bibitem [{\citenamefont {Martinez-Azcona}\ \emph {et~al.}(2025)\citenamefont
  {Martinez-Azcona}, \citenamefont {Shir},\ and\ \citenamefont
  {Chenu}}]{PhysRevB.111.165108}%
  \BibitemOpen
  \bibfield  {author} {\bibinfo {author} {\bibfnamefont {P.}~\bibnamefont
  {Martinez-Azcona}}, \bibinfo {author} {\bibfnamefont {R.}~\bibnamefont
  {Shir}}, \ and\ \bibinfo {author} {\bibfnamefont {A.}~\bibnamefont {Chenu}},\
  }\href {\doibase 10.1103/PhysRevB.111.165108} {\bibfield  {journal} {\bibinfo
   {journal} {Phys. Rev. B}\ }\textbf {\bibinfo {volume} {111}},\ \bibinfo
  {pages} {165108} (\bibinfo {year} {2025})}\BibitemShut {NoStop}%
\bibitem [{\citenamefont {Rabinovici}\ \emph {et~al.}(2021)\citenamefont
  {Rabinovici}, \citenamefont {S\'anchez-Garrido}, \citenamefont {Shir},\ and\
  \citenamefont {Sonner}}]{Rabinovici:2020ryf}%
  \BibitemOpen
  \bibfield  {author} {\bibinfo {author} {\bibfnamefont {E.}~\bibnamefont
  {Rabinovici}}, \bibinfo {author} {\bibfnamefont {A.}~\bibnamefont
  {S\'anchez-Garrido}}, \bibinfo {author} {\bibfnamefont {R.}~\bibnamefont
  {Shir}}, \ and\ \bibinfo {author} {\bibfnamefont {J.}~\bibnamefont
  {Sonner}},\ }\href {\doibase 10.1007/JHEP06(2021)062} {\bibfield  {journal}
  {\bibinfo  {journal} {JHEP}\ }\textbf {\bibinfo {volume} {06}},\ \bibinfo
  {pages} {062} (\bibinfo {year} {2021})},\ \Eprint
  {http://arxiv.org/abs/2009.01862} {arXiv:2009.01862 [hep-th]} \BibitemShut
  {NoStop}%
\bibitem [{\citenamefont {Rabinovici}\ \emph {et~al.}(2022)\citenamefont
  {Rabinovici}, \citenamefont {S\'anchez-Garrido}, \citenamefont {Shir},\ and\
  \citenamefont {Sonner}}]{Rabinovici:2022beu}%
  \BibitemOpen
  \bibfield  {author} {\bibinfo {author} {\bibfnamefont {E.}~\bibnamefont
  {Rabinovici}}, \bibinfo {author} {\bibfnamefont {A.}~\bibnamefont
  {S\'anchez-Garrido}}, \bibinfo {author} {\bibfnamefont {R.}~\bibnamefont
  {Shir}}, \ and\ \bibinfo {author} {\bibfnamefont {J.}~\bibnamefont
  {Sonner}},\ }\href {\doibase 10.1007/JHEP07(2022)151} {\bibfield  {journal}
  {\bibinfo  {journal} {JHEP}\ }\textbf {\bibinfo {volume} {07}},\ \bibinfo
  {pages} {151} (\bibinfo {year} {2022})},\ \Eprint
  {http://arxiv.org/abs/2207.07701} {arXiv:2207.07701 [hep-th]} \BibitemShut
  {NoStop}%
\bibitem [{\citenamefont {Sinai}(1970)}]{Sinai1970}%
  \BibitemOpen
  \bibfield  {author} {\bibinfo {author} {\bibfnamefont {Y.~G.}\ \bibnamefont
  {Sinai}},\ }\href {\doibase 10.1070/RM1970v025n02ABEH003794} {\bibfield
  {journal} {\bibinfo  {journal} {Russian Mathematical Surveys}\ }\textbf
  {\bibinfo {volume} {25}},\ \bibinfo {pages} {137} (\bibinfo {year}
  {1970})}\BibitemShut {NoStop}%
\bibitem [{\citenamefont {Bunimovich}(1974)}]{Bunimovich1974}%
  \BibitemOpen
  \bibfield  {author} {\bibinfo {author} {\bibfnamefont {L.~A.}\ \bibnamefont
  {Bunimovich}},\ }\href {\doibase 10.1007/BF01075700} {\bibfield  {journal}
  {\bibinfo  {journal} {Functional Analysis and Its Applications}\ }\textbf
  {\bibinfo {volume} {8}},\ \bibinfo {pages} {254} (\bibinfo {year}
  {1974})}\BibitemShut {NoStop}%
\bibitem [{\citenamefont {Bunimovich}(1979)}]{Bunimovich1979}%
  \BibitemOpen
  \bibfield  {author} {\bibinfo {author} {\bibfnamefont {L.~A.}\ \bibnamefont
  {Bunimovich}},\ }\href {\doibase 10.1007/BF01197884} {\bibfield  {journal}
  {\bibinfo  {journal} {Communications in Mathematical Physics}\ }\textbf
  {\bibinfo {volume} {65}},\ \bibinfo {pages} {295} (\bibinfo {year}
  {1979})}\BibitemShut {NoStop}%
\bibitem [{\citenamefont {Bunimovich}\ \emph {et~al.}(1991)\citenamefont
  {Bunimovich}, \citenamefont {Sinai},\ and\ \citenamefont
  {Chernov}}]{BunSinChe91}%
  \BibitemOpen
  \bibfield  {author} {\bibinfo {author} {\bibfnamefont {L.~A.}\ \bibnamefont
  {Bunimovich}}, \bibinfo {author} {\bibfnamefont {Y.~G.}\ \bibnamefont
  {Sinai}}, \ and\ \bibinfo {author} {\bibfnamefont {N.~I.}\ \bibnamefont
  {Chernov}},\ }\href {\doibase 10.1070/RM1991v046n04ABEH002827} {\bibfield
  {journal} {\bibinfo  {journal} {Russian Mathematical Surveys}\ }\textbf
  {\bibinfo {volume} {46}},\ \bibinfo {pages} {47} (\bibinfo {year}
  {1991})}\BibitemShut {NoStop}%
\bibitem [{\citenamefont {Benettin}\ and\ \citenamefont
  {Strelcyn}(1978)}]{PhysRevA.17.773}%
  \BibitemOpen
  \bibfield  {author} {\bibinfo {author} {\bibfnamefont {G.}~\bibnamefont
  {Benettin}}\ and\ \bibinfo {author} {\bibfnamefont {J.~M.}\ \bibnamefont
  {Strelcyn}},\ }\href {\doibase 10.1103/PhysRevA.17.773} {\bibfield  {journal}
  {\bibinfo  {journal} {Phys. Rev. A}\ }\textbf {\bibinfo {volume} {17}},\
  \bibinfo {pages} {773} (\bibinfo {year} {1978})}\BibitemShut {NoStop}%
\bibitem [{\citenamefont {Camargo}\ \emph
  {et~al.}(2024{\natexlab{a}})\citenamefont {Camargo}, \citenamefont {Jahnke},
  \citenamefont {Jeong}, \citenamefont {Kim},\ and\ \citenamefont
  {Nishida}}]{Camargo:2023eev}%
  \BibitemOpen
  \bibfield  {author} {\bibinfo {author} {\bibfnamefont {H.~A.}\ \bibnamefont
  {Camargo}}, \bibinfo {author} {\bibfnamefont {V.}~\bibnamefont {Jahnke}},
  \bibinfo {author} {\bibfnamefont {H.-S.}\ \bibnamefont {Jeong}}, \bibinfo
  {author} {\bibfnamefont {K.-Y.}\ \bibnamefont {Kim}}, \ and\ \bibinfo
  {author} {\bibfnamefont {M.}~\bibnamefont {Nishida}},\ }\href {\doibase
  10.1103/PhysRevD.109.046017} {\bibfield  {journal} {\bibinfo  {journal}
  {Phys. Rev. D}\ }\textbf {\bibinfo {volume} {109}},\ \bibinfo {pages}
  {046017} (\bibinfo {year} {2024}{\natexlab{a}})},\ \Eprint
  {http://arxiv.org/abs/2306.11632} {arXiv:2306.11632 [hep-th]} \BibitemShut
  {NoStop}%
\bibitem [{\citenamefont {Bogomolny}\ and\ \citenamefont
  {Schmit}(2004)}]{PhysRevLett.92.244102}%
  \BibitemOpen
  \bibfield  {author} {\bibinfo {author} {\bibfnamefont {E.}~\bibnamefont
  {Bogomolny}}\ and\ \bibinfo {author} {\bibfnamefont {C.}~\bibnamefont
  {Schmit}},\ }\href {\doibase 10.1103/PhysRevLett.92.244102} {\bibfield
  {journal} {\bibinfo  {journal} {Phys. Rev. Lett.}\ }\textbf {\bibinfo
  {volume} {92}},\ \bibinfo {pages} {244102} (\bibinfo {year}
  {2004})}\BibitemShut {NoStop}%
\bibitem [{\citenamefont {Balasubramanian}\ \emph {et~al.}(2025)\citenamefont
  {Balasubramanian}, \citenamefont {Das}, \citenamefont {Erdmenger},\ and\
  \citenamefont {Xian}}]{Balasubramanian:2024ghv}%
  \BibitemOpen
  \bibfield  {author} {\bibinfo {author} {\bibfnamefont {V.}~\bibnamefont
  {Balasubramanian}}, \bibinfo {author} {\bibfnamefont {R.~N.}\ \bibnamefont
  {Das}}, \bibinfo {author} {\bibfnamefont {J.}~\bibnamefont {Erdmenger}}, \
  and\ \bibinfo {author} {\bibfnamefont {Z.-Y.}\ \bibnamefont {Xian}},\ }\href
  {\doibase 10.1088/1742-5468/adba41} {\bibfield  {journal} {\bibinfo
  {journal} {J. Stat. Mech.}\ }\textbf {\bibinfo {volume} {2025}},\ \bibinfo
  {pages} {033202} (\bibinfo {year} {2025})},\ \Eprint
  {http://arxiv.org/abs/2407.11114} {arXiv:2407.11114 [hep-th]} \BibitemShut
  {NoStop}%
\bibitem [{\citenamefont {Zemlyakov}\ and\ \citenamefont
  {Katok}(1975)}]{Zemlyakov:1975}%
  \BibitemOpen
  \bibfield  {author} {\bibinfo {author} {\bibfnamefont {A.~N.}\ \bibnamefont
  {Zemlyakov}}\ and\ \bibinfo {author} {\bibfnamefont {A.~B.}\ \bibnamefont
  {Katok}},\ }\href {\doibase 10.1007/BF01818045} {\bibfield  {journal}
  {\bibinfo  {journal} {Math. Notes}\ }\textbf {\bibinfo {volume} {18}},\
  \bibinfo {pages} {760} (\bibinfo {year} {1975})}\BibitemShut {NoStop}%
\bibitem [{\citenamefont {Caputa}\ \emph
  {et~al.}(2026{\natexlab{a}})\citenamefont {Caputa}, \citenamefont
  {Di~Giulio},\ and\ \citenamefont {Loc}}]{Caputa:2025ozd}%
  \BibitemOpen
  \bibfield  {author} {\bibinfo {author} {\bibfnamefont {P.}~\bibnamefont
  {Caputa}}, \bibinfo {author} {\bibfnamefont {G.}~\bibnamefont {Di~Giulio}}, \
  and\ \bibinfo {author} {\bibfnamefont {T.~Q.}\ \bibnamefont {Loc}},\ }\href
  {\doibase 10.1007/JHEP02(2026)189} {\bibfield  {journal} {\bibinfo  {journal}
  {JHEP}\ }\textbf {\bibinfo {volume} {02}},\ \bibinfo {pages} {189} (\bibinfo
  {year} {2026}{\natexlab{a}})},\ \Eprint {http://arxiv.org/abs/2509.12992}
  {arXiv:2509.12992 [hep-th]} \BibitemShut {NoStop}%
\bibitem [{\citenamefont {Das}\ \emph {et~al.}(2026)\citenamefont {Das},
  \citenamefont {Das}, \citenamefont {Pedraza},\ and\ \citenamefont
  {Qu}}]{Das:2026gko}%
  \BibitemOpen
  \bibfield  {author} {\bibinfo {author} {\bibfnamefont {J.}~\bibnamefont
  {Das}}, \bibinfo {author} {\bibfnamefont {S.}~\bibnamefont {Das}}, \bibinfo
  {author} {\bibfnamefont {J.~F.}\ \bibnamefont {Pedraza}}, \ and\ \bibinfo
  {author} {\bibfnamefont {L.-C.}\ \bibnamefont {Qu}},\ }\href@noop {} {\
  (\bibinfo {year} {2026})},\ \Eprint {http://arxiv.org/abs/2608.19346}
  {arXiv:2608.19346 [hep-th]} \BibitemShut {NoStop}%
\bibitem [{\citenamefont {Samaj}\ and\ \citenamefont
  {Bajnok}(2013)}]{Samaj:2013yva}%
  \BibitemOpen
  \bibfield  {author} {\bibinfo {author} {\bibfnamefont {L.}~\bibnamefont
  {Samaj}}\ and\ \bibinfo {author} {\bibfnamefont {Z.}~\bibnamefont {Bajnok}},\
  }\href@noop {} {\emph {\bibinfo {title} {{Introduction to the statistical
  physics of integrable many-body systems}}}}\ (\bibinfo  {publisher}
  {Cambridge University Press},\ \bibinfo {address} {Cambridge},\ \bibinfo
  {year} {2013})\BibitemShut {NoStop}%
\bibitem [{\citenamefont {Gubin}\ and\ \citenamefont
  {F.~Santos}(2012)}]{Gubin_2012}%
  \BibitemOpen
  \bibfield  {author} {\bibinfo {author} {\bibfnamefont {A.}~\bibnamefont
  {Gubin}}\ and\ \bibinfo {author} {\bibfnamefont {L.}~\bibnamefont
  {F.~Santos}},\ }\href {\doibase 10.1119/1.3671068} {\bibfield  {journal}
  {\bibinfo  {journal} {American Journal of Physics}\ }\textbf {\bibinfo
  {volume} {80}},\ \bibinfo {pages} {246{\^a}} (\bibinfo {year}
  {2012})}\BibitemShut {NoStop}%
\bibitem [{\citenamefont {Camargo}\ \emph
  {et~al.}(2024{\natexlab{b}})\citenamefont {Camargo}, \citenamefont {Huh},
  \citenamefont {Jahnke}, \citenamefont {Jeong}, \citenamefont {Kim},\ and\
  \citenamefont {Nishida}}]{Camargo:2024deu}%
  \BibitemOpen
  \bibfield  {author} {\bibinfo {author} {\bibfnamefont {H.~A.}\ \bibnamefont
  {Camargo}}, \bibinfo {author} {\bibfnamefont {K.-B.}\ \bibnamefont {Huh}},
  \bibinfo {author} {\bibfnamefont {V.}~\bibnamefont {Jahnke}}, \bibinfo
  {author} {\bibfnamefont {H.-S.}\ \bibnamefont {Jeong}}, \bibinfo {author}
  {\bibfnamefont {K.-Y.}\ \bibnamefont {Kim}}, \ and\ \bibinfo {author}
  {\bibfnamefont {M.}~\bibnamefont {Nishida}},\ }\href {\doibase
  10.1007/JHEP08(2024)241} {\bibfield  {journal} {\bibinfo  {journal} {JHEP}\
  }\textbf {\bibinfo {volume} {08}},\ \bibinfo {pages} {241} (\bibinfo {year}
  {2024}{\natexlab{b}})},\ \Eprint {http://arxiv.org/abs/2405.11254}
  {arXiv:2405.11254 [hep-th]} \BibitemShut {NoStop}%
\bibitem [{Note2()}]{Note2}%
  \BibitemOpen
  \bibinfo {note} {While the normalized saturation value $1/2$ is accurately
  recovered in the chaotic XXZ chain, the integrable case exhibits small
  fluctuations around the plateau due to spectral degeneracies~\cite
  {Camargo:2024deu}.}\BibitemShut {Stop}%
\bibitem [{\citenamefont {Caputa}\ \emph
  {et~al.}(2026{\natexlab{b}})\citenamefont {Caputa}, \citenamefont {Creed},
  \citenamefont {Das}, \citenamefont {Demulder},\ and\ \citenamefont
  {McLoughlin}}]{Caputa:2026hvh}%
  \BibitemOpen
  \bibfield  {author} {\bibinfo {author} {\bibfnamefont {P.}~\bibnamefont
  {Caputa}}, \bibinfo {author} {\bibfnamefont {B.}~\bibnamefont {Creed}},
  \bibinfo {author} {\bibfnamefont {R.~N.}\ \bibnamefont {Das}}, \bibinfo
  {author} {\bibfnamefont {S.}~\bibnamefont {Demulder}}, \ and\ \bibinfo
  {author} {\bibfnamefont {T.}~\bibnamefont {McLoughlin}},\ }\href@noop {} {\
  (\bibinfo {year} {2026}{\natexlab{b}})},\ \Eprint
  {http://arxiv.org/abs/2606.18351} {arXiv:2606.18351 [hep-th]} \BibitemShut
  {NoStop}%
\bibitem [{\citenamefont {Sachdev}\ and\ \citenamefont
  {Ye}(1993)}]{Sachdev:1992fk}%
  \BibitemOpen
  \bibfield  {author} {\bibinfo {author} {\bibfnamefont {S.}~\bibnamefont
  {Sachdev}}\ and\ \bibinfo {author} {\bibfnamefont {J.}~\bibnamefont {Ye}},\
  }\href {\doibase 10.1103/PhysRevLett.70.3339} {\bibfield  {journal} {\bibinfo
   {journal} {Phys. Rev. Lett.}\ }\textbf {\bibinfo {volume} {70}},\ \bibinfo
  {pages} {3339} (\bibinfo {year} {1993})},\ \Eprint
  {http://arxiv.org/abs/cond-mat/9212030} {arXiv:cond-mat/9212030} \BibitemShut
  {NoStop}%
\bibitem [{\citenamefont {Chowdhury}\ \emph {et~al.}(2022)\citenamefont
  {Chowdhury}, \citenamefont {Georges}, \citenamefont {Parcollet},\ and\
  \citenamefont {Sachdev}}]{RevModPhys.94.035004}%
  \BibitemOpen
  \bibfield  {author} {\bibinfo {author} {\bibfnamefont {D.}~\bibnamefont
  {Chowdhury}}, \bibinfo {author} {\bibfnamefont {A.}~\bibnamefont {Georges}},
  \bibinfo {author} {\bibfnamefont {O.}~\bibnamefont {Parcollet}}, \ and\
  \bibinfo {author} {\bibfnamefont {S.}~\bibnamefont {Sachdev}},\ }\href
  {\doibase 10.1103/RevModPhys.94.035004} {\bibfield  {journal} {\bibinfo
  {journal} {Rev. Mod. Phys.}\ }\textbf {\bibinfo {volume} {94}},\ \bibinfo
  {pages} {035004} (\bibinfo {year} {2022})}\BibitemShut {NoStop}%
\bibitem [{\citenamefont {Huh}\ \emph {et~al.}(2025)\citenamefont {Huh},
  \citenamefont {Jeong}, \citenamefont {Pando~Zayas},\ and\ \citenamefont
  {Pedraza}}]{Huh:2024ytz}%
  \BibitemOpen
  \bibfield  {author} {\bibinfo {author} {\bibfnamefont {K.-B.}\ \bibnamefont
  {Huh}}, \bibinfo {author} {\bibfnamefont {H.-S.}\ \bibnamefont {Jeong}},
  \bibinfo {author} {\bibfnamefont {L.~A.}\ \bibnamefont {Pando~Zayas}}, \ and\
  \bibinfo {author} {\bibfnamefont {J.~F.}\ \bibnamefont {Pedraza}},\ }\href
  {\doibase 10.1103/gmy7-dn7l} {\bibfield  {journal} {\bibinfo  {journal}
  {Phys. Rev. D}\ }\textbf {\bibinfo {volume} {111}},\ \bibinfo {pages}
  {L121902} (\bibinfo {year} {2025})},\ \Eprint
  {http://arxiv.org/abs/2412.04963} {arXiv:2412.04963 [hep-th]} \BibitemShut
  {NoStop}%
\bibitem [{\citenamefont {Kanazawa}\ and\ \citenamefont
  {Wettig}(2017)}]{Kanazawa:2017dpd}%
  \BibitemOpen
  \bibfield  {author} {\bibinfo {author} {\bibfnamefont {T.}~\bibnamefont
  {Kanazawa}}\ and\ \bibinfo {author} {\bibfnamefont {T.}~\bibnamefont
  {Wettig}},\ }\href {\doibase 10.1007/JHEP09(2017)050} {\bibfield  {journal}
  {\bibinfo  {journal} {JHEP}\ }\textbf {\bibinfo {volume} {09}},\ \bibinfo
  {pages} {050} (\bibinfo {year} {2017})},\ \Eprint
  {http://arxiv.org/abs/1706.03044} {arXiv:1706.03044 [hep-th]} \BibitemShut
  {NoStop}%
\bibitem [{\citenamefont {Garc{\'\i}a-Garc{\'\i}a}\ and\ \citenamefont
  {Verbaarschot}(2016)}]{Garcia-Garcia:2016mno}%
  \BibitemOpen
  \bibfield  {author} {\bibinfo {author} {\bibfnamefont {A.~M.}\ \bibnamefont
  {Garc{\'\i}a-Garc{\'\i}a}}\ and\ \bibinfo {author} {\bibfnamefont {J.~J.~M.}\
  \bibnamefont {Verbaarschot}},\ }\href {\doibase 10.1103/PhysRevD.94.126010}
  {\bibfield  {journal} {\bibinfo  {journal} {Phys. Rev. D}\ }\textbf {\bibinfo
  {volume} {94}},\ \bibinfo {pages} {126010} (\bibinfo {year} {2016})},\
  \Eprint {http://arxiv.org/abs/1610.03816} {arXiv:1610.03816 [hep-th]}
  \BibitemShut {NoStop}%
\bibitem [{\citenamefont {M{\"u}ck}\ and\ \citenamefont
  {Yang}(2022)}]{Muck:2022xfc}%
  \BibitemOpen
  \bibfield  {author} {\bibinfo {author} {\bibfnamefont {W.}~\bibnamefont
  {M{\"u}ck}}\ and\ \bibinfo {author} {\bibfnamefont {Y.}~\bibnamefont
  {Yang}},\ }\href {\doibase 10.1016/j.nuclphysb.2022.115948} {\bibfield
  {journal} {\bibinfo  {journal} {Nucl. Phys. B}\ }\textbf {\bibinfo {volume}
  {984}},\ \bibinfo {pages} {115948} (\bibinfo {year} {2022})},\ \Eprint
  {http://arxiv.org/abs/2205.12815} {arXiv:2205.12815 [hep-th]} \BibitemShut
  {NoStop}%
\bibitem [{\citenamefont {Balasubramanian}\ \emph
  {et~al.}(2026{\natexlab{a}})\citenamefont {Balasubramanian}, \citenamefont
  {Das}, \citenamefont {Erdmenger}, \citenamefont {Karl},\ and\ \citenamefont
  {Verlinde}}]{Balasubramanian:2026azk}%
  \BibitemOpen
  \bibfield  {author} {\bibinfo {author} {\bibfnamefont {V.}~\bibnamefont
  {Balasubramanian}}, \bibinfo {author} {\bibfnamefont {R.~N.}\ \bibnamefont
  {Das}}, \bibinfo {author} {\bibfnamefont {J.}~\bibnamefont {Erdmenger}},
  \bibinfo {author} {\bibfnamefont {J.}~\bibnamefont {Karl}}, \ and\ \bibinfo
  {author} {\bibfnamefont {H.}~\bibnamefont {Verlinde}},\ }\href@noop {} {\
  (\bibinfo {year} {2026}{\natexlab{a}})},\ \Eprint
  {http://arxiv.org/abs/2602.02645} {arXiv:2602.02645 [hep-th]} \BibitemShut
  {NoStop}%
\bibitem [{\citenamefont {Balasubramanian}\ \emph
  {et~al.}(2026{\natexlab{b}})\citenamefont {Balasubramanian}, \citenamefont
  {Caputa},\ and\ \citenamefont {Sim{\'o}n}}]{Balasubramanian:2025xkj}%
  \BibitemOpen
  \bibfield  {author} {\bibinfo {author} {\bibfnamefont {V.}~\bibnamefont
  {Balasubramanian}}, \bibinfo {author} {\bibfnamefont {P.}~\bibnamefont
  {Caputa}}, \ and\ \bibinfo {author} {\bibfnamefont {J.}~\bibnamefont
  {Sim{\'o}n}},\ }\href {\doibase 10.1007/JHEP04(2026)172} {\bibfield
  {journal} {\bibinfo  {journal} {JHEP}\ }\textbf {\bibinfo {volume} {04}},\
  \bibinfo {pages} {172} (\bibinfo {year} {2026}{\natexlab{b}})},\ \Eprint
  {http://arxiv.org/abs/2511.03775} {arXiv:2511.03775 [hep-th]} \BibitemShut
  {NoStop}%
\bibitem [{\citenamefont {Favard}(1935)}]{favard1935polynomes}%
  \BibitemOpen
  \bibfield  {author} {\bibinfo {author} {\bibfnamefont {J.}~\bibnamefont
  {Favard}},\ }\href {https://gallica.bnf.fr/ark:/12148/bpt6k31525/f2052.item}
  {\bibfield  {journal} {\bibinfo  {journal} {C. R. Acad. Sci. Paris}\ }\textbf
  {\bibinfo {volume} {200}},\ \bibinfo {pages} {2052} (\bibinfo {year}
  {1935})}\BibitemShut {NoStop}%
\bibitem [{\citenamefont {Shohat}(1938)}]{shohat1938polynomes}%
  \BibitemOpen
  \bibfield  {author} {\bibinfo {author} {\bibfnamefont {J.}~\bibnamefont
  {Shohat}},\ }\href {https://zbmath.org/0019.40503} {\bibfield  {journal}
  {\bibinfo  {journal} {C. R. Acad. Sci. Paris}\ }\textbf {\bibinfo {volume}
  {207}},\ \bibinfo {pages} {556} (\bibinfo {year} {1938})}\BibitemShut
  {NoStop}%
\bibitem [{\citenamefont {Apostol}(1976)}]{apostol1976introduction}%
  \BibitemOpen
  \bibfield  {author} {\bibinfo {author} {\bibfnamefont {T.~M.}\ \bibnamefont
  {Apostol}},\ }\href {\doibase 10.1007/978-1-4757-5579-4} {\emph {\bibinfo
  {title} {Introduction to Analytic Number Theory}}},\ Undergraduate Texts in
  Mathematics\ (\bibinfo  {publisher} {Springer-Verlag},\ \bibinfo {address}
  {New York},\ \bibinfo {year} {1976})\BibitemShut {NoStop}%
\bibitem [{\citenamefont {Christoffel}(1858)}]{christoffel1858ueber}%
  \BibitemOpen
  \bibfield  {author} {\bibinfo {author} {\bibfnamefont {E.~B.}\ \bibnamefont
  {Christoffel}},\ }\href {\doibase 10.1515/crll.1858.55.61} {\bibfield
  {journal} {\bibinfo  {journal} {J. Reine Angew. Math.}\ }\textbf {\bibinfo
  {volume} {1858}},\ \bibinfo {pages} {61} (\bibinfo {year}
  {1858})}\BibitemShut {NoStop}%
\bibitem [{\citenamefont {Darboux}(1878)}]{darboux1878memoire}%
  \BibitemOpen
  \bibfield  {author} {\bibinfo {author} {\bibfnamefont {G.}~\bibnamefont
  {Darboux}},\ }\href {http://www.numdam.org/item/JMPA_1878_3_4__5_0/}
  {\bibfield  {journal} {\bibinfo  {journal} {J. Math. Pures Appl.}\ }\bibinfo
  {series} {3},\ \textbf {\bibinfo {volume} {4}},\ \bibinfo {pages} {5}
  (\bibinfo {year} {1878})}\BibitemShut {NoStop}%
\bibitem [{\citenamefont {Szeg{\H{o}}}(1939)}]{szego1939orthogonal}%
  \BibitemOpen
  \bibfield  {author} {\bibinfo {author} {\bibfnamefont {G.}~\bibnamefont
  {Szeg{\H{o}}}},\ }\href {https://bookstore.ams.org/coll-23} {\emph {\bibinfo
  {title} {Orthogonal Polynomials}}},\ \bibinfo {series} {Colloquium
  Publications}, Vol.~\bibinfo {volume} {23}\ (\bibinfo  {publisher} {American
  Mathematical Society},\ \bibinfo {address} {Providence, RI},\ \bibinfo {year}
  {1939})\BibitemShut {NoStop}%
\bibitem [{\citenamefont {Qu}(2026{\natexlab{a}})}]{Qu:2025lgo}%
  \BibitemOpen
  \bibfield  {author} {\bibinfo {author} {\bibfnamefont {L.-C.}\ \bibnamefont
  {Qu}},\ }\href {\doibase 10.1007/JHEP05(2026)225} {\bibfield  {journal}
  {\bibinfo  {journal} {JHEP}\ }\textbf {\bibinfo {volume} {05}},\ \bibinfo
  {pages} {225} (\bibinfo {year} {2026}{\natexlab{a}})},\ \Eprint
  {http://arxiv.org/abs/2512.15857} {arXiv:2512.15857 [hep-th]} \BibitemShut
  {NoStop}%
\bibitem [{\citenamefont {Pedraza}\ and\ \citenamefont
  {Qu}(2026)}]{Pedraza:2026zji}%
  \BibitemOpen
  \bibfield  {author} {\bibinfo {author} {\bibfnamefont {J.~F.}\ \bibnamefont
  {Pedraza}}\ and\ \bibinfo {author} {\bibfnamefont {L.-C.}\ \bibnamefont
  {Qu}},\ }\href@noop {} {\  (\bibinfo {year} {2026})},\ \Eprint
  {http://arxiv.org/abs/2608.10072} {arXiv:2608.10072 [hep-th]} \BibitemShut
  {NoStop}%
\bibitem [{\citenamefont {Qu}(2026{\natexlab{b}})}]{Qu:2026dmv}%
  \BibitemOpen
  \bibfield  {author} {\bibinfo {author} {\bibfnamefont {L.-C.}\ \bibnamefont
  {Qu}},\ }\href@noop {} {\  (\bibinfo {year} {2026}{\natexlab{b}})},\ \Eprint
  {http://arxiv.org/abs/2608.16982} {arXiv:2608.16982 [hep-th]} \BibitemShut
  {NoStop}%
\bibitem [{\citenamefont {Aguilar-Gutierrez}\ \emph {et~al.}(2026)\citenamefont
  {Aguilar-Gutierrez}, \citenamefont {Das}, \citenamefont {Erdmenger},\ and\
  \citenamefont {Xian}}]{Aguilar-Gutierrez:2026jjv}%
  \BibitemOpen
  \bibfield  {author} {\bibinfo {author} {\bibfnamefont {S.~E.}\ \bibnamefont
  {Aguilar-Gutierrez}}, \bibinfo {author} {\bibfnamefont {R.~N.}\ \bibnamefont
  {Das}}, \bibinfo {author} {\bibfnamefont {J.}~\bibnamefont {Erdmenger}}, \
  and\ \bibinfo {author} {\bibfnamefont {Z.-Y.}\ \bibnamefont {Xian}},\ }\href
  {\doibase 10.1007/JHEP06(2026)109} {\bibfield  {journal} {\bibinfo  {journal}
  {JHEP}\ }\textbf {\bibinfo {volume} {06}},\ \bibinfo {pages} {109} (\bibinfo
  {year} {2026})},\ \Eprint {http://arxiv.org/abs/2601.09801} {arXiv:2601.09801
  [hep-th]} \BibitemShut {NoStop}%
\bibitem [{\citenamefont {Bhattacharya}\ \emph {et~al.}(2023)\citenamefont
  {Bhattacharya}, \citenamefont {Das}, \citenamefont {Dey},\ and\ \citenamefont
  {Erdmenger}}]{Bhattacharya:2023yec}%
  \BibitemOpen
  \bibfield  {author} {\bibinfo {author} {\bibfnamefont {A.}~\bibnamefont
  {Bhattacharya}}, \bibinfo {author} {\bibfnamefont {R.~N.}\ \bibnamefont
  {Das}}, \bibinfo {author} {\bibfnamefont {B.}~\bibnamefont {Dey}}, \ and\
  \bibinfo {author} {\bibfnamefont {J.}~\bibnamefont {Erdmenger}},\ }\href@noop
  {} {\  (\bibinfo {year} {2023})},\ \Eprint {http://arxiv.org/abs/2312.11635}
  {arXiv:2312.11635 [hep-th]} \BibitemShut {NoStop}%
\bibitem [{\citenamefont {Nandy}\ \emph
  {et~al.}(2025{\natexlab{b}})\citenamefont {Nandy}, \citenamefont {Pathak},
  \citenamefont {Xian},\ and\ \citenamefont {Erdmenger}}]{Nandy:2024mml}%
  \BibitemOpen
  \bibfield  {author} {\bibinfo {author} {\bibfnamefont {P.}~\bibnamefont
  {Nandy}}, \bibinfo {author} {\bibfnamefont {T.}~\bibnamefont {Pathak}},
  \bibinfo {author} {\bibfnamefont {Z.-Y.}\ \bibnamefont {Xian}}, \ and\
  \bibinfo {author} {\bibfnamefont {J.}~\bibnamefont {Erdmenger}},\ }\href
  {\doibase 10.1103/PhysRevB.111.064203} {\bibfield  {journal} {\bibinfo
  {journal} {Phys. Rev. B}\ }\textbf {\bibinfo {volume} {111}},\ \bibinfo
  {pages} {064203} (\bibinfo {year} {2025}{\natexlab{b}})},\ \Eprint
  {http://arxiv.org/abs/2411.09309} {arXiv:2411.09309 [quant-ph]} \BibitemShut
  {NoStop}%
\bibitem [{\citenamefont {den Brinker}(2021)}]{denBrinker2020Controlled}%
  \BibitemOpen
  \bibfield  {author} {\bibinfo {author} {\bibfnamefont {A.~C.}\ \bibnamefont
  {den Brinker}},\ }in\ \href {\doibase 10.23919/Eusipco47968.2020.9287544}
  {\emph {\bibinfo {booktitle} {2020 28th European Signal Processing Conference
  (EUSIPCO)}}}\ (\bibinfo {year} {2021})\ pp.\ \bibinfo {pages}
  {2279--2283}\BibitemShut {NoStop}%
\bibitem [{\citenamefont {Guhr}\ \emph
  {et~al.}(1999{\natexlab{b}})\citenamefont {Guhr}, \citenamefont {Ma},
  \citenamefont {Meyer},\ and\ \citenamefont {Walke}}]{Guhr1999Semiclassical}%
  \BibitemOpen
  \bibfield  {author} {\bibinfo {author} {\bibfnamefont {T.}~\bibnamefont
  {Guhr}}, \bibinfo {author} {\bibfnamefont {J.-Z.}\ \bibnamefont {Ma}},
  \bibinfo {author} {\bibfnamefont {S.}~\bibnamefont {Meyer}}, \ and\ \bibinfo
  {author} {\bibfnamefont {T.}~\bibnamefont {Walke}},\ }\href {\doibase
  10.1103/PhysRevD.59.054501} {\bibfield  {journal} {\bibinfo  {journal}
  {Physical Review D}\ }\textbf {\bibinfo {volume} {59}},\ \bibinfo {pages}
  {054501} (\bibinfo {year} {1999}{\natexlab{b}})}\BibitemShut {NoStop}%
\bibitem [{\citenamefont {G{\'o}mez}\ \emph
  {et~al.}(2002{\natexlab{b}})\citenamefont {G{\'o}mez}, \citenamefont
  {Molina}, \citenamefont {Rela{\~n}o},\ and\ \citenamefont
  {Retamosa}}]{Gomez2002Misleading}%
  \BibitemOpen
  \bibfield  {author} {\bibinfo {author} {\bibfnamefont {J.~M.~G.}\
  \bibnamefont {G{\'o}mez}}, \bibinfo {author} {\bibfnamefont {R.~A.}\
  \bibnamefont {Molina}}, \bibinfo {author} {\bibfnamefont {A.}~\bibnamefont
  {Rela{\~n}o}}, \ and\ \bibinfo {author} {\bibfnamefont {J.}~\bibnamefont
  {Retamosa}},\ }\href {\doibase 10.1103/PhysRevE.66.036209} {\bibfield
  {journal} {\bibinfo  {journal} {Physical Review E}\ }\textbf {\bibinfo
  {volume} {66}},\ \bibinfo {pages} {036209} (\bibinfo {year}
  {2002}{\natexlab{b}})}\BibitemShut {NoStop}%
\bibitem [{\citenamefont {Morales}\ \emph
  {et~al.}(2011{\natexlab{c}})\citenamefont {Morales}, \citenamefont {Landa},
  \citenamefont {Str{\'a}nsk{\'y}},\ and\ \citenamefont
  {Frank}}]{Morales2011Improved}%
  \BibitemOpen
  \bibfield  {author} {\bibinfo {author} {\bibfnamefont {I.~O.}\ \bibnamefont
  {Morales}}, \bibinfo {author} {\bibfnamefont {E.}~\bibnamefont {Landa}},
  \bibinfo {author} {\bibfnamefont {P.}~\bibnamefont {Str{\'a}nsk{\'y}}}, \
  and\ \bibinfo {author} {\bibfnamefont {A.}~\bibnamefont {Frank}},\ }\href
  {\doibase 10.1103/PhysRevE.84.016203} {\bibfield  {journal} {\bibinfo
  {journal} {Physical Review E}\ }\textbf {\bibinfo {volume} {84}},\ \bibinfo
  {pages} {016203} (\bibinfo {year} {2011}{\natexlab{c}})}\BibitemShut
  {NoStop}%
\bibitem [{\citenamefont {Abul-Magd}\ and\ \citenamefont
  {Abul-Magd}(2014)}]{ABULMAGD2014185}%
  \BibitemOpen
  \bibfield  {author} {\bibinfo {author} {\bibfnamefont {A.~A.}\ \bibnamefont
  {Abul-Magd}}\ and\ \bibinfo {author} {\bibfnamefont {A.~Y.}\ \bibnamefont
  {Abul-Magd}},\ }\href {\doibase https://doi.org/10.1016/j.physa.2013.11.012}
  {\bibfield  {journal} {\bibinfo  {journal} {Physica A: Statistical Mechanics
  and its Applications}\ }\textbf {\bibinfo {volume} {396}},\ \bibinfo {pages}
  {185} (\bibinfo {year} {2014})}\BibitemShut {NoStop}%
\bibitem [{\citenamefont {Timberlake}(2006)}]{10.1119/1.2198883}%
  \BibitemOpen
  \bibfield  {author} {\bibinfo {author} {\bibfnamefont {T.}~\bibnamefont
  {Timberlake}},\ }\href {\doibase 10.1119/1.2198883} {\bibfield  {journal}
  {\bibinfo  {journal} {American Journal of Physics}\ }\textbf {\bibinfo
  {volume} {74}},\ \bibinfo {pages} {547} (\bibinfo {year} {2006})},\ \Eprint
  {http://arxiv.org/abs/https://pubs.aip.org/aapt/ajp/article-pdf/74/6/547/13098337/547\_1\_online.pdf}
  {https://pubs.aip.org/aapt/ajp/article-pdf/74/6/547/13098337/547\_1\_online.pdf}
  \BibitemShut {NoStop}%
\end{thebibliography}%
\end{document}